\documentclass[%
 aip,
 amsmath,amssymb,
 reprint,%
 citeautoscript,
]{revtex4-2}
\usepackage{amsmath}
\usepackage{amsthm}
\usepackage{amssymb}
\usepackage{mathrsfs}
\usepackage{xcolor}
\usepackage{chemformula}
\usepackage{multirow}
\usepackage{graphicx} 

\newcommand{\ourmethod}{\textbf{CHEM}}
\newcommand{\ourmethodU}{U_{\textrm{\textbf{CHEM}}}}
\usepackage{physics}
\usepackage{braket}
\definecolor{mygreen}{RGB}{0,168,0}
\newcommand\green[1]{{\color{mygreen} #1}}

\usepackage[ruled, linesnumbered]{algorithm2e}

\begin{document}

\title{Towards chemical accuracy with shallow quantum circuits: A Clifford-based Hamiltonian engineering approach}
\author{Jiace Sun}
\affiliation{Division of Chemistry and Chemical Engineering, California Institute of Technology, Pasadena, CA 91125, USA}
\author{Lixue Cheng}
\affiliation{Division of Chemistry and Chemical Engineering, California Institute of Technology, Pasadena, CA 91125, USA}
\affiliation{Microsoft Research AI4Science Lab, Berlin 10178, Germany}
\author{Weitang Li}
\email{liw31@gmail.com}
\affiliation{Tencent Quantum Lab, Shenzhen, 518057, China}

\begin{abstract}
    Achieving chemical accuracy with shallow quantum circuits is a significant challenge in quantum computational chemistry, particularly for near-term quantum devices. In this work, we present a Clifford-based Hamiltonian engineering algorithm, namely \ourmethod{}, that addresses the trade-off between circuit depth and accuracy. Based on variational quantum eigensolver and hardware-efficient ansatz, our method designs Clifford-based Hamiltonian transformation that (1) ensures a set of initial circuit parameters corresponding to the Hartree--Fock energy can be generated, (2) effectively maximizes the initial energy gradient with respect to circuit parameters, (3) imposes negligible overhead for classical processing and does not require additional quantum resources, and (4) is compatible with any circuit topology. 
We demonstrate the efficacy of our approach using a quantum hardware emulator, achieving chemical accuracy for systems as large as 12 qubits with fewer than 30 two-qubit gates. Our Clifford-based Hamiltonian engineering approach offers a promising avenue for practical quantum computational chemistry on near-term quantum devices.
\end{abstract}

\maketitle

\section{Introduction}
In the past decade, quantum computing has been on an extraordinary trajectory of growth and development, paving the way for the quantum simulations of molecular properties~\cite{cao2019quantum, bauer2020quantum, mcardle2020quantum, motta2022emerging, liu2022quantum, ma2023multiscale}. A groundbreaking study in 2014 successfully simulated the \ch{HeH+} molecule using a 2-qubit photonic quantum processor~\cite{peruzzo2014variational}. This pioneering work introduced the Variational Quantum Eigensolver (VQE) algorithm~\cite{cerezo2021variational, tilly2022variational}, which has become the most widely adopted algorithm for the quantum simulation of molecules in the noisy intermediate-scale quantum (NISQ) era~\cite{Preskill18, bharti2022noisy}.
Building on this foundation, a research team from IBM in 2017 expanded the scope of quantum simulations to larger molecules, such as \ch{BeH2}, by employing a 6-qubit superconducting quantum processor~\cite{kandala2017hardware}. Subsequent progress in this field unlocked the potential for quantum computation to profile the chemical reaction for diazene isomerization and cyclobutene ring opening using the 54-qubit ``Sycamore'' quantum processor~\cite{google2020hartree, o2022purification}. However, hindered by the limited circuit depth and hardware noise, the accuracy of these studies remained suboptimal.
A recent attempt to fill this gap has reached the chemical accuracy of the \ch{LiH} molecule by an optimized unitary coupled cluster(UCC) ansatz on superconducting qubits~\cite{guo2022experimental}. These achievements show the potential of quantum computing to reshape computational chemistry. As quantum hardware and algorithms evolve rapidly, it is envisioned that quantum computers will play an increasingly significant role to enhance our understanding of molecular systems and to expedite the discovery of novel materials and drugs~\cite{barkoutsos2021quantum, gao2021applications, li2022toward, malone2022towards, motta2023quantum}.

The VQE algorithm is a powerful method for estimating the ground state energy of a given molecular system, leveraging the Rayleigh-Ritz variational principle~\cite{peruzzo2014variational, mcclean2016theory, fedorov2022vqe}.
In this approach, a parameterized quantum circuit, also referred to as an ansatz, is constructed to approximate the true ground state wavefunction. 
A quantum computer evaluates the energy expectation for a given set of circuit parameters at each iteration, 
and then a classical optimizer modifies the circuit parameters to minimize the energy.
A critical challenge in the successful implementation of the VQE algorithm lies in the design of ansätze that strike a balance between flexibility and simplicity~\cite{lee2018generalized, gard2020efficient, anselmetti2021local, xiao2023physics}. 
Specifically, ansätze must be capable of representing the wavefunction of molecular systems while remaining sufficiently simple to be efficiently implemented on quantum computers.

\begin{figure*}[thb]
    \centering
    \includegraphics[width=1\textwidth]{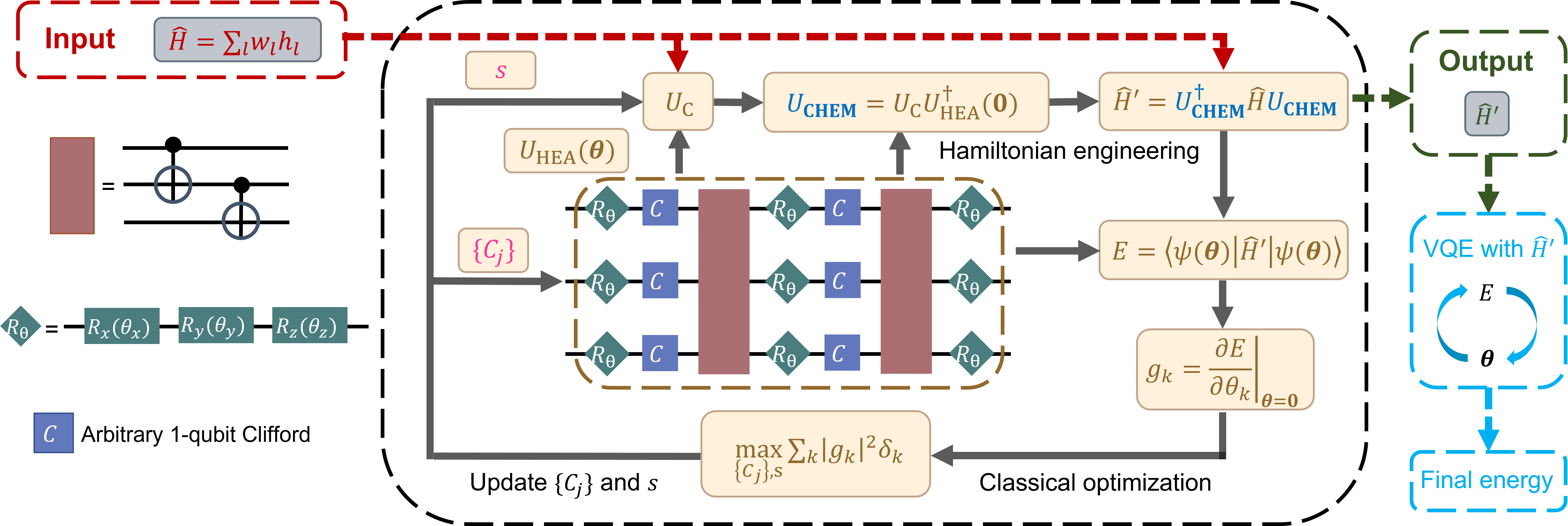}
    \caption{Schematic diagram for \ourmethod{} approach. The overall objective is to find Clifford Hamiltonian transformation $\ourmethodU{}$ to maximize the reward function $R$ (Eq.~\ref{eq:reward}). 
    The parameters to optimize in the workflow are $\{C_j\}$ and $s$, which are shown on the left of the flowchart and marked in magenta. 
    $C_j$ are single-qubit Clifford gates in the HEA circuit $U_{\textrm{HEA}}(\boldsymbol{\theta})$.  $s$ is a permutation defined by Eq.~\ref{eq:permutation} that determines the Clifford unitary $U_{\textrm{C}}$ based on $U_{\textrm{HEA}}(\boldsymbol{0})$ and system Hamiltonian $\hat H$.
    $\ourmethodU{}$ is constructed according to Eq.~\ref{eq:ourmethodu}, which is then used to transform the Hamiltonian $\hat H$. $g_k$, the energy gradient with respect to the engineered Hamiltonian $\hat H'$ at $\boldsymbol{\theta} = \boldsymbol{0}$, is calculated according to Eq.~\ref{eq:final_g}, which in turn determines $R$. 
    The values of $\{C_j\}$ and $s$ are optimized by a discrete optimizer based on $R$. The optimal values are used to construct the engineered Hamiltonian $\hat H'$ fed into the downstream VQE task.}
    \label{fig:workflow}
\end{figure*}

The UCC ansatz is initially proposed due to its clear chemical picture and close resemblance to the traditional Coupled Cluster (CC) theory~\cite{peruzzo2014variational, anand2022quantum}.
The ansatz starts with a Hartree--Fock (HF) initial state and then applies excitations to capture the correlation energy. 
Despite its high accuracy with only single and double excitations (UCCSD), most quantum hardware experiments are conducted using the Hardware-Efficient Ansatz (HEA)
~\cite{ kandala2017hardware, kandala2019error, gao2021applications, gao2021computational, kirsopp2022quantum} or substantially simplified UCC variants~\cite{peruzzo2014variational, o2016scalable, nam2020ground, google2020hartree, o2022purification, guo2022experimental} due to the deep circuit depth of UCC.
The HEA family of ansätze is specifically designed for simple implementation on NISQ devices with shallow circuits. 
However, the HEA approach presents its own set of challenges, often related to optimization difficulties~\cite{mcclean2018barren, choy2023molecular}, which result in a lower accuracy compared to the UCC ansatz. 
For instance, random initial guesses are typically employed due to the lack of systematic methods to appropriately initialize HEA circuit parameters. 
Unlike the UCC ansatz, where setting all parameters to zero corresponds to initiating the VQE optimization from the HF state, selecting circuit parameters that yield the HF state as the output is nontrivial for the HEA approach. 
Compounding this issue, random HEA circuits are known to suffer from the ``barren plateau'' phenomenon, wherein gradients vanish as the number of qubits and the circuit depth increases~\cite{mcclean2018barren}.
Consequently, for complex molecules, it becomes challenging for HEA to even reach the HF accuracy.

There is a growing interest in designing quantum algorithms to reach the best of both worlds, i.e., to achieve chemical accuracy with shallow quantum circuits.
The Adaptive Derivative-Assembled Pseudo Trotter for VQE (ADAPT-VQE)~\cite{grimsley2019adaptive} and its variants~\cite{yordanov2021qubit,tang2021qubit} are promising approaches that balance the accuracy and circuit simplicity. 
These methods construct the ansatz iteratively based on the information obtained from previous VQE runs.
In parallel, orbital-optimized approaches enhance the quality of simplified UCC ans\"atz by optimizing molecular orbitals classically~\cite{mizukami2020orbital, sokolov2020quantum, bierman2023improving}.
While both ADAPT-VQE and orbital-optimized approaches  demonstrate improvements in accuracy and circuit depth, they may inadvertently  increase the measurement cost associated with VQE.

In this study, we present a Hamiltonian engineering approach to address the trade-off between circuit depth and accuracy in quantum computational chemistry. Our method exploits the property that any unitary transformation applied to the Hamiltonian preserves its eigenspectrum.
Based on HEA circuits, 
the Hamiltonian transformation is designed to 
1) ensure that a set of initial circuit parameters that correspond to the HF energy can be generated;
2) maximize the energy gradient with respect to circuit parameters;
3) impose negligible overhead for classical processing without input from quantum computers; and
4) be compatible with any HEA circuit topology.
As these objectives are achieved by virtue of the Gottesman–Knill theorem for Clifford circuits \cite{gottesman1998heisenberg,aaronson2004improved},
we term this Clifford-based Hamiltonian Engineering approach for Molecules as \ourmethod{}.
Experimental results obtained from a quantum hardware emulator provide compelling evidence of our method's efficacy. Remarkably, it achieves chemical accuracy for systems as large as 12 qubits, using quantum circuits with fewer than 30 two-qubit gates and linear qubit connectivity. To the best of our knowledge, such a level of performance has not been previously reported in the literature.

\section{Theory and Method}

In this section, we first provide an overview of the basic concepts of VQE, HEA and Clifford group in Sec.~\ref{sec:vqe-hea}, and present our proposed Hamiltonian engineering approach for accurately estimating molecular energy with shallow circuits in Sec.~\ref{sec:chem}-Sec.~\ref{sec:numerical}.

\subsection{VQE, HEA, and Clifford Circuit}
\label{sec:vqe-hea}
In this work, we consider the molecular
electronic structure Hamiltonian in the second-quantized form:
\begin{equation}
\label{eq:ham-abinit}
    \hat H = \sum_{pq}v_{pq} \hat a^\dagger_p \hat a_q + \frac{1}{2}\sum_{pqrs}v_{pqrs}\hat a^\dagger_p \hat a^\dagger_q \hat a_r \hat a_s,
\end{equation}
where $v_{pq}$ and $v_{pqrs}$ are one-electron integrals and two-electron integrals, respectively. $\hat a^\dagger, \hat a$ are fermionic creation and annihilation operators.
In order to measure the expectation of $\hat H$ on a quantum computer,
the Hamiltonian is converted into a summation of $N$-qubit Pauli string
\begin{equation}
\label{eq:ham}
    \hat H = \sum_l^{N_H} w_l h_l,
\end{equation}
where $w_j$ are real coefficients, $N_H$ is the total number of terms,
and $h_l = \sigma_{l_1}\otimes \sigma_{l_2}\otimes\ldots \sigma_{l_N}$ with $\sigma_{l_k} \in \{X, Y, Z, I\}$.

The wavefunction of the molecular system, represented by a parameterized quantum circuit $U(\boldsymbol{\theta})$, can be written as
\begin{equation}
    \ket{\psi(\boldsymbol{\theta})} = U(\boldsymbol{\theta}) \ket{\phi},
\end{equation}
where the initial state $\ket{\phi}$ is usually a product state.
The molecular energy with given parameters
$E(\boldsymbol{\theta}) = \braket{\psi(\boldsymbol{\theta})|\hat H|\psi(\boldsymbol{\theta})}$
can then be computed using a quantum computer.
The goal of VQE is to find a set of optimal parameters $\boldsymbol{\theta}^*$ that minimizes $E$ using classical optimizers.

HEA is a family of versatile and widely-used ansatz that is particularly suitable for execution on near-term quantum hardware. \cite{leone2022practical}
Compared with UCC type of ansatz, HEA type ansatz has a lower circuit depth by utilizing hardware connectivity and maximizing the number of operations at each layer.
Typically, HEA is composed of interleaved single-qubit rotation gate layers $L_{\rm{rot}}^{(l)}(\{\theta_{lj}\})$ and entangler gate layers $L_{\rm{ent}}^{(l)}$. An HEA circuit with $p$ layers can be expressed as:
\begin{equation}
\label{eq:hea}
    U_{\textrm{HEA}}(\{\theta_{lj}\}) :=  \prod_{l=p}^1 \left [ L_{\rm{rot}}^{(l)}\{\theta_{lj}\}) L_{\rm{ent}}^{(l)} \right ] L_{\rm{rot}}^{(0)}(\{\theta_{0j}\}). 
\end{equation}
Note that the product $\prod_{l=p}^{1}$ is written in a reverse order to match the direction of the circuit, i.e., the gates with smaller indices are applied first.
Single qubit rotation layer $L_{\rm{rot}}(\{\theta_{nj}\})$ can generally be written as
$L_{\rm{rot}}^{(l)}(\{\theta_{lj}\}) = \prod_{j=M}^1R_j(\theta_{lj})$
with $R_{j}(\theta)=\exp(-i\frac{\theta}{2}P_{j})$ and $P_j \in \{X, Y, Z\}$.
It follows that $L_{\rm{rot}}^{(l)}(\{\theta_{lj}=0\}) = I$.
The form of $L_{\rm{ent}}^{(l)}$ is rather flexible,
and an example with consecutive CNOT gates and arbitrary Clifford gates $\{C_j\}$ is illustrated in the brown dashed box in Fig.~\ref{fig:workflow}.
A key feature of $L_{\rm{ent}}^{(l)}$ is that they are typically Clifford circuits,
which enables the Clifford-based Hamiltonian engineering approach for molecules (\ourmethod{}) described below.

Clifford circuits are a special class of quantum circuits that consist of gates from the Clifford group, which is a subgroup of the unitary group. 
The Clifford group is generated by three gates: the Hadamard gate ($H$), the phase gate ($S$), and the controlled-NOT gate (CNOT).
If $C$ is a Clifford operator, then for any Pauli string $P$, $C^\dagger P C$ is still a Pauli string.
The ability to classically simulate Clifford circuits stems from the Gottesman-Knill theorem, which states that any quantum circuit composed solely of Clifford gates and measurements in the computational basis can be efficiently simulated on a classical computer using the stabilizer formalism\cite{gottesman1998heisenberg,aaronson2004improved}. 
The classical simulation complexity of Clifford circuits without measurements is $\mathcal{O}(N)$ in the number of qubits $N$, as opposed to the exponential complexity associated with simulating general quantum circuits. 

\begin{figure*}[thb]
    \centering
    \includegraphics[width=0.7\textwidth]{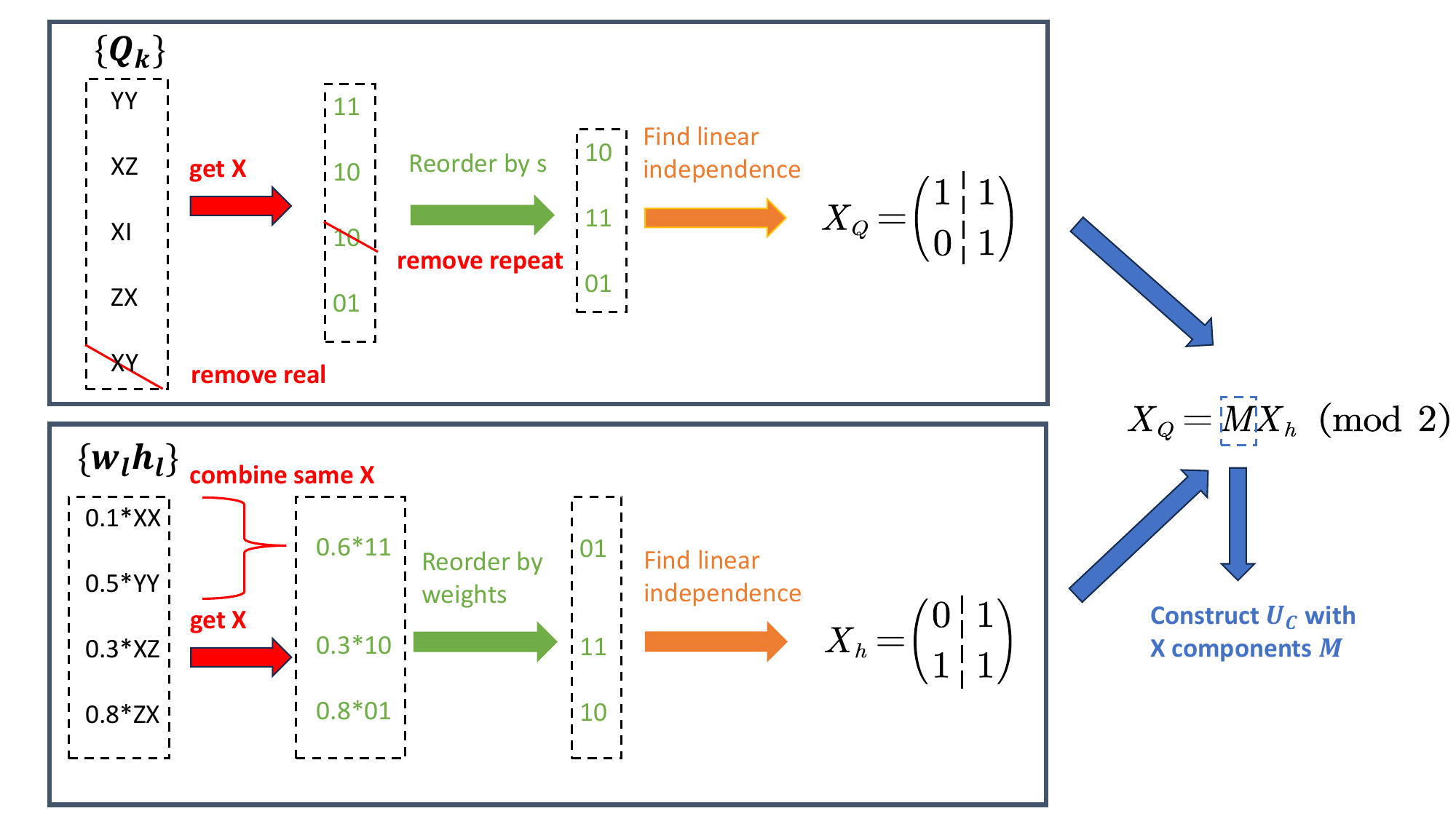}
    \caption{A simple 2-qubit example demonstrating the greedy algorithm to find Clifford transformation $U_C$. $\{h_l\}$ and $\{w_l\}$ are the Pauli terms and weights of the Hamiltonian respectively, while $\{Q_k\}$ are constructed by Eq.~\ref{eq:Q_expression}. The first step (marked in red) is to make some simplifications for $\{h_l\}$ and $\{Q_k\}$ by (1) removing $Q_k$ with real phases or repeated $X$ components and (2) combining $\{h_l\}$ with the same $X$ components. We note that the X component of $0.1XX+0.5YY$ is $0.4$ since $YY=(-iZX)(-iZX)$ has an opposite sign with $XX$. The second step (marked in green) is to reorder the X components of $\{Q_k\}$ and $\{h_l\}$ by permutation $s$ and weights $w_l$, respectively. The third step (marked in orange) is to use Gaussian elimination to find linear independent terms to construct $X_h$ and $X_Q$. Finally we solve $M$ and construct $U_C$ with X components $M$.}
    \label{fig:greedy}
\end{figure*}

\subsection{Clifford-based Hamiltonian Engineering}
\label{sec:chem}
In this section, we introduce how to design the Hamiltonian transformation $\ourmethodU{}$ to facilitate the optimization of HEA circuits.
The overall workflow is depicted in Fig.~\ref{fig:workflow}.

The transformed Hamiltonian $\hat H'$ is defined as
\begin{equation}
\label{eq:transform_H}
    \hat H' = \ourmethodU{}^\dagger \hat H \ourmethodU{}.
\end{equation}
$\hat H'$ shares the same eigenspectrum with $\hat H$ but with different eigenstates.
We restrict $\ourmethodU{}$ to be Clifford operation, so that $\hat H'$ can be efficiently processed on a classical computer and $\hat H'$ has the same number of terms with $\hat H$.
We first let $\ourmethodU{} = U_{\textrm{HEA}}^\dagger(\bf{0})$
and set the initial state of the circuit $\ket{\phi}$ as the Hartree--Fock state $\ket{\psi_{\textrm{HF}}}$.
Then, the energy expectation at $\boldsymbol{\theta} = \boldsymbol{0}$ is 
\begin{equation}
\label{eq:che-hf}
    \braket{\psi(\boldsymbol{0})|\hat H'|\psi(\boldsymbol{0})} = \braket{\psi_{\textrm{HF}}|\hat H|\psi_{\textrm{HF}}}.
\end{equation}
Within this transformation, the HEA circuit reaches HF energy by setting $\boldsymbol{\theta}=\boldsymbol{0}$, which serves as a good starting point for subsequent parameter optimization.

Then we take one step further by introducing an extra freedom $U_C$ to $\ourmethodU{}$: 
\begin{equation}
\label{eq:ourmethodu}
    \ourmethodU{} = U_{\textrm{C}} U_{\textrm{HEA}}^\dagger( {\bf 0} ),
\end{equation}
where $U_{\textrm{C}}$ is an arbitrary Clifford operator with the only restriction that it is invariant to the HF state,
i.e., $U_{\textrm{C}}|\psi_{\textrm{HF}}\rangle=e^{i\alpha}|\psi_{\textrm{HF}}\rangle$ up to some global phase factor $e^{i\alpha}$.
This is necessary for Eq.~\ref{eq:che-hf} to still hold. 
Here $U_C$ is allowed to be dependent on the structure of $U_{\textrm{HEA}}$.

In order to find better estimations of the ground state energy,
we construct $U_{\text{C}}$ by maximizing the energy gradients at $\boldsymbol{\theta}=\boldsymbol{0}$. 
Maximizing energy gradients has been proven to be a reliable approach towards compact circuit ansatz and efficient Hamiltonian transformation~\cite{ryabinkin2018qubit, grimsley2019adaptive}.

To simplify notation, we rewrite the HEA in Eq.~\ref{eq:hea} as
\begin{equation}
\label{eq:hea2}
    U_{\textrm{HEA}}(\boldsymbol{\theta}) =  \prod_{j=M}^{1}R_{j}(\theta_{j})B_{j},
\end{equation}
where $B_j$ is a (multi-qubit) Clifford operator (could also be identity).
We note that Eq.~\ref{eq:hea2} is not restricted to the HEA structure in Fig.~\ref{fig:workflow}, thus our method can also be applied to other types of HEA as well.
According to the simplified indices in Eq.~\ref{eq:hea2}, the energy derivative is written as
\begin{equation}
    g_{k} =\partial_{\theta_{k}}E(\boldsymbol{0}) = \left. \pdv{E}{\theta_k} \right|_{\boldsymbol{\theta}=\bf{0}},
\end{equation}
The reward function that we seek to maximize is defined as
\begin{equation}
\label{eq:reward}
    R  = \sum_k |g_k|^2 \delta_k,
\end{equation}
where $\delta_k=0$ or 1 indicates whether the gradient is redundant due to linear dependencies in $\{\partial_{\theta_{k}}|\psi(\boldsymbol{0})\rangle\}$.
For example, if two of the tangent vectors are linearly dependent, the corresponding parameters are redundant during the optimization. Thus only one of them should be kept in the reward function.
In principle, higher-order derivatives, such as $
h_{kl} =\partial_{\theta_{k}}\partial_{\theta_{l}}E(\boldsymbol{0})$, can also be included in $R$,
which shall be the subject of our future work.

Next, we present simplified expressions for $g_k$ and $\delta_k$.
To begin with, we define Pauli string $Q_{k}$ that is transformed from the Pauli operator $P_k$ associated with the $k$th rotation gate $R_k$:
\begin{equation}
\label{eq:Q_expression}Q_{k}=\prod_{j=1}^{k}B^\dagger_{j}P_{k}\prod_{j=k}^{1}B_{j}.
\end{equation}
$Q_k$ can be efficiently calculated on a classical computer since $B_j$ is Clifford.
The wavefunction derivatives are then written as
\begin{equation} \label{eq:gradient_psi}\partial_{\theta_{k}}|\psi(\boldsymbol{0})\rangle 
=-\frac{i}{2}U_{\textrm{HEA}}(\boldsymbol{0}) Q_{k}|\psi_{\text{HF}}\rangle.
\end{equation}

Using Eq.~\ref{eq:gradient_psi},
$\delta_k$ in Eq.~\ref{eq:reward} is determined as follows.
The overlaps between $\{\partial_{\theta_{k}}|\psi(\boldsymbol{0})\rangle\}$ is
\begin{equation}
\label{eq:ovlp}
\begin{aligned}
     s_{kl} &= \langle \partial_{\theta_{k}}\psi(\boldsymbol{0}) | \partial_{\theta_{l}}\psi(\boldsymbol{0})\rangle \\
     &= \frac{1}{4} \langle \psi_{\text{HF}} |Q_{k}Q_{l}|\psi_{\text{HF}}\rangle \in \{\pm \frac{1}{4}, \pm \frac{i}{4}, 0\}   
\end{aligned}
\end{equation}
Thus we let $\delta_k = 0$ if $s_{kl} = \pm \frac{1}{4}$ for some $l < k$, and $\delta_k = 1$ otherwise.

On the other hand, the energy derivatives are 
\begin{equation} \label{eq:g_expression}
g_{k}
 =\sum_{l=1}^{N_{H}}w_l \text{Im}[\langle\psi_{\text{HF}}|h'_{l}Q_{k}|\psi_{\text{HF}}\rangle].
\end{equation}
with $h'_l = U^\dagger_{\textrm{C}}h_{l}U_{\textrm{C}}$.
Formally, Eq.~\ref{eq:g_expression} reduces to the expression of energy gradients iterative qubit coupled cluster (iQCC) approach~\cite{ryabinkin2020iterative, ryabinkin2021posteriori} when the circuit is simply $e^{-i\theta Q_k/2}$.
However, unlike iQCC, in which case the goal is to find a series of $\{Q_k\}$ and construct the circuit based on $\{Q_k\}$, our method targets $U_{\textrm{C}}$ to avoid breaking the HEA circuit topology.

Based on Eq.~\ref{eq:gradient_psi} and Eq.~\ref{eq:g_expression}, it is straight-forward to calculate $R$  once $U_{\textrm{C}}$ is known.
However, the reversed task of finding $U_\textrm{C}$ that maximizes $R$, given $\hat H$ and $\{Q_k\}$, is considerably more complex.
In this work, we propose a heuristic greedy algorithm designed to address this challenge.
The details of the algorithm extend beyond the fundamental concept of \ourmethod{} outlined above
and will be presented in the subsequent section.

To summarize this section, we briefly recap the key idea of \ourmethod{} by analyzing how \ourmethod{} changes the landscape of HEA.
Both traditional HEA and HEA with \ourmethod{} can be viewed as a parameterized ansatz that maps to a subspace of the entire wavefunction space.
However, traditional HEA effectively chooses a random subspace, while \ourmethod{} tries to find the optimal subspace by maximizing the energy derivatives at the HF state.
Since the subspace dimension is exponentially small compared with the full space, randomly choosing the subspace can be expected to give only exponentially small energy improvements.
In contrast, a subspace with large energy derivatives ensures that at least a decent local minimum can be reached.
To ensure a global minimum, we need to consider the infinite order of derivatives at $\boldsymbol{\theta}=\bf{0}$.
The current approach of using only gradients can be viewed as a first-order approximation.

\subsection{An efficient algorithm for finding $U_{\textrm{C}}$}
In this section, we describe a greedy algorithm for finding $U_{\textrm{C}}$, which proves effective in our numerical experiments.
We will start by simplifying the expression of $R$ using the binary representation of Pauli strings.
Next, we propose a greedy algorithm that constructs $U_{\textrm{C}}$ based on the HEA circuit structure and the Hamiltonian.
The whole algorithm is outlined in Algorithm~\ref{algo:algo}.
We also present a simple 2-qubit example to demonstrate the algorithm in Fig.~\ref{fig:greedy}.
We note that the whole process is carried out efficiently on classical computers without any input from quantum computers, and a complexity analysis is provided at the end of the section.

Without loss of generality, we assume $|\psi_{\text{HF}}\rangle = |0\rangle^{\otimes n}$, then the expectation in Eq.~\ref{eq:g_expression} can be simplified by
\begin{equation} \label{eq:hQ_sim}
\langle\psi_{\text{HF}}|PQ|\psi_{\text{HF}}\rangle=\begin{cases}
(-1)^{z_Q \cdot x_Q} \phi_{P} \phi_Q  & \text{if } x_P = x_Q \\
0 & \text{else}
\end{cases}
\end{equation}
for arbitrary Pauli operators $P$ and $Q$, 
where $x$, $z$ and $\phi$ are binary representations of Pauli operators. 
Specifically, for Pauli operator $P$ we have $P=\phi_{P}\prod_{i}(Z[i])^{z[i]_{P}}(X[i])^{x[i]_{P}}$, where $\phi_{P}\in\{1,i,-1,-i\}$, $x[i]_{P} \in \{0,1\}$, $z[i]_{P} \in \{0,1\}$.
The index in the square parenthesis is the qubit index. $x_{P}$ and $z_{P}$ are the vector representations of $\{x[i]_{P}\}$ and $\{z[i]_{P}\}$.
Using this representation, $PQ$ can be expressed as $\phi_{P} \phi_Q \prod_{i}(Z[i])^{z[i]_{P}}(X[i])^{x[i]_{P}}(Z[i])^{z[i]_{Q}}(X[i])^{x[i]_{Q}}$.
With the anticommutation relationships of $X$ and $Z$, Eq.~\ref{eq:hQ_sim} then follows naturally.

Eq.~\ref{eq:hQ_sim} allows us to simplify the calculation of the reward function $R$ in Eq.~\ref{eq:reward} from the following aspects one by one.
The first aspect involves screening out terms with zero contributions in the summation in Eq.~\ref{eq:reward}.
Since $\phi_{h'}$ is always real, $\langle\psi_{\text{HF}}|h'_{l}Q_{k}|\psi_{\text{HF}}\rangle$ in Eq.~\ref{eq:g_expression}
has imaginary part only when $Q_k$ has imaginary part.
Thus, indices $k$ with $\phi_{Q_k} \neq \pm i$ can be dropped in Eq.~\ref{eq:reward} since the corresponding $g_k$ is zero. 
Second, from Eq.~\ref{eq:ovlp}, $s_{kl}=\pm \frac{1}{4}$ is equivalent to $x_{Q_k}=x_{Q_l}$.
Thus, the $\delta_k$ factor in Eq.~\ref{eq:reward} essentially removes repeating $x_Q$ vectors and in the following we may assume $x_Q$ is unique.
The third aspect is more for notational simplicity. Specifically,
we combine those $h'_l$ with the same $x_{h'_l}$ and keep only one of them.
Meanwhile, the weights $w_l$ are summed accordingly with the phase information from $\phi_{h'_l}$ taken into account. 
This is viable because $z_P$ does not appear in Eq.~\ref{eq:hQ_sim} and $h'_l$ with the same $x_{h'_l}$ has the same $\langle\psi_{\text{HF}}|h'_{l}Q_{k}|\psi_{\text{HF}}\rangle$.
Consequently, we may also assume $x_{h'}$ is unique.
The treatment is similar to the Hamiltonian partitioning process in the iQCC algorithm~\cite{ryabinkin2020iterative, ryabinkin2021posteriori}, but simpler due to the stronger $|\psi_{\text{HF}}\rangle = |0\rangle^{\otimes n}$ assumption we have.
After these operations, $g_k$ and $R$ becomes
\begin{align}
g_k &= 
\begin{cases}
\pm w_l & \text{if } x_{Q_{k}} = x_{h'_l} \text{ for some } l \\
0 & \text{otherwise}
\end{cases} \label{eq:final_g}
\\
R &= \sum_{l\in\{l|x_{h'_{l}}\in \{x_{Q_k}\}\}}|w_{l}|^{2}, \label{eq:final_R}
\end{align}
where $|w_l|$, $x_{h'_l}$ and $x_{Q_k}$ are after the above simplifications (the same hereinafter).
Note that there is at most one $x_{h'_l}$ that satisfies $x_{Q_{k}} = x_{h'_l}$.
$g_k$ and $R$ only depend on $|w_l|$, $x_{h'_l}$ and $x_{Q_k}$. 
The theoretical upper limit of $R$ is $R_\text{max}=\sum_{l}|w_{l}|^{2}$, which is achieved when $\{x_{h'_l}\} \subseteq \{x_{Q_k}\}$.

\begin{algorithm}
\DontPrintSemicolon
\caption{A greedy algorithm for finding $U_{\textrm{C}}$}
\label{algo:algo}
\SetKw{Break}{break}
\SetKw{Continue}{continue}
\SetKwFunction{GaussianElimination}{GaussianElimination}
\SetKwFunction{Rank}{Rank}
\SetKwFunction{Min}{Min}
\SetKwFunction{PseudoInverse}{PseudoInverse}
\SetKwInOut{Input}{input}\SetKwInOut{Output}{output}

\Input{Hamiltonian $\hat H$ (Eq.~\ref{eq:ham}) and quantum circuit $U_{\textrm{HEA}}(\bf{0})$ (Eq.~\ref{eq:hea2})}
\Output{Clifford unitary transformation $U_{\textrm{C}}$}
\BlankLine
Construct $X_h$ from $\hat H$ \;
Combine repeated $x_{h_l}$ in $X_h$ with summed weights $w_l$ \;
Reorder columns of $X_h$ based on weights $|w_l|$ \;
$X_h$ = \GaussianElimination{$X_h$} \;
$X_h^r=X_h[:, :r]$ \tcc*{The first $r$ columns}
\While{not converged}{
    Generate new $s$ and $\{C_j\}$ \;
    \tcc{$Q_k$ is determined by the quantum circuit}
    Construct $\{Q_k\}$ from $U_{\textrm{HEA}}(\bf{0})$ and $\{C_j\}$ based on Eq.~\ref{eq:Q_expression}  \;
    Drop $Q_k$ with real $\phi_{Q_k}$ or repeated $x_{Q_k}$ \;
    Construct $X_Q$ from $\{Q_k\}$ \;
    $r$ = \Min{\Rank{$X_h$}, \Rank{$X_Q$}} \;
    Reorder columns of $X_Q$ based on $s$ \;
    $X_Q$ = \GaussianElimination{$X_Q$} \;
    $X_Q^r=X_Q[:, :r]$ \tcc*{The first $r$ columns}
    $M=X_Q^r\cdot$\PseudoInverse{$X_h^r$} \;
    Calculate $R$ based on Eq.~\ref{eq:final_R} \;
} 
Construct $U_{\textrm{C}}$ based on $M$ \;
\end{algorithm}

\begin{figure*}[htbp]
    \centering    \includegraphics[width=0.8\textwidth]{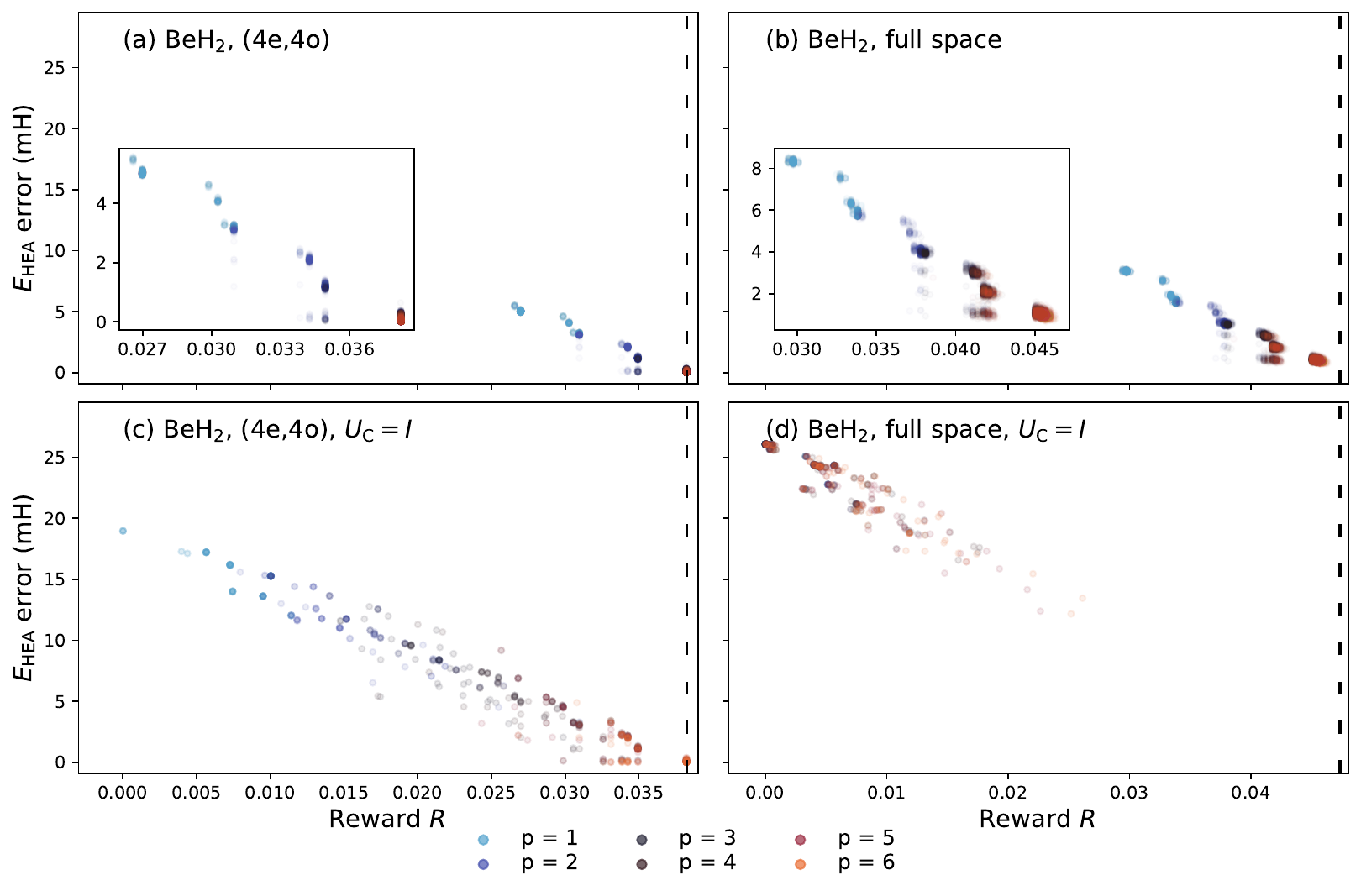}
    \caption{Errors of optimized HEA energies as a function of the reward $R$ with and without Clifford-based Hamiltonian engineering using \ch{BeH2} systems as examples. In (a) and (c) the (4e, 4o) active space is employed and in (b) and (d) the full space is employed.
    In the top panels (a) and (b) the Hamiltonian engineering is performed and in bottom panels (c) and (d) the Hamiltonian engineering is not performed. Optimized HEA energy errors are computed with the corresponding reference FCI energies. The dashed lines represent the theoretical highest reward $R_\text{max}$ that the corresponding molecular system can reach. In each panel, for each circuit depth $p$, the results of 1000 trials with the corresponding labeled settings are plotted.}
    \label{fig:score_vs_e}
\end{figure*}

Next, we present an efficient algorithm to find $U_{\textrm{C}}$ that approximately maximizes $R$.
Under the stabilizer formalism, a Clifford operator is uniquely specified up to a global phase by its transformation result of $\{Z[1],Z[2],...,Z[N]\}$ and $\{X[1],X[2],...,X[N]\}$. 
Therefore, requiring $U_{\textrm{C}}|\psi_{\textrm{HF}}\rangle=e^{i\alpha}|\psi_{\textrm{HF}}\rangle$  means  that $U_{\textrm{C}}^{\dagger}Z[i]U_{\textrm{C}}$ only has $Z$ components in the binary representation. 
As a consequence, $U_{\textrm{C}}^{\dagger}X[i]U_{\textrm{C}}$ must have only $X$ components since Clifford transformation preserves commutation/anticommutation relations.  
Since we don't care about the signs, $U_C$ can be specified by $M\in\mathbb{Z}_{2}^{N\times N}$ such that $U_{\textrm{C}}^{\dagger}X[i]U_{\textrm{C}}$ has $X$ component $M_i$ (the $i$th row of $M$) in the binary representation.
This allows us to relate $x_{h'_l}$ and $x_{h_l}$ by 
\begin{equation}
x_{h'_l}=Mx_{h_l}.
\end{equation}
Supposing the matrix forms of $x_{h}$ and $x_Q$ are 
 $X_{h}\in\mathbb{Z}_{2}^{N\times N_{h}}$ and $X_{Q}\in\mathbb{Z}_{2}^{N\times N_{Q}}$, where $N_{h}$ and $N_{Q}$ are the number of $h$ and $Q$ respectively,
maximizing Eq.~\ref{eq:final_R} is equivalent to
finding $M$ such that $X_{h}^{\prime} = M X_{h}$
has the most repeated columns with $X_{Q}$ weighted by $|w_{l}|^2$.
On the other hand, the transformation for $\{Z[1],Z[2],...,Z[N]\}$, which does not affect $R$, is determined by $M$ up to a phase because Clifford transformation preserves commutation/anticommutation relations.

Although the problem of finding $M$ described above is more tractable than ``finding the best $U_{\textrm{C}}$'', it still represents a discrete optimization problem, which is generally difficult to solve.
In the following, we describe a greedy algorithm for this maximization problem, which proves effective in our numerical simulation.
The algorithm is motivated by the fact that any two full-rank $n\times r$ matrices can be related by an $n\times n$ invertible transformation $M$. 
Specifically, let the rank of $X_{h}$ be $r_{h}$, the rank of $X_{Q}$ be $r_{Q}$, and $r=\min(r_{h},r_{Q})$.
Then we \emph{greedily} find $r$ linearly independent columns of $X_{h}$ as $X_{h}^{r}\in\mathbb{Z}_{2}^{N\times r}$ with the largest weights $|w_l|^2$, and find $r$ linearly independent columns of $X_{Q}$ as $X_{Q}^{r}\in\mathbb{Z}_{2}^{N\times r}$.
The searching of linearly independent columns can be done using Gaussian elimination with modulo 2.
Since $X_{h}^{r}$ and $X_{Q}^{r}$ are full-rank, we can easily construct $M\in\mathbb{Z}_{2}^{N\times N}$ such that $MX_{h}^{r}=X_{Q}^{r}$.
Thus, it is guaranteed to achieve $R\geq\sum_{i=1}^{r}|w_{h_{i}}^{r}|^{2}$, where $w_{h}^{r}$ is the corresponding weights of $X_{h}^{r}$. 
In Algorithm~\ref{algo:algo} we present the pseudo-code for constructing $M$ based on the greedy algorithm.

Since the $r$ linearly independent columns of $X_{Q}$ are not unique,
we may optimize the columns chosen such that  $R$ is maximized.
To do this, we reorder the columns of $X_Q$ by a permutation $s$ 
\begin{equation}
\label{eq:permutation} 
    s: x_{Q_i} \rightarrow x_{Q_{s(i)}}
\end{equation}
before the Gaussian elimination and allow $s$ to change freely.
Furthermore, we permit the single-qubit Clifford gates $\{C_j\}$ shown in Fig.~\ref{fig:workflow} to vary freely among all 24 different single-qubit Clifford operators.
Although these gates do not generate entanglement in the quantum state, they allow $Q_k$ to traverse the Pauli string space,
and eventually, help to shape the optimization landscape such that a high-quality local minimum can be reached easily.
The parameters $s$ and $\{C_j\}$ can be optimized by some classical discrete optimizer.
The overall workflow for Clifford-based Hamiltonian engineering, illustrated in Fig.~\ref{fig:workflow}, is as follows. 
At the beginning of each optimization step, we input a list of single-qubit Clifford gates $\{C_{j}\}$ and permutation $s$. 
Subsequently, we obtain $\{Q_{k}\}$ using Eq. \ref{eq:Q_expression}. 
$\{Q_k\}$ and $s$, combined with the Hamiltonian $H=\sum w_{l}h_{l}$, are then used to generate the Clifford transformation $U_{\textrm{C}}$ by the aforementioned greedy algorithm. 
After that, the Hamiltonian is transformed according to Eq.~\ref{eq:transform_H}. 
Finally, we calculate the reward function $R$ using Eq.~\ref{eq:final_R} and update $\{C_{j}\}$ and $s$ for the next iteration.
After the optimization is complete, VQE with the engineered $\hat H'$ is performed to calculate the molecular energy.
It is worth noting that the discrete optimization over $s$ and $C_j$ is not the key component of \ourmethod{}.
Rather, the greedy algorithm, which constructs $U_\textrm{C}$ from given $s$ and $\{C_j\}$, is the primary driver behind the success of \ourmethod{}.
In Table~\ref{tab:chem_no_opt} we show that based on random $s$ \ourmethod{} significantly improves the $R_y$ ansatz.

Lastly, we discuss the computational cost complexity of this algorithm in each iteration.
We recall that $p$ is the number of layers so there are $O(Np)$ single-qubit rotations and CNOT gates.
To calculate $Q_k$ we need to do $O(Np)$ Clifford evolutions, which have cost $O(N^2p)$.
The cost of performing the Gaussian eliminations for $X_Q$ is $O(N^3p)$.
We note that Gaussian elimination for $X_h$ is only needed to be performed once so is not counted in the cost of each iteration here.
Thus, the total cost is finally $O(N^3p)$.
We note that a direct calculation of Eq.~\ref{eq:g_expression} is $O(N^4)$ for each $g_k$, 
which would lead to an overall scaling of $O(N^5p)$.
By simplifying Eq.~\ref{eq:g_expression} to Eq.~\ref{eq:final_g}, the computational complexity for each $g_k$ is reduced to $O(\log N^4) < O(N)$ through a binary search.
Therefore, its contribution to the final scaling can be neglected. 
Given that the entire iteration is executed on classical computers with low scaling with $N$, the \ourmethod{} method accelerates the subsequent VQE with minimal additional preprocessing overhead.

\subsection{Relation with Previous Works}
\label{sec:comparison}
\ourmethod{} engineers the Hamiltonian via Clifford similarity transformation and outputs a Clifford-initialized circuit for VQE.
Here we compare and discuss the relation of \ourmethod{} with previous works that either use Hamiltonian similarity transformation or Clifford initialization techniques for VQE problems, respectively.

Both the iQCC approach ~\cite{ryabinkin2020iterative,lang2020unitary, genin2022estimating,ryabinkin2023efficient} and \ourmethod{} aims at reducing the circuit depth by maximizing the first-order energy derivatives in Eq.~\ref{eq:g_expression}.
However, the iQCC approach selects the best $\{Q_k\}$ and constructs the circuit with the form $e^{-i \theta Q_k / 2}$, which generally decomposes to $\sim 2N$ (where $N$ is the number of qubits) linearly connected CNOT gates and does not fully utilize the HEA topology.
Compared with iQCC, \ourmethod{} focus on finding the best Clifford transformation $\ourmethodU{}$ instead of $\{Q_k\}$, which does not require any decomposition of multi-qubit gates while allowing user-specified or hardware-specific HEA circuit topology.
Additionally, the iQCC approach involves an iterative transformation of the Hamiltonian using multiqubit rotation gates, resulting in a growth of terms in the Hamiltonian and thus a higher measurement cost.
\ourmethod{} does not have this issue since the transformation is fully Clifford.
Furthermore, since Clifford transformations preserve commutation properties, the number of mutually commuting groups for the Hamiltonian remains the same after the transformation. 
As a result, assuming entangled measurements, the number of measurement shots does not increase.
On the other hand, it is possible that the Pauli string becomes more non-local if the Bravyi-Kitaev transformation is employed,
which will result in a deeper measurement circuit.

A recent study employed hierarchical Clifford similarity transformations to reduce entanglement in the wavefunction~\cite{mishmash2023hierarchical}.
The method is tested for VQE tasks and shows promising results.
In contrast, \ourmethod{} engineers Clifford similarity transformation by directly targeting optimized VQE energies. Clifford transformation has also been utilized to entangle independent quantum circuits for VQE~\cite{schleich2023partitioning}.
In all aforementioned methods, the transformation over $\hat H$ is built iteratively using information from expensive trial VQE runs.
In contrast, \ourmethod{} constructs the overall transformation $\ourmethodU{}$ on a classical computer using the efficient algorithm proposed in Sec.~\ref{sec:chem}.

\ourmethod{} optimizes $\{C_j\}$ in the HEA circuit,
which shows resemblance to the Clifford circuit initialization methods~\cite{cheng2022clifford, mitarai2022quadratic}.
These approaches aim to improve VQE performance by optimizing circuit architecture classically, requiring the initial circuit to be composed of Clifford gates.
The optimization objective is either the energy or its derivatives.
However, without a proper engineering of $U_{\textrm{C}}$, the number of required trials exponentially increases with increasing the number of qubits to find a set of optimal $\{C_j\}$ with non-zero $g_k$ by brute force search.
To see this, we can assume that the Pauli operator $Q$ calculated from Eq. \ref{eq:Q_expression} is randomly distributed in all $N$-qubit Pauli operators, then the probability that $g=\text{Im}[\langle\psi_{\text{HF}}|hQ|\psi_{\text{HF}}\rangle]$ gives nonzero value for some fixed $h$ decays as $2^{-N}$.
Thus, it is crucial to match $x_h$ and $x_Q$ via the Clifford Hamiltonian engineering proposed in Sec.~\ref{sec:chem}.
Failing to do so would merely transfer the difficulty in VQE optimization to the challenge of optimizing Clifford gates.

\begin{table}[hbpt]
\caption{Molecular systems tested in this work. $d$ is the bond length and $\phi$ is the bond angle. STO-3G basis set is used throughout.}
\label{tab:mols}
\begin{tabular}{ccc}
\hline
Molecule & Geometry & Active space\\
\hline
\multirow{2}{*}{\ch{H4}} & \multirow{2}{*}{$d=0.8000$ \AA} & (2e, 3o), $1a_{1u}, 2a_{1g}, 2a_{1u}$     \\
                         &                     & full space   \\
\multirow{2}{*}{\ch{LiH}} & \multirow{2}{*}{$d=1.600$ \AA} & (2e, 3o), $2a_1, 3a_1, 4a_1$     \\
                         &                     & full space   \\
\multirow{2}{*}{\ch{BeH2}} &  \multirow{2}{*}{$d=1.000$ \AA} & (4e, 4o), $2a_{1g}, 1a_{1u}, 3a_{1g}, 2a_{1u}$     \\
                         &                     & full space    \\
\multirow{2}{*}{\ch{H2O}} & $d=0.9584$  \AA, & \multirow{2}{*}{(4e, 4o), $3a_1, 1b_2, 4a_1, 2b_1$} \\
                           & $\phi=104.45$° & \\
\hline
\end{tabular}
\end{table}

\begin{figure}[htb]
    \centering
    \includegraphics[width=1\linewidth]{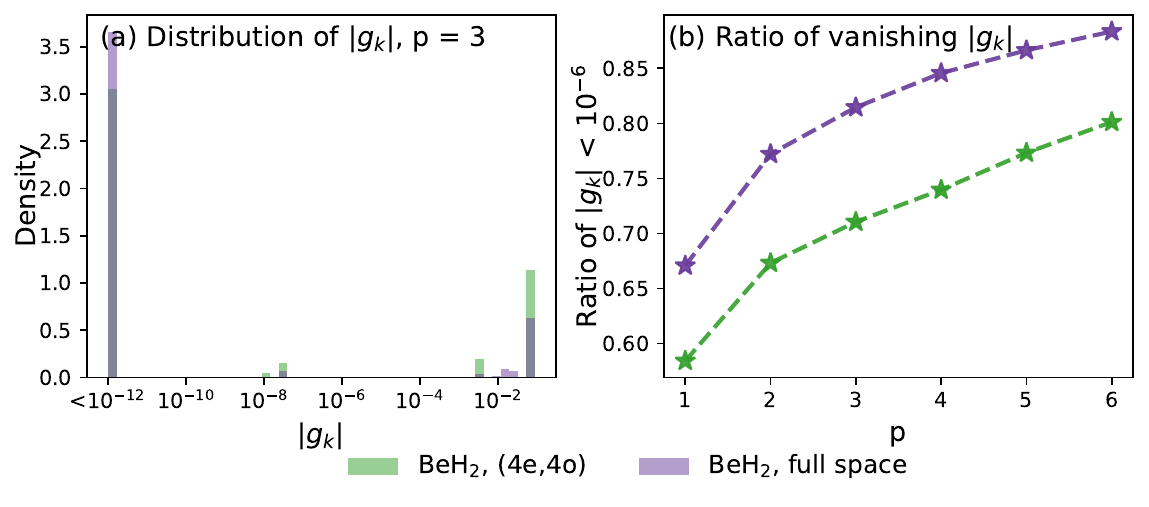}
    \caption{Distributions of gradients $|g_k|$ obtained from \ourmethod{} for \ch{BeH2} systems calculated with (4e, 4o) and full space with different circuit depths. (a) The density histograms of the absolute values of the gradients with $p=3$ are shown on the plots ($|g_k|$). To better display the small values, we plot the $x$-axes on logarithm scales. Addition histograms of $|g_k|$ collected with different $p$ values are shown in Supporting Information Figure S2. Panel (b) displays the ratio of $|g_k|$ values smaller than $10^{-6}$ versus the depth of the circuits $p$ for two systems. }
    \label{fig:grad}
\end{figure}

\begin{figure*}[htb]
    \centering    \includegraphics[width=0.85\textwidth]{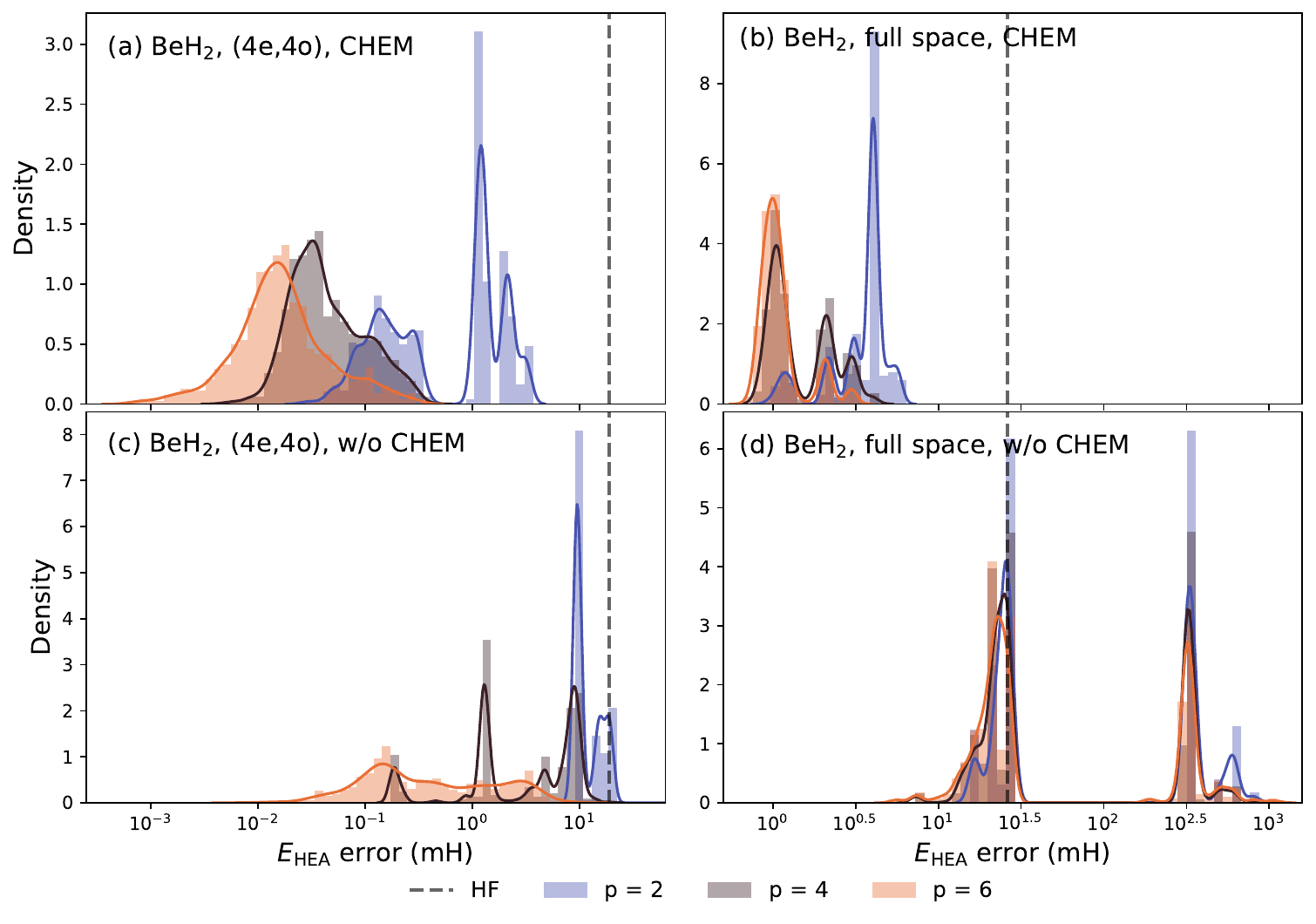}
    \caption{Optimized energy distributions using \ourmethod{} protocol and $R_y$ ansatz with random initializations for (a) \ch{BeH2} (4e,4o) and (b) \ch{BeH2} full space. 
    In (c) and (d) the corresponding results for the $R_y$ ansatz without \ourmethod{} are shown. 
    The gray dashed vertical lines represent the corresponding errors for HF energy. The solid curves are from kernel density estimation (KDE).}
    \label{fig:opt_e_dist}
\end{figure*}
\subsection{Numerical Details}\label{sec:numerical}
In this study, all the HEA simulations are performed using \textsc{TenCirChem} package \cite{li2023tencirchem} with inputs computed using \ourmethod{} package (\url{https://github.com/sherrylixuecheng/CHEM}). 
The molecular integrals and reference energies based on classical computational chemistry are obtained via the \textsc{PySCF} package~\cite{sun2018pyscf}.
The parity transformation~\cite{bravyi2002fermionic, seeley2012bravyi} with two-qubit reduction is employed to convert the Hamiltonian defined by Eq.~\ref{eq:ham-abinit} into a summation of Pauli strings.
The calculations of $g_k$ and $R$ are implemented by JAX \cite{jax2018github} with just-in-time compilations.
The discrete values of ${C_j}$ and $s$ are optimized by the simulated annealing (SA) method.
Practically, we use $\sqrt{R}$ instead of $R$ as the objective function.
In each step of SA, we make a trial move by randomly changing one value of ${C_j}$ and randomly switching two elements in $s$.
For a given circuit with $N$ qubits and $p$ layers, we perform the SA optimizations by $50*N*p$ iterations, and the corresponding temperatures in the SA optimizations decrease from 0.05 to 0.002 evenly on a logarithmic scale.
Due to the efficient algorithm proposed in Sec.~\ref{sec:chem}, the whole iteration terminates within seconds for the systems tested in this work.
To obtain the final HEA energies, we use the L-BFGS-B optimizer implemented in \textsc{SciPy} package \cite{scipy} with the default settings to optimize the HEA circuits engineered by \ourmethod{}.

\section{Results and Discussion}
\label{sec:results}
We show the ability of \ourmethod{} to reach high accuracy with shallow circuits by simulating four molecules and seven electronic structure systems numerically. The molecular systems tested in this study are summarized in Table~\ref{tab:mols}. In this work, the results collected on \ch{BeH2} systems are demonstrated and discussed extensively.
With STO-3G basis set and parity transformation, \ch{BeH2} corresponds to 7 spatial orbitals and 12 qubits.
We also employ a (4e, 4o) active space description of the \ch{BeH2} with all valence electrons and $A_1$ orbitals.

\subsection{Validity of reward function $R$ and necessity of $U_C$}
Fig.~\ref{fig:score_vs_e} demonstrates how engineering the Hamiltonian can result in lower VQE-optimized energy by maximizing the reward $R$.
The data points are obtained from 1000 random runs of simulated annealing.
Across all sub-figures in Fig.~\ref{fig:score_vs_e}, the optimized energy errors are well correlated with the reward $R$ for a Hamiltonian transformation $\ourmethodU{}$,
indicating that the design of $R$ is a valid proxy to the quality of the corresponding VQE problem.
That is, the better $R$ we achieve, the lower energy we can reach in the subsequent VQE.
Furthermore, compared with the approach in which $U_{\textrm{C}}$ is assumed to be identity and only $\{C_j\}$ is optimized, which is shown in Fig.~\ref{fig:score_vs_e}(c) and (d), our engineering of $U_{\textrm{C}}$ significantly increases the reward $R$ and reduces the error of VQE energy (Fig.~\ref{fig:score_vs_e}(a) and (b)).
Thus, the crucial aspect of our approach is the engineering of the Hamiltonian to maximize the initial energy gradients.

In Sec.~\ref{sec:comparison}, we have shown that na\"ive Clifford circuit optimization without proper engineering of $U_C$ becomes inefficient as $N$ increases.
This trend is manifested by comparing Fig.~\ref{fig:score_vs_e}(c) and (d),
in which extending from the active space treatment to full space treatment introduces significant challenges to maximizing $R$.
Meanwhile,
although in Fig.~\ref{fig:score_vs_e}(d) most of the cases have nearly zero $R$ values, 
in Fig.~\ref{fig:score_vs_e}(b) most of the data points are close to the theoretical highest reward $R_\text{max}$,
and the energy errors are significantly lower than the ones in Fig.~\ref{fig:score_vs_e}(d).
Such findings underscore the necessity of performing Hamiltonian engineering.
Additionally, \ourmethod{} scales well with the number of layers $p$ in the HEA circuit.
In all the panels, the rewards tend to increase and lead to lower VQE energies with increasing $p$ values.

\begin{table}[hbpt]
\caption{Errors of optimized HEA energies with respect to FCI for the \ch{BeH2} molecule with full space.
The number of layers is set to 3.
The three columns are for the $R_y$ ansatz, \ourmethod{}, and \ourmethod{} without SA optimization from 1000 trials, respectively.}
\label{tab:chem_no_opt}
\begin{tabular}{cccc}
\hline
& w/o \ourmethod{}  & \ourmethod{}  & SA-free \ourmethod{}\\
\hline
MIN$^\dagger$ (mH) & 16.61 & 0.7954 & 0.914 \\
MAE$^*$ (mH) & 85.50 & 2.240 & 7.844     \\
STD$^{\#}$ (mH) & 141.02 & 1.012 & 0.930     \\
\hline
\end{tabular}
\\$^\dagger$ MIN: Minimum error.
\\$^*$ MAE: Mean absolute error.
\\$^{\#}$ STD: standard deviation of the error.
\end{table}

\subsection{Potential reduction of VQE parameters}

Another advantage of the \ourmethod{} framework is that most of the $g_k$ is zero as a result of Eq.~\ref{eq:hQ_sim}, which implies their impacts on the energy minimization procedure may be neglected.
The density histograms of $g_k$ collected from 1000 \ourmethod{} trials with different $p$ are shown in Fig.~\ref{fig:grad}. 
The histograms show that a small fraction of the parameters contributes to most of the reward function $R$ and more than half of the parameters have nearly zero ($|g_k|<10^{-6}$) gradients.
Fig. ~\ref{fig:grad}(b) plots the ratios of parameters with vanishing gradients ($|g_k| < 10^{-6}$) as a function of circuit depth $p$. This ratio increases when the number of qubits $N$ or the circuit depth $p$ increases by comparing the results of the simulations with (4e, 4o) and full space, and the results of the simulations with different $p$ values, respectively. 
By fixing the parameters with zero gradients, a reduction of the dimensionality in the optimization problem can be achieved to significantly accelerate VQE optimizations. 
Incorporating this fact with more advanced optimization techniques, such as Bayesian optimizations\cite{cheng2023error}, could also enable efficient and noise-robust simulations on the near-term quantum devices, which is a future direction and is not taken into account in this work for simplicity.

\begin{table*}[htpb]
\scriptsize
\caption{Errors of optimized HEA energies and required average numbers of iterations using random initialization and \ourmethod{} for different systems. The errors are computed using the FCI energies as references for the corresponding systems. Errors below chemical accuracy (about 1.6mH) are marked in green.}
\label{tab:stats}
\begin{tabular}{lc|c|ccc|ccc}
\hline
System & No. of & HF error & \multicolumn{3}{c|}{w/o \ourmethod{} $\times$1000 ($p$=3)} & \multicolumn{3}{c}{\ourmethod{} $\times$1000 ($p$=3)} \\
 & qubits & (mH) & MIN$^*$ (mH) & MAE$^\dagger$ (mH) & ITER$^{\#}$   & MIN (mH) & MAE (mH) & ITER  \\
\hline
\ch{H4}, (2e, 3o) & 4 & 20.81 & \green{3.755e-06} & \green{4.763e-02} & 118.6 & \green{1.371e-07} & \green{2.412e-02} & 26.02 \\
\ch{LiH}, (2e, 3o) & 4 & 19.21 & \green{2.656e-06} & \green{0.1514} & 63.89 & \green{2.608e-05} & \green{5.807e-02} & 48.31 \\
\ch{BeH2}, (4e, 4o) & 6 & 18.97 & 7.242 & 10.57 & 128.8 & \green{1.115e-02} & \green{0.1445} & 28.39 \\
\ch{H2O}, (4e, 4o) & 6 & 7.451 & \green{1.353} & 5.367 & 85.56 & \green{1.427e-05} & \green{2.824e-02} & 15.97 \\
\ch{H4}, full space & 6 & 46.17 & 17.61 & 22.05 & 179.9 & \green{1.876e-02} & \green{0.8048} & 96.10 \\
\ch{LiH}, full space & 10 & 20.46 & \green{1.405} & 36.83 & 113.6 & \green{0.1828} & \green{0.4897} & 56.20 \\
\ch{BeH2}, full space & 12 & 26.07 & 16.61 & 85.50 & 156.1 & \green{0.7954} & 2.240 & 59.53 \\
\hline
\end{tabular}
\\$^*$ MIN: Minimum error of the 1000 trials.
\\$^\dagger$ MAE: Mean absolute error of the 1000 trials.
\\$^{\#}$ ITER: Averaged number of iterations for VQE optimization
\end{table*}

\subsection{Comparison to HEA without CHEM}

Next, we compare the optimized energy from \ourmethod{} with traditional VQE based on the $R_y$ ansatz.
The $R_y$ ansatz is a special form of HEA that allows only the $R_y$ gates in the rotation layers $L_{\textrm{rot}}$~\cite{gao2021computational, gao2021applications,mihalikova2022cost, choy2023molecular}.
The ansatz is popular for molecular electronic structure simulations because it enforces real amplitude in the wavefunction.
Due to optimization challenges associated with traditional HEA circuits, it is common to run VQE multiple times with different initial guesses in order to obtain meaningful results.
In the following, the lowest energy obtained from 1000 independent trials is reported for the $R_y$ ansatz.
Since \ourmethod{} deterministically starts the optimization from $\boldsymbol{\theta}=\boldsymbol{0}$, we also run \ourmethod{} 1000 times with random SA, corresponding to  1000 independent VQE optimizations, to ensure a fair comparison.

The distributions of the VQE optimized energies with $p=2, 4, 6$ are illustrated in Fig.~\ref{fig:opt_e_dist}.
The results with $p=1,3,5$ are included in the Supporting Information Fig. S5.
In all cases considered, the results from \ourmethod{} outperform the results from the $R_y$ ansatz by a considerable margin, as demonstrated in Fig.~\ref{fig:opt_e_dist}(b) for the full space calculation of the \ch{BeH2} molecule.
Since with \ourmethod{} the VQE optimization starts from the HF energy,
the energy errors of \ourmethod{} are consistently lower than the ones of HF.
The same is not true for traditional HEA, in which case the largest error is around 1 Hartree due to the poor initial guess.
Additionally, the accuracy of \ourmethod{} monotonically increases as $p$ increases from 2 to 6.
In both panels of Fig.~\ref{fig:opt_e_dist}, it is visible that \ourmethod{} can reach chemical accuracy (error < 1.6 mH) when $p=2$.
Intriguingly, we have observed unusual distributions of the optimized parameters for the $R_y$ ansatz over the 1000 VQE runs, which is plotted in the Supporting Information Fig. S1.
More specifically, most of the optimized parameters are close to integer or half-integer times of $\pi$, which turns the corresponding rotation gates $R_j(\theta_j)$ into Clifford gates.
In the Supporting Information, we discuss its potential connection to the remarkable efficiency of our method.

In Table~\ref{tab:chem_no_opt}, we present a comparison of error statistics between \ourmethod{} and the $R_y$ ansatz.
Consistent with the results shown in Fig.~\ref{fig:opt_e_dist}, \ourmethod{} significantly outperforms the 
$R_y$ ansatz in terms of minimum error, absolute error, and standard deviation of error.
Additionally, we include results from \ourmethod{} without SA optimization in Table~\ref{tab:chem_no_opt}.
In this case, we employ the $R_y$ ansatz and construct $U_\textrm{C}$ from randomly chosen $X_Q$ columns.
While the accuracy of the SA-free method is expectedly lower than \ourmethod{}, the mean absolute error is only 10\% of the $R_y$ ansatz, and the minimum error is only 15\% higher than \ourmethod{}.
Thus, the strength of \ourmethod{} lies in Hamiltonian engineering and the SA optimization serves as an improvement component with very little additional cost (see Fig.~\ref{fig:stats}).

\subsection{Benchmark of optimized VQE energies}

\begin{figure}[thbp]
    \centering
    \includegraphics[width=.95\linewidth]{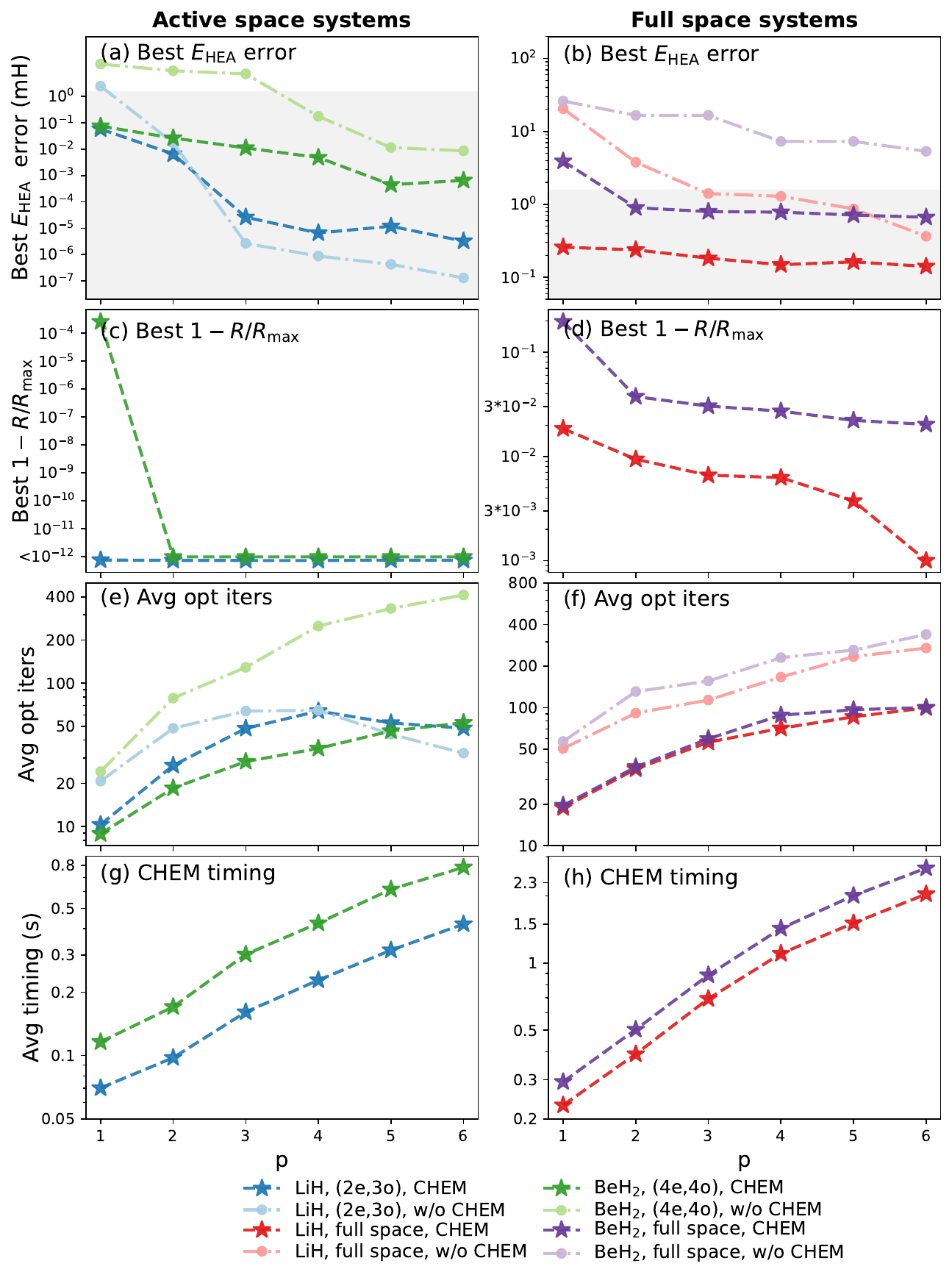}
    \caption{Optimized HEA energies (in (a) and (b)), reward function (in (c) and (d)), optimization costs (in (e) and (f)), and the wall time (in (g) and (h)) for \ourmethod{} with and without active space treatments of the three molecules. Intransparent lines with star markers are for \ourmethod{} and transparent lines with circle markers are for the VQE based on the Hamiltonian without Clifford transformation. 
    All $y$ axes are plotted on logarithm scales.
    The shaded areas in (a) and (b) correspond to the chemical accuracy of 1 kcal/mol (or 1.6 mH). }
    \label{fig:stats}
\end{figure}

We then examine the performance of \ourmethod{} by extensive benchmark over 7 different electronic structure problems, ranging from 4 qubits to 12 qubits.
In Fig.~\ref{fig:stats} (a) and (b) we show the best error over 1000 different initializations based on both \ourmethod{} and the $R_y$ ansatz.
For simpler systems such as \ch{LiH} with (2e, 3o) active space shown in Fig.~\ref{fig:stats} (a), both methods can achieve high accuracy.
For larger systems such as \ch{BeH2} without active space approximation, \ourmethod{} consistently outperforms the $R_y$ ansatz.
While the $R_y$ ansatz without Hamiltonian engineering struggles to decrease the error below 10 mH,
\ourmethod{} reaches chemical accuracy with $p=2$.
Since the number of CNOT gates in the circuit is $(N-1)p$ in our setting (Fig.~\ref{fig:workflow}), \ourmethod{} drives the error below 1 mH with only 22 CNOT gates in the circuit.
Note that the circuit used in Fig.~\ref{fig:workflow} has linear qubit connectivity
which is the minimal requirement for the quantum hardware architecture.
Upon increasing $p$, the best error from \ourmethod{} decreases at a slow pace.
We conjecture that the origin of the slow pace is the saturation of the reward function $R$ to $R_\text{max}$.
As shown in Fig.~\ref{fig:stats}(c) and (d), the optimized $R$ is very close to the theoretical maximum value. 
For smaller systems employing active space approximation, the reward function $R$ hardly changes with $p$, which is a possible reason for the fluctuation of best errors shown in Fig.~\ref{fig:stats}(a).
Therefore, a possible direction to further improve \ourmethod{} when $p$ becomes large is to design a more sophisticated $R$ that acts as a more faithful proxy to the VQE-optimized energy.
A straightforward approach without much additional complexity is to consider higher-order energy derivatives.
One may notice that the circuit used for \ourmethod{} has more parameters than the $R_y$ ansatz with a given number of layers $p$
and expect the VQE optimization of \ourmethod{} to be more difficult than that of the $R_y$ ansatz without \ourmethod{}.
However, as shown in Fig.~\ref{fig:stats}(e) and (f), the optimization of \ourmethod{} takes fewer steps to converge due to the large initial gradients.
For the LiH (2e, 3o) system, the benefits of our method, CHEM, are not as pronounced for larger $p$ values due to the saturation of the reward function. However, it's important to note that this phenomenon is only observed in smaller systems where the optimization process does not present a significant challenge. Consequently, we believe that the impact of this occurrence on the practical application of VQE is minimal.
Also, as shown in Fig.~\ref{fig:grad}, many parameters have zero gradients and can potentially be set as frozen during parameter optimization, which further reduces the optimization difficulty.
Lastly, we note that it is almost instant to perform \ourmethod{} in all cases reported (Fig.~\ref{fig:stats}(g) and (h)).
Besides, the wall time scales linearly with $p$, in agreement with the formal scaling of 
$\mathcal{O}(N^2(N+p))$.

In Table~\ref{tab:stats} we list the statistics of errors including the minimum error (MIN) and the mean absolute error (MAE) over the 7 different systems at $p=3$.
Due to the lack of a proper initial guess for traditional HEA, it is not trivial for the $R_y$ ansatz to reach the HF energy. For example, the MAEs of full space \ch{LiH} and \ch{BeH2} are higher than the HF error when the $R_y$ ansatz is employed, although their minimum errors are lower.
On the other hand, \ourmethod{} starts the optimization from the HF energy with large energy gradients. As a result, the MAE of \ourmethod{} is much smaller than the HF error and reaches chemical accuracy except for the \ch{BeH2} system.
And if the minimum error is considered, chemical accuracy is reached for all systems studied in this work.
The distributions of the optimized \ourmethod{} energies of the systems are included in the Supporting Information (Fig. S3-S6).

\section{Conclusion and Outlook}
In conclusion, we have developed a Clifford Hamiltonian Engineering for Molecules (\ourmethod{}) approach for highly accurate molecular energy estimation with shallow quantum circuits. By designing Hamiltonian transformation $\ourmethodU{}$ that maximizes the energy gradient, \ourmethod{} remarkably alleviates the optimization difficulties associated with hardware-efficient ansatz without breaking the arbitrariness of the circuit topology. 
Based on the efficient algorithmic implementation, \ourmethod{} bears minimal classical computing overhead, with a computational complexity of $\mathcal{O}(N^2(N+p))$,
and no additional quantum resources are needed. 
The numerical experimental results for a variety of molecular systems show that our approach scales well with both $N$ and $p$. 
In particular, we find \ourmethod{} reaches chemical accuracy for these molecules
using quantum circuits with fewer than 30 two-qubit gates and linear qubit connectivity.
This level of performance is unprecedented in the literature and highlights the potential of our method for practical applications in quantum computational chemistry.

Looking forward, there are several avenues for further research and development. First, by designing a more sophisticated reward function $R$ and a better algorithm to match $x_h$ and $x_Q$, it is possible to further improve the performance of \ourmethod{}. Second, integrating \ourmethod{} with other optimization techniques and ansätze could lead to even more efficient and accurate quantum algorithms. Finally, the implementation of our Clifford Hamiltonian engineering approach on real quantum hardware would provide valuable insights into its performance and robustness in the presence of hardware noise.

It is intriguing to contemplate the potential application of \ourmethod{} to long-term quantum algorithms, such as quantum phase estimation (QPE) and adiabatic state preparation (ASP). 
In the context of QPE, increasing the overlap between the trial wavefunction and the ground state presents a significant challenge. In this scenario, both VQE and \ourmethod{} could provide good input for QPE by generating accurate trial wavefunctions.
On the other hand, the success of ASP is dependent on maintaining a large energy gap along the evolution path. Since \ourmethod{} preserves the eigenspectrum, an effective strategy to utilize \ourmethod{} for ASP deserves further investigation.

Overall, our work contributes to the ongoing efforts to develop efficient and accurate quantum algorithms for chemistry applications and brings us one step closer to unleashing the full potential of quantum computational chemistry on near-term quantum devices.

\section*{Data availability}
All graphs and results presented in this study are shared on a GitHub repository (\url{https://github.com/sherrylixuecheng/CHEM}). The additional figures for test results are shown in the Supporting Information, and full accuracy statistics are included in a separate excel (Supplementary Data).

\section*{Code availability}
The entire CHEM framework is available on GitHub (\url{https://github.com/sherrylixuecheng/CHEM}) with example jupyter notebooks and all the testing results. The example codes to perform HEA simulations available on \textsc{TenCirChem} (\url{https://github.com/tencent-quantum-lab/tencirchem})\cite{li2023tencirchem}.

\section*{Author contributions}
J.S. and L.W. developed the idea and implemented the codes. L.C. performed the numerical experiments. J.S., L.C., and L.W. analyzed the results and wrote the manuscript.

\section*{Acknowledgements}
Jiace Sun thanks the support from Hongyan Scholarship.

\section*{Supporting Information}
The Supporting Information is available free of charge online, including
distributions of the optimized parameters of the $R_y$ ansatz and distributions of the energy gradients and optimized energies of the \ourmethod{} method.

\bibliography{main}

\begin{thebibliography}{61}%
\makeatletter
\providecommand \@ifxundefined [1]{%
 \@ifx{#1\undefined}
}%
\providecommand \@ifnum [1]{%
 \ifnum #1\expandafter \@firstoftwo
 \else \expandafter \@secondoftwo
 \fi
}%
\providecommand \@ifx [1]{%
 \ifx #1\expandafter \@firstoftwo
 \else \expandafter \@secondoftwo
 \fi
}%
\providecommand \natexlab [1]{#1}%
\providecommand \enquote  [1]{``#1''}%
\providecommand \bibnamefont  [1]{#1}%
\providecommand \bibfnamefont [1]{#1}%
\providecommand \citenamefont [1]{#1}%
\providecommand \href@noop [0]{\@secondoftwo}%
\providecommand \href [0]{\begingroup \@sanitize@url \@href}%
\providecommand \@href[1]{\@@startlink{#1}\@@href}%
\providecommand \@@href[1]{\endgroup#1\@@endlink}%
\providecommand \@sanitize@url [0]{\catcode `\\12\catcode `\$12\catcode
  `\&12\catcode `\#12\catcode `\^12\catcode `\_12\catcode `\%12\relax}%
\providecommand \@@startlink[1]{}%
\providecommand \@@endlink[0]{}%
\providecommand \url  [0]{\begingroup\@sanitize@url \@url }%
\providecommand \@url [1]{\endgroup\@href {#1}{\urlprefix }}%
\providecommand \urlprefix  [0]{URL }%
\providecommand \Eprint [0]{\href }%
\providecommand \doibase [0]{https://doi.org/}%
\providecommand \selectlanguage [0]{\@gobble}%
\providecommand \bibinfo  [0]{\@secondoftwo}%
\providecommand \bibfield  [0]{\@secondoftwo}%
\providecommand \translation [1]{[#1]}%
\providecommand \BibitemOpen [0]{}%
\providecommand \bibitemStop [0]{}%
\providecommand \bibitemNoStop [0]{.\EOS\space}%
\providecommand \EOS [0]{\spacefactor3000\relax}%
\providecommand \BibitemShut  [1]{\csname bibitem#1\endcsname}%
\let\auto@bib@innerbib\@empty
\bibitem [{\citenamefont {Cao}\ \emph {et~al.}(2019)\citenamefont {Cao},
  \citenamefont {Romero}, \citenamefont {Olson}, \citenamefont {Degroote},
  \citenamefont {Johnson}, \citenamefont {Kieferov{\'a}}, \citenamefont
  {Kivlichan}, \citenamefont {Menke}, \citenamefont {Peropadre}, \citenamefont
  {Sawaya}, \citenamefont {Sim}, \citenamefont {Veis},\ and\ \citenamefont
  {Aspuru-Guzik}}]{cao2019quantum}%
  \BibitemOpen
  \bibfield  {author} {\bibinfo {author} {\bibfnamefont {Y.}~\bibnamefont
  {Cao}}, \bibinfo {author} {\bibfnamefont {J.}~\bibnamefont {Romero}},
  \bibinfo {author} {\bibfnamefont {J.~P.}\ \bibnamefont {Olson}}, \bibinfo
  {author} {\bibfnamefont {M.}~\bibnamefont {Degroote}}, \bibinfo {author}
  {\bibfnamefont {P.~D.}\ \bibnamefont {Johnson}}, \bibinfo {author}
  {\bibfnamefont {M.}~\bibnamefont {Kieferov{\'a}}}, \bibinfo {author}
  {\bibfnamefont {I.~D.}\ \bibnamefont {Kivlichan}}, \bibinfo {author}
  {\bibfnamefont {T.}~\bibnamefont {Menke}}, \bibinfo {author} {\bibfnamefont
  {B.}~\bibnamefont {Peropadre}}, \bibinfo {author} {\bibfnamefont {N.~P.}\
  \bibnamefont {Sawaya}}, \bibinfo {author} {\bibfnamefont {S.}~\bibnamefont
  {Sim}}, \bibinfo {author} {\bibfnamefont {L.}~\bibnamefont {Veis}},\ and\
  \bibinfo {author} {\bibfnamefont {A.}~\bibnamefont {Aspuru-Guzik}},\
  }\bibfield  {title} {\enquote {\bibinfo {title} {Quantum chemistry in the age
  of quantum computing},}\ }\href@noop {} {\bibfield  {journal} {\bibinfo
  {journal} {Chem. Rev.}\ }\textbf {\bibinfo {volume} {119}},\ \bibinfo {pages}
  {10856--10915} (\bibinfo {year} {2019})}\BibitemShut {NoStop}%
\bibitem [{\citenamefont {Bauer}\ \emph {et~al.}(2020)\citenamefont {Bauer},
  \citenamefont {Bravyi}, \citenamefont {Motta},\ and\ \citenamefont
  {Chan}}]{bauer2020quantum}%
  \BibitemOpen
  \bibfield  {author} {\bibinfo {author} {\bibfnamefont {B.}~\bibnamefont
  {Bauer}}, \bibinfo {author} {\bibfnamefont {S.}~\bibnamefont {Bravyi}},
  \bibinfo {author} {\bibfnamefont {M.}~\bibnamefont {Motta}},\ and\ \bibinfo
  {author} {\bibfnamefont {G.~K.-L.}\ \bibnamefont {Chan}},\ }\bibfield
  {title} {\enquote {\bibinfo {title} {Quantum algorithms for quantum chemistry
  and quantum materials science},}\ }\href@noop {} {\bibfield  {journal}
  {\bibinfo  {journal} {Chem. Rev.}\ }\textbf {\bibinfo {volume} {120}},\
  \bibinfo {pages} {12685--12717} (\bibinfo {year} {2020})}\BibitemShut
  {NoStop}%
\bibitem [{\citenamefont {McArdle}\ \emph {et~al.}(2020)\citenamefont
  {McArdle}, \citenamefont {Endo}, \citenamefont {Aspuru-Guzik}, \citenamefont
  {Benjamin},\ and\ \citenamefont {Yuan}}]{mcardle2020quantum}%
  \BibitemOpen
  \bibfield  {author} {\bibinfo {author} {\bibfnamefont {S.}~\bibnamefont
  {McArdle}}, \bibinfo {author} {\bibfnamefont {S.}~\bibnamefont {Endo}},
  \bibinfo {author} {\bibfnamefont {A.}~\bibnamefont {Aspuru-Guzik}}, \bibinfo
  {author} {\bibfnamefont {S.~C.}\ \bibnamefont {Benjamin}},\ and\ \bibinfo
  {author} {\bibfnamefont {X.}~\bibnamefont {Yuan}},\ }\bibfield  {title}
  {\enquote {\bibinfo {title} {Quantum computational chemistry},}\ }\href@noop
  {} {\bibfield  {journal} {\bibinfo  {journal} {Rev. Mod. Phys.}\ }\textbf
  {\bibinfo {volume} {92}},\ \bibinfo {pages} {015003} (\bibinfo {year}
  {2020})}\BibitemShut {NoStop}%
\bibitem [{\citenamefont {Motta}\ and\ \citenamefont
  {Rice}(2022)}]{motta2022emerging}%
  \BibitemOpen
  \bibfield  {author} {\bibinfo {author} {\bibfnamefont {M.}~\bibnamefont
  {Motta}}\ and\ \bibinfo {author} {\bibfnamefont {J.~E.}\ \bibnamefont
  {Rice}},\ }\bibfield  {title} {\enquote {\bibinfo {title} {Emerging quantum
  computing algorithms for quantum chemistry},}\ }\href@noop {} {\bibfield
  {journal} {\bibinfo  {journal} {Wiley Interdiscip. Rev. Comput. Mol. Sci.}\
  }\textbf {\bibinfo {volume} {12}},\ \bibinfo {pages} {e1580} (\bibinfo {year}
  {2022})}\BibitemShut {NoStop}%
\bibitem [{\citenamefont {Liu}\ \emph {et~al.}(2022)\citenamefont {Liu},
  \citenamefont {Fan}, \citenamefont {Li},\ and\ \citenamefont
  {Yang}}]{liu2022quantum}%
  \BibitemOpen
  \bibfield  {author} {\bibinfo {author} {\bibfnamefont {J.}~\bibnamefont
  {Liu}}, \bibinfo {author} {\bibfnamefont {Y.}~\bibnamefont {Fan}}, \bibinfo
  {author} {\bibfnamefont {Z.}~\bibnamefont {Li}},\ and\ \bibinfo {author}
  {\bibfnamefont {J.}~\bibnamefont {Yang}},\ }\bibfield  {title} {\enquote
  {\bibinfo {title} {Quantum algorithms for electronic structures: basis sets
  and boundary conditions},}\ }\href@noop {} {\bibfield  {journal} {\bibinfo
  {journal} {Chem. Soc. Rev.}\ ,\ \bibinfo {pages} {3263--3279}} (\bibinfo
  {year} {2022})}\BibitemShut {NoStop}%
\bibitem [{\citenamefont {Ma}\ \emph {et~al.}(2023)\citenamefont {Ma},
  \citenamefont {Liu}, \citenamefont {Shang}, \citenamefont {Fan},
  \citenamefont {Li},\ and\ \citenamefont {Yang}}]{ma2023multiscale}%
  \BibitemOpen
  \bibfield  {author} {\bibinfo {author} {\bibfnamefont {H.}~\bibnamefont
  {Ma}}, \bibinfo {author} {\bibfnamefont {J.}~\bibnamefont {Liu}}, \bibinfo
  {author} {\bibfnamefont {H.}~\bibnamefont {Shang}}, \bibinfo {author}
  {\bibfnamefont {Y.}~\bibnamefont {Fan}}, \bibinfo {author} {\bibfnamefont
  {Z.}~\bibnamefont {Li}},\ and\ \bibinfo {author} {\bibfnamefont
  {J.}~\bibnamefont {Yang}},\ }\bibfield  {title} {\enquote {\bibinfo {title}
  {Multiscale quantum algorithms for quantum chemistry},}\ }\href@noop {}
  {\bibfield  {journal} {\bibinfo  {journal} {Chem. Sci.}\ }\textbf {\bibinfo
  {volume} {14}},\ \bibinfo {pages} {3190--3205} (\bibinfo {year}
  {2023})}\BibitemShut {NoStop}%
\bibitem [{\citenamefont {Peruzzo}\ \emph {et~al.}(2014)\citenamefont
  {Peruzzo}, \citenamefont {McClean}, \citenamefont {Shadbolt}, \citenamefont
  {Yung}, \citenamefont {Zhou}, \citenamefont {Love}, \citenamefont
  {Aspuru-Guzik},\ and\ \citenamefont {O’brien}}]{peruzzo2014variational}%
  \BibitemOpen
  \bibfield  {author} {\bibinfo {author} {\bibfnamefont {A.}~\bibnamefont
  {Peruzzo}}, \bibinfo {author} {\bibfnamefont {J.}~\bibnamefont {McClean}},
  \bibinfo {author} {\bibfnamefont {P.}~\bibnamefont {Shadbolt}}, \bibinfo
  {author} {\bibfnamefont {M.-H.}\ \bibnamefont {Yung}}, \bibinfo {author}
  {\bibfnamefont {X.-Q.}\ \bibnamefont {Zhou}}, \bibinfo {author}
  {\bibfnamefont {P.~J.}\ \bibnamefont {Love}}, \bibinfo {author}
  {\bibfnamefont {A.}~\bibnamefont {Aspuru-Guzik}},\ and\ \bibinfo {author}
  {\bibfnamefont {J.~L.}\ \bibnamefont {O’brien}},\ }\bibfield  {title}
  {\enquote {\bibinfo {title} {A variational eigenvalue solver on a photonic
  quantum processor},}\ }\href@noop {} {\bibfield  {journal} {\bibinfo
  {journal} {Nat. Commun.}\ }\textbf {\bibinfo {volume} {5}},\ \bibinfo {pages}
  {4213} (\bibinfo {year} {2014})}\BibitemShut {NoStop}%
\bibitem [{\citenamefont {Cerezo}\ \emph {et~al.}(2021)\citenamefont {Cerezo},
  \citenamefont {Arrasmith}, \citenamefont {Babbush}, \citenamefont {Benjamin},
  \citenamefont {Endo}, \citenamefont {Fujii}, \citenamefont {McClean},
  \citenamefont {Mitarai}, \citenamefont {Yuan}, \citenamefont {Cincio},\ and\
  \citenamefont {Coles}}]{cerezo2021variational}%
  \BibitemOpen
  \bibfield  {author} {\bibinfo {author} {\bibfnamefont {M.}~\bibnamefont
  {Cerezo}}, \bibinfo {author} {\bibfnamefont {A.}~\bibnamefont {Arrasmith}},
  \bibinfo {author} {\bibfnamefont {R.}~\bibnamefont {Babbush}}, \bibinfo
  {author} {\bibfnamefont {S.~C.}\ \bibnamefont {Benjamin}}, \bibinfo {author}
  {\bibfnamefont {S.}~\bibnamefont {Endo}}, \bibinfo {author} {\bibfnamefont
  {K.}~\bibnamefont {Fujii}}, \bibinfo {author} {\bibfnamefont {J.~R.}\
  \bibnamefont {McClean}}, \bibinfo {author} {\bibfnamefont {K.}~\bibnamefont
  {Mitarai}}, \bibinfo {author} {\bibfnamefont {X.}~\bibnamefont {Yuan}},
  \bibinfo {author} {\bibfnamefont {L.}~\bibnamefont {Cincio}},\ and\ \bibinfo
  {author} {\bibfnamefont {P.~J.}\ \bibnamefont {Coles}},\ }\bibfield  {title}
  {\enquote {\bibinfo {title} {Variational quantum algorithms},}\ }\href@noop
  {} {\bibfield  {journal} {\bibinfo  {journal} {Nat. Rev. Phys.}\ }\textbf
  {\bibinfo {volume} {3}},\ \bibinfo {pages} {625--644} (\bibinfo {year}
  {2021})}\BibitemShut {NoStop}%
\bibitem [{\citenamefont {Tilly}\ \emph {et~al.}(2022)\citenamefont {Tilly},
  \citenamefont {Chen}, \citenamefont {Cao}, \citenamefont {Picozzi},
  \citenamefont {Setia}, \citenamefont {Li}, \citenamefont {Grant},
  \citenamefont {Wossnig}, \citenamefont {Rungger}, \citenamefont {Booth},\
  and\ \citenamefont {Tennyson}}]{tilly2022variational}%
  \BibitemOpen
  \bibfield  {author} {\bibinfo {author} {\bibfnamefont {J.}~\bibnamefont
  {Tilly}}, \bibinfo {author} {\bibfnamefont {H.}~\bibnamefont {Chen}},
  \bibinfo {author} {\bibfnamefont {S.}~\bibnamefont {Cao}}, \bibinfo {author}
  {\bibfnamefont {D.}~\bibnamefont {Picozzi}}, \bibinfo {author} {\bibfnamefont
  {K.}~\bibnamefont {Setia}}, \bibinfo {author} {\bibfnamefont
  {Y.}~\bibnamefont {Li}}, \bibinfo {author} {\bibfnamefont {E.}~\bibnamefont
  {Grant}}, \bibinfo {author} {\bibfnamefont {L.}~\bibnamefont {Wossnig}},
  \bibinfo {author} {\bibfnamefont {I.}~\bibnamefont {Rungger}}, \bibinfo
  {author} {\bibfnamefont {G.~H.}\ \bibnamefont {Booth}},\ and\ \bibinfo
  {author} {\bibfnamefont {J.}~\bibnamefont {Tennyson}},\ }\bibfield  {title}
  {\enquote {\bibinfo {title} {The variational quantum eigensolver: a review of
  methods and best practices},}\ }\href@noop {} {\bibfield  {journal} {\bibinfo
   {journal} {Phys. Rep.}\ }\textbf {\bibinfo {volume} {986}},\ \bibinfo
  {pages} {1--128} (\bibinfo {year} {2022})}\BibitemShut {NoStop}%
\bibitem [{\citenamefont {Preskill}(2018)}]{Preskill18}%
  \BibitemOpen
  \bibfield  {author} {\bibinfo {author} {\bibfnamefont {J.}~\bibnamefont
  {Preskill}},\ }\bibfield  {title} {\enquote {\bibinfo {title} {Quantum
  {C}omputing in the {NISQ} era and beyond},}\ }\href
  {https://doi.org/10.22331/q-2018-08-06-79} {\bibfield  {journal} {\bibinfo
  {journal} {{Quantum}}\ }\textbf {\bibinfo {volume} {2}},\ \bibinfo {pages}
  {79} (\bibinfo {year} {2018})}\BibitemShut {NoStop}%
\bibitem [{\citenamefont {Bharti}\ \emph {et~al.}(2022)\citenamefont {Bharti},
  \citenamefont {Cervera-Lierta}, \citenamefont {Kyaw}, \citenamefont {Haug},
  \citenamefont {Alperin-Lea}, \citenamefont {Anand}, \citenamefont {Degroote},
  \citenamefont {Heimonen}, \citenamefont {Kottmann}, \citenamefont {Menke},
  \citenamefont {Mok}, \citenamefont {Sim}, \citenamefont {Kwek},\ and\
  \citenamefont {Aspuru-Guzik}}]{bharti2022noisy}%
  \BibitemOpen
  \bibfield  {author} {\bibinfo {author} {\bibfnamefont {K.}~\bibnamefont
  {Bharti}}, \bibinfo {author} {\bibfnamefont {A.}~\bibnamefont
  {Cervera-Lierta}}, \bibinfo {author} {\bibfnamefont {T.~H.}\ \bibnamefont
  {Kyaw}}, \bibinfo {author} {\bibfnamefont {T.}~\bibnamefont {Haug}}, \bibinfo
  {author} {\bibfnamefont {S.}~\bibnamefont {Alperin-Lea}}, \bibinfo {author}
  {\bibfnamefont {A.}~\bibnamefont {Anand}}, \bibinfo {author} {\bibfnamefont
  {M.}~\bibnamefont {Degroote}}, \bibinfo {author} {\bibfnamefont
  {H.}~\bibnamefont {Heimonen}}, \bibinfo {author} {\bibfnamefont {J.~S.}\
  \bibnamefont {Kottmann}}, \bibinfo {author} {\bibfnamefont {T.}~\bibnamefont
  {Menke}}, \bibinfo {author} {\bibfnamefont {W.-K.}\ \bibnamefont {Mok}},
  \bibinfo {author} {\bibfnamefont {S.}~\bibnamefont {Sim}}, \bibinfo {author}
  {\bibfnamefont {L.-C.}\ \bibnamefont {Kwek}},\ and\ \bibinfo {author}
  {\bibfnamefont {A.}~\bibnamefont {Aspuru-Guzik}},\ }\bibfield  {title}
  {\enquote {\bibinfo {title} {Noisy intermediate-scale quantum algorithms},}\
  }\href@noop {} {\bibfield  {journal} {\bibinfo  {journal} {Rev. Mod. Phys.}\
  }\textbf {\bibinfo {volume} {94}},\ \bibinfo {pages} {015004} (\bibinfo
  {year} {2022})}\BibitemShut {NoStop}%
\bibitem [{\citenamefont {Kandala}\ \emph {et~al.}(2017)\citenamefont
  {Kandala}, \citenamefont {Mezzacapo}, \citenamefont {Temme}, \citenamefont
  {Takita}, \citenamefont {Brink}, \citenamefont {Chow},\ and\ \citenamefont
  {Gambetta}}]{kandala2017hardware}%
  \BibitemOpen
  \bibfield  {author} {\bibinfo {author} {\bibfnamefont {A.}~\bibnamefont
  {Kandala}}, \bibinfo {author} {\bibfnamefont {A.}~\bibnamefont {Mezzacapo}},
  \bibinfo {author} {\bibfnamefont {K.}~\bibnamefont {Temme}}, \bibinfo
  {author} {\bibfnamefont {M.}~\bibnamefont {Takita}}, \bibinfo {author}
  {\bibfnamefont {M.}~\bibnamefont {Brink}}, \bibinfo {author} {\bibfnamefont
  {J.~M.}\ \bibnamefont {Chow}},\ and\ \bibinfo {author} {\bibfnamefont
  {J.~M.}\ \bibnamefont {Gambetta}},\ }\bibfield  {title} {\enquote {\bibinfo
  {title} {Hardware-efficient variational quantum eigensolver for small
  molecules and quantum magnets},}\ }\href@noop {} {\bibfield  {journal}
  {\bibinfo  {journal} {Nature}\ }\textbf {\bibinfo {volume} {549}},\ \bibinfo
  {pages} {242--246} (\bibinfo {year} {2017})}\BibitemShut {NoStop}%
\bibitem [{\citenamefont {{Google AI Quantum and Collaborators}}\ \emph
  {et~al.}(2020)\citenamefont {{Google AI Quantum and Collaborators}},
  \citenamefont {Arute}, \citenamefont {Arya}, \citenamefont {Babbush},
  \citenamefont {Bacon}, \citenamefont {Bardin}, \citenamefont {Barends},
  \citenamefont {Boixo}, \citenamefont {Broughton}, \citenamefont {Buckley},
  \citenamefont {Buell}, \citenamefont {Burkett}, \citenamefont {Bushnell},
  \citenamefont {Chen}, \citenamefont {Chen}, \citenamefont {Chiaro},
  \citenamefont {Collins}, \citenamefont {Courtney}, \citenamefont {Demura},
  \citenamefont {Dunsworth}, \citenamefont {Farhi}, \citenamefont {Fowler},
  \citenamefont {Foxen}, \citenamefont {Gidney}, \citenamefont {Giustina},
  \citenamefont {Graff}, \citenamefont {Habegger}, \citenamefont {Harrigan},
  \citenamefont {Ho}, \citenamefont {Hong}, \citenamefont {Huang},
  \citenamefont {Huggins}, \citenamefont {Ioffe}, \citenamefont {Isakov},
  \citenamefont {Jeffrey}, \citenamefont {Jiang}, \citenamefont {Jones},
  \citenamefont {Kafri}, \citenamefont {Kechedzhi}, \citenamefont {Kelly},
  \citenamefont {Kim}, \citenamefont {Klimov}, \citenamefont {Korotkov},
  \citenamefont {Kostritsa}, \citenamefont {Landhuis}, \citenamefont {Laptev},
  \citenamefont {Lindmark}, \citenamefont {Lucero}, \citenamefont {Martin},
  \citenamefont {Martinis}, \citenamefont {McClean}, \citenamefont {McEwen},
  \citenamefont {Megrant}, \citenamefont {Mi}, \citenamefont {Mohseni},
  \citenamefont {Mruczkiewicz}, \citenamefont {Mutus}, \citenamefont {Naaman},
  \citenamefont {Neeley}, \citenamefont {Neill}, \citenamefont {Neven},
  \citenamefont {Niu}, \citenamefont {O’Brien}, \citenamefont {Ostby},
  \citenamefont {Petukhov}, \citenamefont {Putterman}, \citenamefont
  {Quintana}, \citenamefont {Roushan}, \citenamefont {Rubin}, \citenamefont
  {Sank}, \citenamefont {Satzinger}, \citenamefont {Smelyanskiy}, \citenamefont
  {Strain}, \citenamefont {Sung}, \citenamefont {Szalay}, \citenamefont
  {Takeshita}, \citenamefont {Vainsencher}, \citenamefont {White},
  \citenamefont {Wiebe}, \citenamefont {Yao}, \citenamefont {Yeh},\ and\
  \citenamefont {Zalcman}}]{google2020hartree}%
  \BibitemOpen
  \bibfield  {author} {\bibinfo {author} {\bibnamefont {{Google AI Quantum and
  Collaborators}}}, \bibinfo {author} {\bibfnamefont {F.}~\bibnamefont
  {Arute}}, \bibinfo {author} {\bibfnamefont {K.}~\bibnamefont {Arya}},
  \bibinfo {author} {\bibfnamefont {R.}~\bibnamefont {Babbush}}, \bibinfo
  {author} {\bibfnamefont {D.}~\bibnamefont {Bacon}}, \bibinfo {author}
  {\bibfnamefont {J.~C.}\ \bibnamefont {Bardin}}, \bibinfo {author}
  {\bibfnamefont {R.}~\bibnamefont {Barends}}, \bibinfo {author} {\bibfnamefont
  {S.}~\bibnamefont {Boixo}}, \bibinfo {author} {\bibfnamefont
  {M.}~\bibnamefont {Broughton}}, \bibinfo {author} {\bibfnamefont {B.~B.}\
  \bibnamefont {Buckley}}, \bibinfo {author} {\bibfnamefont {D.~A.}\
  \bibnamefont {Buell}}, \bibinfo {author} {\bibfnamefont {B.}~\bibnamefont
  {Burkett}}, \bibinfo {author} {\bibfnamefont {N.}~\bibnamefont {Bushnell}},
  \bibinfo {author} {\bibfnamefont {Y.}~\bibnamefont {Chen}}, \bibinfo {author}
  {\bibfnamefont {Z.}~\bibnamefont {Chen}}, \bibinfo {author} {\bibfnamefont
  {B.}~\bibnamefont {Chiaro}}, \bibinfo {author} {\bibfnamefont
  {R.}~\bibnamefont {Collins}}, \bibinfo {author} {\bibfnamefont
  {W.}~\bibnamefont {Courtney}}, \bibinfo {author} {\bibfnamefont
  {S.}~\bibnamefont {Demura}}, \bibinfo {author} {\bibfnamefont
  {A.}~\bibnamefont {Dunsworth}}, \bibinfo {author} {\bibfnamefont
  {E.}~\bibnamefont {Farhi}}, \bibinfo {author} {\bibfnamefont
  {A.}~\bibnamefont {Fowler}}, \bibinfo {author} {\bibfnamefont
  {B.}~\bibnamefont {Foxen}}, \bibinfo {author} {\bibfnamefont
  {C.}~\bibnamefont {Gidney}}, \bibinfo {author} {\bibfnamefont
  {M.}~\bibnamefont {Giustina}}, \bibinfo {author} {\bibfnamefont
  {R.}~\bibnamefont {Graff}}, \bibinfo {author} {\bibfnamefont
  {S.}~\bibnamefont {Habegger}}, \bibinfo {author} {\bibfnamefont {M.~P.}\
  \bibnamefont {Harrigan}}, \bibinfo {author} {\bibfnamefont {A.}~\bibnamefont
  {Ho}}, \bibinfo {author} {\bibfnamefont {S.}~\bibnamefont {Hong}}, \bibinfo
  {author} {\bibfnamefont {T.}~\bibnamefont {Huang}}, \bibinfo {author}
  {\bibfnamefont {W.~J.}\ \bibnamefont {Huggins}}, \bibinfo {author}
  {\bibfnamefont {L.}~\bibnamefont {Ioffe}}, \bibinfo {author} {\bibfnamefont
  {S.~V.}\ \bibnamefont {Isakov}}, \bibinfo {author} {\bibfnamefont
  {E.}~\bibnamefont {Jeffrey}}, \bibinfo {author} {\bibfnamefont
  {Z.}~\bibnamefont {Jiang}}, \bibinfo {author} {\bibfnamefont
  {C.}~\bibnamefont {Jones}}, \bibinfo {author} {\bibfnamefont
  {D.}~\bibnamefont {Kafri}}, \bibinfo {author} {\bibfnamefont
  {K.}~\bibnamefont {Kechedzhi}}, \bibinfo {author} {\bibfnamefont
  {J.}~\bibnamefont {Kelly}}, \bibinfo {author} {\bibfnamefont
  {S.}~\bibnamefont {Kim}}, \bibinfo {author} {\bibfnamefont {P.~V.}\
  \bibnamefont {Klimov}}, \bibinfo {author} {\bibfnamefont {A.}~\bibnamefont
  {Korotkov}}, \bibinfo {author} {\bibfnamefont {F.}~\bibnamefont {Kostritsa}},
  \bibinfo {author} {\bibfnamefont {D.}~\bibnamefont {Landhuis}}, \bibinfo
  {author} {\bibfnamefont {P.}~\bibnamefont {Laptev}}, \bibinfo {author}
  {\bibfnamefont {M.}~\bibnamefont {Lindmark}}, \bibinfo {author}
  {\bibfnamefont {E.}~\bibnamefont {Lucero}}, \bibinfo {author} {\bibfnamefont
  {O.}~\bibnamefont {Martin}}, \bibinfo {author} {\bibfnamefont {J.~M.}\
  \bibnamefont {Martinis}}, \bibinfo {author} {\bibfnamefont {J.~R.}\
  \bibnamefont {McClean}}, \bibinfo {author} {\bibfnamefont {M.}~\bibnamefont
  {McEwen}}, \bibinfo {author} {\bibfnamefont {A.}~\bibnamefont {Megrant}},
  \bibinfo {author} {\bibfnamefont {X.}~\bibnamefont {Mi}}, \bibinfo {author}
  {\bibfnamefont {M.}~\bibnamefont {Mohseni}}, \bibinfo {author} {\bibfnamefont
  {W.}~\bibnamefont {Mruczkiewicz}}, \bibinfo {author} {\bibfnamefont
  {J.}~\bibnamefont {Mutus}}, \bibinfo {author} {\bibfnamefont
  {O.}~\bibnamefont {Naaman}}, \bibinfo {author} {\bibfnamefont
  {M.}~\bibnamefont {Neeley}}, \bibinfo {author} {\bibfnamefont
  {C.}~\bibnamefont {Neill}}, \bibinfo {author} {\bibfnamefont
  {H.}~\bibnamefont {Neven}}, \bibinfo {author} {\bibfnamefont {M.~Y.}\
  \bibnamefont {Niu}}, \bibinfo {author} {\bibfnamefont {T.~E.}\ \bibnamefont
  {O’Brien}}, \bibinfo {author} {\bibfnamefont {E.}~\bibnamefont {Ostby}},
  \bibinfo {author} {\bibfnamefont {A.}~\bibnamefont {Petukhov}}, \bibinfo
  {author} {\bibfnamefont {H.}~\bibnamefont {Putterman}}, \bibinfo {author}
  {\bibfnamefont {C.}~\bibnamefont {Quintana}}, \bibinfo {author}
  {\bibfnamefont {P.}~\bibnamefont {Roushan}}, \bibinfo {author} {\bibfnamefont
  {N.~C.}\ \bibnamefont {Rubin}}, \bibinfo {author} {\bibfnamefont
  {D.}~\bibnamefont {Sank}}, \bibinfo {author} {\bibfnamefont {K.~J.}\
  \bibnamefont {Satzinger}}, \bibinfo {author} {\bibfnamefont {V.}~\bibnamefont
  {Smelyanskiy}}, \bibinfo {author} {\bibfnamefont {D.}~\bibnamefont {Strain}},
  \bibinfo {author} {\bibfnamefont {K.~J.}\ \bibnamefont {Sung}}, \bibinfo
  {author} {\bibfnamefont {M.}~\bibnamefont {Szalay}}, \bibinfo {author}
  {\bibfnamefont {T.~Y.}\ \bibnamefont {Takeshita}}, \bibinfo {author}
  {\bibfnamefont {A.}~\bibnamefont {Vainsencher}}, \bibinfo {author}
  {\bibfnamefont {T.}~\bibnamefont {White}}, \bibinfo {author} {\bibfnamefont
  {N.}~\bibnamefont {Wiebe}}, \bibinfo {author} {\bibfnamefont {Z.~J.}\
  \bibnamefont {Yao}}, \bibinfo {author} {\bibfnamefont {P.}~\bibnamefont
  {Yeh}},\ and\ \bibinfo {author} {\bibfnamefont {A.}~\bibnamefont {Zalcman}},\
  }\bibfield  {title} {\enquote {\bibinfo {title} {Hartree-fock on a
  superconducting qubit quantum computer},}\ }\href@noop {} {\bibfield
  {journal} {\bibinfo  {journal} {Science}\ }\textbf {\bibinfo {volume}
  {369}},\ \bibinfo {pages} {1084--1089} (\bibinfo {year} {2020})}\BibitemShut
  {NoStop}%
\bibitem [{\citenamefont {O'Brien}\ \emph {et~al.}(2022)\citenamefont
  {O'Brien}, \citenamefont {Anselmetti}, \citenamefont {Gkritsis},
  \citenamefont {Elfving}, \citenamefont {Polla}, \citenamefont {Huggins},
  \citenamefont {Oumarou}, \citenamefont {Kechedzhi}, \citenamefont {Abanin},
  \citenamefont {Acharya}, \citenamefont {Aleiner}, \citenamefont {Allen},
  \citenamefont {Andersen}, \citenamefont {Anderson}, \citenamefont {Ansmann},
  \citenamefont {Arute}, \citenamefont {Arya}, \citenamefont {Asfaw},
  \citenamefont {Atalaya}, \citenamefont {Bardin}, \citenamefont {Bengtsson},
  \citenamefont {Bortoli}, \citenamefont {Bourassa}, \citenamefont {Bovaird},
  \citenamefont {Brill}, \citenamefont {Broughton}, \citenamefont {Buckley},
  \citenamefont {Buell}, \citenamefont {Burger}, \citenamefont {Burkett},
  \citenamefont {Bushnell}, \citenamefont {Campero}, \citenamefont {Chen},
  \citenamefont {Chiaro}, \citenamefont {Chik}, \citenamefont {Cogan},
  \citenamefont {Collins}, \citenamefont {Conner}, \citenamefont {Courtney},
  \citenamefont {Crook}, \citenamefont {Curtin}, \citenamefont {Debroy},
  \citenamefont {Demura}, \citenamefont {Drozdov}, \citenamefont {Dunsworth},
  \citenamefont {Erickson}, \citenamefont {Faoro}, \citenamefont {Farhi},
  \citenamefont {Fatemi}, \citenamefont {Ferreira}, \citenamefont
  {Flores~Burgos}, \citenamefont {Forati}, \citenamefont {Fowler},
  \citenamefont {Foxen}, \citenamefont {Giang}, \citenamefont {Gidney},
  \citenamefont {Gilboa}, \citenamefont {Giustina}, \citenamefont {Gosula},
  \citenamefont {Grajales~Dau}, \citenamefont {Gross}, \citenamefont
  {Habegger}, \citenamefont {Hamilton}, \citenamefont {Hansen}, \citenamefont
  {Harrigan}, \citenamefont {Harrington}, \citenamefont {Heu}, \citenamefont
  {Hoffmann}, \citenamefont {Hong}, \citenamefont {Huang}, \citenamefont
  {Huff}, \citenamefont {Ioffe}, \citenamefont {Isakov}, \citenamefont
  {Iveland}, \citenamefont {Jeffrey}, \citenamefont {Jiang}, \citenamefont
  {Jones}, \citenamefont {Juhas}, \citenamefont {Kafri}, \citenamefont
  {Khattar}, \citenamefont {Khezri}, \citenamefont {Kieferov{\'a}},
  \citenamefont {Kim}, \citenamefont {Klimov}, \citenamefont {Klots},
  \citenamefont {Korotkov}, \citenamefont {Kostritsa}, \citenamefont
  {Kreikebaum}, \citenamefont {Landhuis}, \citenamefont {Laptev}, \citenamefont
  {Lau}, \citenamefont {Laws}, \citenamefont {Lee}, \citenamefont {Lee},
  \citenamefont {Lester}, \citenamefont {Lill}, \citenamefont {Liu},
  \citenamefont {Livingston}, \citenamefont {Locharla}, \citenamefont {Malone},
  \citenamefont {Mandr{\`a}}, \citenamefont {Martin}, \citenamefont {Martin},
  \citenamefont {McClean}, \citenamefont {McCourt}, \citenamefont {McEwen},
  \citenamefont {Mi}, \citenamefont {Mieszala}, \citenamefont {Miao},
  \citenamefont {Mohseni}, \citenamefont {Montazeri}, \citenamefont {Morvan},
  \citenamefont {Movassagh}, \citenamefont {Mruczkiewicz}, \citenamefont
  {Naaman}, \citenamefont {Neeley}, \citenamefont {Neill}, \citenamefont
  {Nersisyan}, \citenamefont {Newman}, \citenamefont {Ng}, \citenamefont
  {Nguyen}, \citenamefont {Nguyen}, \citenamefont {Niu}, \citenamefont
  {Omonije}, \citenamefont {Opremcak}, \citenamefont {Petukhov}, \citenamefont
  {Potter}, \citenamefont {Pryadko}, \citenamefont {Quintana}, \citenamefont
  {Rocque}, \citenamefont {Roushan}, \citenamefont {Saei}, \citenamefont
  {Sank}, \citenamefont {Sankaragomathi}, \citenamefont {Satzinger},
  \citenamefont {Schurkus}, \citenamefont {Schuster}, \citenamefont {Shearn},
  \citenamefont {Shorter}, \citenamefont {Shutty}, \citenamefont {Shvarts},
  \citenamefont {Skruzny}, \citenamefont {Smith}, \citenamefont {Somma},
  \citenamefont {Sterling}, \citenamefont {Strain}, \citenamefont {Szalay},
  \citenamefont {Thor}, \citenamefont {Torres}, \citenamefont {Vidal},
  \citenamefont {Villalonga}, \citenamefont {Vollgraff~Heidweiller},
  \citenamefont {White}, \citenamefont {Woo}, \citenamefont {Xing},
  \citenamefont {Yao}, \citenamefont {Yeh}, \citenamefont {Yoo}, \citenamefont
  {Young}, \citenamefont {Zalcman}, \citenamefont {Zhang}, \citenamefont {Zhu},
  \citenamefont {Zobrist}, \citenamefont {Bacon}, \citenamefont {Boixo},
  \citenamefont {Chen}, \citenamefont {Hilton}, \citenamefont {Kelly},
  \citenamefont {Lucero}, \citenamefont {Megrant}, \citenamefont {Neven},
  \citenamefont {Smelyanskiy}, \citenamefont {Gogolin}, \citenamefont
  {Babbush},\ and\ \citenamefont {Rubin}}]{o2022purification}%
  \BibitemOpen
  \bibfield  {author} {\bibinfo {author} {\bibfnamefont {T.~E.}\ \bibnamefont
  {O'Brien}}, \bibinfo {author} {\bibfnamefont {G.}~\bibnamefont {Anselmetti}},
  \bibinfo {author} {\bibfnamefont {F.}~\bibnamefont {Gkritsis}}, \bibinfo
  {author} {\bibfnamefont {V.~E.}\ \bibnamefont {Elfving}}, \bibinfo {author}
  {\bibfnamefont {S.}~\bibnamefont {Polla}}, \bibinfo {author} {\bibfnamefont
  {W.~J.}\ \bibnamefont {Huggins}}, \bibinfo {author} {\bibfnamefont
  {O.}~\bibnamefont {Oumarou}}, \bibinfo {author} {\bibfnamefont
  {K.}~\bibnamefont {Kechedzhi}}, \bibinfo {author} {\bibfnamefont
  {D.}~\bibnamefont {Abanin}}, \bibinfo {author} {\bibfnamefont
  {R.}~\bibnamefont {Acharya}}, \bibinfo {author} {\bibfnamefont
  {I.}~\bibnamefont {Aleiner}}, \bibinfo {author} {\bibfnamefont
  {R.}~\bibnamefont {Allen}}, \bibinfo {author} {\bibfnamefont {T.~I.}\
  \bibnamefont {Andersen}}, \bibinfo {author} {\bibfnamefont {K.}~\bibnamefont
  {Anderson}}, \bibinfo {author} {\bibfnamefont {M.}~\bibnamefont {Ansmann}},
  \bibinfo {author} {\bibfnamefont {F.}~\bibnamefont {Arute}}, \bibinfo
  {author} {\bibfnamefont {K.}~\bibnamefont {Arya}}, \bibinfo {author}
  {\bibfnamefont {A.}~\bibnamefont {Asfaw}}, \bibinfo {author} {\bibfnamefont
  {J.}~\bibnamefont {Atalaya}}, \bibinfo {author} {\bibfnamefont {J.~C.}\
  \bibnamefont {Bardin}}, \bibinfo {author} {\bibfnamefont {A.}~\bibnamefont
  {Bengtsson}}, \bibinfo {author} {\bibfnamefont {G.}~\bibnamefont {Bortoli}},
  \bibinfo {author} {\bibfnamefont {A.}~\bibnamefont {Bourassa}}, \bibinfo
  {author} {\bibfnamefont {J.}~\bibnamefont {Bovaird}}, \bibinfo {author}
  {\bibfnamefont {L.}~\bibnamefont {Brill}}, \bibinfo {author} {\bibfnamefont
  {M.}~\bibnamefont {Broughton}}, \bibinfo {author} {\bibfnamefont
  {B.}~\bibnamefont {Buckley}}, \bibinfo {author} {\bibfnamefont {D.~A.}\
  \bibnamefont {Buell}}, \bibinfo {author} {\bibfnamefont {T.}~\bibnamefont
  {Burger}}, \bibinfo {author} {\bibfnamefont {B.}~\bibnamefont {Burkett}},
  \bibinfo {author} {\bibfnamefont {N.}~\bibnamefont {Bushnell}}, \bibinfo
  {author} {\bibfnamefont {J.}~\bibnamefont {Campero}}, \bibinfo {author}
  {\bibfnamefont {Z.}~\bibnamefont {Chen}}, \bibinfo {author} {\bibfnamefont
  {B.}~\bibnamefont {Chiaro}}, \bibinfo {author} {\bibfnamefont
  {D.}~\bibnamefont {Chik}}, \bibinfo {author} {\bibfnamefont {J.}~\bibnamefont
  {Cogan}}, \bibinfo {author} {\bibfnamefont {R.}~\bibnamefont {Collins}},
  \bibinfo {author} {\bibfnamefont {P.}~\bibnamefont {Conner}}, \bibinfo
  {author} {\bibfnamefont {W.}~\bibnamefont {Courtney}}, \bibinfo {author}
  {\bibfnamefont {A.~L.}\ \bibnamefont {Crook}}, \bibinfo {author}
  {\bibfnamefont {B.}~\bibnamefont {Curtin}}, \bibinfo {author} {\bibfnamefont
  {D.~M.}\ \bibnamefont {Debroy}}, \bibinfo {author} {\bibfnamefont
  {S.}~\bibnamefont {Demura}}, \bibinfo {author} {\bibfnamefont
  {I.}~\bibnamefont {Drozdov}}, \bibinfo {author} {\bibfnamefont
  {A.}~\bibnamefont {Dunsworth}}, \bibinfo {author} {\bibfnamefont
  {C.}~\bibnamefont {Erickson}}, \bibinfo {author} {\bibfnamefont
  {L.}~\bibnamefont {Faoro}}, \bibinfo {author} {\bibfnamefont
  {E.}~\bibnamefont {Farhi}}, \bibinfo {author} {\bibfnamefont
  {R.}~\bibnamefont {Fatemi}}, \bibinfo {author} {\bibfnamefont {V.~S.}\
  \bibnamefont {Ferreira}}, \bibinfo {author} {\bibfnamefont {L.}~\bibnamefont
  {Flores~Burgos}}, \bibinfo {author} {\bibfnamefont {E.}~\bibnamefont
  {Forati}}, \bibinfo {author} {\bibfnamefont {A.~G.}\ \bibnamefont {Fowler}},
  \bibinfo {author} {\bibfnamefont {B.}~\bibnamefont {Foxen}}, \bibinfo
  {author} {\bibfnamefont {W.}~\bibnamefont {Giang}}, \bibinfo {author}
  {\bibfnamefont {C.}~\bibnamefont {Gidney}}, \bibinfo {author} {\bibfnamefont
  {D.}~\bibnamefont {Gilboa}}, \bibinfo {author} {\bibfnamefont
  {M.}~\bibnamefont {Giustina}}, \bibinfo {author} {\bibfnamefont
  {R.}~\bibnamefont {Gosula}}, \bibinfo {author} {\bibfnamefont
  {A.}~\bibnamefont {Grajales~Dau}}, \bibinfo {author} {\bibfnamefont {J.~A.}\
  \bibnamefont {Gross}}, \bibinfo {author} {\bibfnamefont {S.}~\bibnamefont
  {Habegger}}, \bibinfo {author} {\bibfnamefont {M.~C.}\ \bibnamefont
  {Hamilton}}, \bibinfo {author} {\bibfnamefont {M.}~\bibnamefont {Hansen}},
  \bibinfo {author} {\bibfnamefont {M.~P.}\ \bibnamefont {Harrigan}}, \bibinfo
  {author} {\bibfnamefont {S.~D.}\ \bibnamefont {Harrington}}, \bibinfo
  {author} {\bibfnamefont {P.}~\bibnamefont {Heu}}, \bibinfo {author}
  {\bibfnamefont {M.~R.}\ \bibnamefont {Hoffmann}}, \bibinfo {author}
  {\bibfnamefont {S.}~\bibnamefont {Hong}}, \bibinfo {author} {\bibfnamefont
  {T.}~\bibnamefont {Huang}}, \bibinfo {author} {\bibfnamefont
  {A.}~\bibnamefont {Huff}}, \bibinfo {author} {\bibfnamefont {L.~B.}\
  \bibnamefont {Ioffe}}, \bibinfo {author} {\bibfnamefont {S.~V.}\ \bibnamefont
  {Isakov}}, \bibinfo {author} {\bibfnamefont {J.}~\bibnamefont {Iveland}},
  \bibinfo {author} {\bibfnamefont {E.}~\bibnamefont {Jeffrey}}, \bibinfo
  {author} {\bibfnamefont {Z.}~\bibnamefont {Jiang}}, \bibinfo {author}
  {\bibfnamefont {C.}~\bibnamefont {Jones}}, \bibinfo {author} {\bibfnamefont
  {P.}~\bibnamefont {Juhas}}, \bibinfo {author} {\bibfnamefont
  {D.}~\bibnamefont {Kafri}}, \bibinfo {author} {\bibfnamefont
  {T.}~\bibnamefont {Khattar}}, \bibinfo {author} {\bibfnamefont
  {M.}~\bibnamefont {Khezri}}, \bibinfo {author} {\bibfnamefont
  {M.}~\bibnamefont {Kieferov{\'a}}}, \bibinfo {author} {\bibfnamefont
  {S.}~\bibnamefont {Kim}}, \bibinfo {author} {\bibfnamefont {P.~V.}\
  \bibnamefont {Klimov}}, \bibinfo {author} {\bibfnamefont {A.~R.}\
  \bibnamefont {Klots}}, \bibinfo {author} {\bibfnamefont {A.~N.}\ \bibnamefont
  {Korotkov}}, \bibinfo {author} {\bibfnamefont {F.}~\bibnamefont {Kostritsa}},
  \bibinfo {author} {\bibfnamefont {J.~M.}\ \bibnamefont {Kreikebaum}},
  \bibinfo {author} {\bibfnamefont {D.}~\bibnamefont {Landhuis}}, \bibinfo
  {author} {\bibfnamefont {P.}~\bibnamefont {Laptev}}, \bibinfo {author}
  {\bibfnamefont {K.-M.}\ \bibnamefont {Lau}}, \bibinfo {author} {\bibfnamefont
  {L.}~\bibnamefont {Laws}}, \bibinfo {author} {\bibfnamefont {J.}~\bibnamefont
  {Lee}}, \bibinfo {author} {\bibfnamefont {K.}~\bibnamefont {Lee}}, \bibinfo
  {author} {\bibfnamefont {B.~J.}\ \bibnamefont {Lester}}, \bibinfo {author}
  {\bibfnamefont {A.~T.}\ \bibnamefont {Lill}}, \bibinfo {author}
  {\bibfnamefont {W.}~\bibnamefont {Liu}}, \bibinfo {author} {\bibfnamefont
  {W.~P.}\ \bibnamefont {Livingston}}, \bibinfo {author} {\bibfnamefont
  {A.}~\bibnamefont {Locharla}}, \bibinfo {author} {\bibfnamefont {F.~D.}\
  \bibnamefont {Malone}}, \bibinfo {author} {\bibfnamefont {S.}~\bibnamefont
  {Mandr{\`a}}}, \bibinfo {author} {\bibfnamefont {O.}~\bibnamefont {Martin}},
  \bibinfo {author} {\bibfnamefont {S.}~\bibnamefont {Martin}}, \bibinfo
  {author} {\bibfnamefont {J.~R.}\ \bibnamefont {McClean}}, \bibinfo {author}
  {\bibfnamefont {T.}~\bibnamefont {McCourt}}, \bibinfo {author} {\bibfnamefont
  {M.}~\bibnamefont {McEwen}}, \bibinfo {author} {\bibfnamefont
  {X.}~\bibnamefont {Mi}}, \bibinfo {author} {\bibfnamefont {A.}~\bibnamefont
  {Mieszala}}, \bibinfo {author} {\bibfnamefont {K.~C.}\ \bibnamefont {Miao}},
  \bibinfo {author} {\bibfnamefont {M.}~\bibnamefont {Mohseni}}, \bibinfo
  {author} {\bibfnamefont {S.}~\bibnamefont {Montazeri}}, \bibinfo {author}
  {\bibfnamefont {A.}~\bibnamefont {Morvan}}, \bibinfo {author} {\bibfnamefont
  {R.}~\bibnamefont {Movassagh}}, \bibinfo {author} {\bibfnamefont
  {W.}~\bibnamefont {Mruczkiewicz}}, \bibinfo {author} {\bibfnamefont
  {O.}~\bibnamefont {Naaman}}, \bibinfo {author} {\bibfnamefont
  {M.}~\bibnamefont {Neeley}}, \bibinfo {author} {\bibfnamefont
  {C.}~\bibnamefont {Neill}}, \bibinfo {author} {\bibfnamefont
  {A.}~\bibnamefont {Nersisyan}}, \bibinfo {author} {\bibfnamefont
  {M.}~\bibnamefont {Newman}}, \bibinfo {author} {\bibfnamefont {J.~H.}\
  \bibnamefont {Ng}}, \bibinfo {author} {\bibfnamefont {A.}~\bibnamefont
  {Nguyen}}, \bibinfo {author} {\bibfnamefont {M.}~\bibnamefont {Nguyen}},
  \bibinfo {author} {\bibfnamefont {M.~Y.}\ \bibnamefont {Niu}}, \bibinfo
  {author} {\bibfnamefont {S.}~\bibnamefont {Omonije}}, \bibinfo {author}
  {\bibfnamefont {A.}~\bibnamefont {Opremcak}}, \bibinfo {author}
  {\bibfnamefont {A.}~\bibnamefont {Petukhov}}, \bibinfo {author}
  {\bibfnamefont {R.}~\bibnamefont {Potter}}, \bibinfo {author} {\bibfnamefont
  {L.~P.}\ \bibnamefont {Pryadko}}, \bibinfo {author} {\bibfnamefont
  {C.}~\bibnamefont {Quintana}}, \bibinfo {author} {\bibfnamefont
  {C.}~\bibnamefont {Rocque}}, \bibinfo {author} {\bibfnamefont
  {P.}~\bibnamefont {Roushan}}, \bibinfo {author} {\bibfnamefont
  {N.}~\bibnamefont {Saei}}, \bibinfo {author} {\bibfnamefont {D.}~\bibnamefont
  {Sank}}, \bibinfo {author} {\bibfnamefont {K.}~\bibnamefont
  {Sankaragomathi}}, \bibinfo {author} {\bibfnamefont {K.~J.}\ \bibnamefont
  {Satzinger}}, \bibinfo {author} {\bibfnamefont {H.~F.}\ \bibnamefont
  {Schurkus}}, \bibinfo {author} {\bibfnamefont {C.}~\bibnamefont {Schuster}},
  \bibinfo {author} {\bibfnamefont {M.~J.}\ \bibnamefont {Shearn}}, \bibinfo
  {author} {\bibfnamefont {A.}~\bibnamefont {Shorter}}, \bibinfo {author}
  {\bibfnamefont {N.}~\bibnamefont {Shutty}}, \bibinfo {author} {\bibfnamefont
  {V.}~\bibnamefont {Shvarts}}, \bibinfo {author} {\bibfnamefont
  {J.}~\bibnamefont {Skruzny}}, \bibinfo {author} {\bibfnamefont {W.~C.}\
  \bibnamefont {Smith}}, \bibinfo {author} {\bibfnamefont {R.~D.}\ \bibnamefont
  {Somma}}, \bibinfo {author} {\bibfnamefont {G.}~\bibnamefont {Sterling}},
  \bibinfo {author} {\bibfnamefont {D.}~\bibnamefont {Strain}}, \bibinfo
  {author} {\bibfnamefont {M.}~\bibnamefont {Szalay}}, \bibinfo {author}
  {\bibfnamefont {D.}~\bibnamefont {Thor}}, \bibinfo {author} {\bibfnamefont
  {A.}~\bibnamefont {Torres}}, \bibinfo {author} {\bibfnamefont
  {G.}~\bibnamefont {Vidal}}, \bibinfo {author} {\bibfnamefont
  {B.}~\bibnamefont {Villalonga}}, \bibinfo {author} {\bibfnamefont
  {C.}~\bibnamefont {Vollgraff~Heidweiller}}, \bibinfo {author} {\bibfnamefont
  {T.}~\bibnamefont {White}}, \bibinfo {author} {\bibfnamefont {B.~W.~K.}\
  \bibnamefont {Woo}}, \bibinfo {author} {\bibfnamefont {C.}~\bibnamefont
  {Xing}}, \bibinfo {author} {\bibfnamefont {Z.~J.}\ \bibnamefont {Yao}},
  \bibinfo {author} {\bibfnamefont {P.}~\bibnamefont {Yeh}}, \bibinfo {author}
  {\bibfnamefont {J.}~\bibnamefont {Yoo}}, \bibinfo {author} {\bibfnamefont
  {G.}~\bibnamefont {Young}}, \bibinfo {author} {\bibfnamefont
  {A.}~\bibnamefont {Zalcman}}, \bibinfo {author} {\bibfnamefont
  {Y.}~\bibnamefont {Zhang}}, \bibinfo {author} {\bibfnamefont
  {N.}~\bibnamefont {Zhu}}, \bibinfo {author} {\bibfnamefont {N.}~\bibnamefont
  {Zobrist}}, \bibinfo {author} {\bibfnamefont {D.}~\bibnamefont {Bacon}},
  \bibinfo {author} {\bibfnamefont {S.}~\bibnamefont {Boixo}}, \bibinfo
  {author} {\bibfnamefont {Y.}~\bibnamefont {Chen}}, \bibinfo {author}
  {\bibfnamefont {J.}~\bibnamefont {Hilton}}, \bibinfo {author} {\bibfnamefont
  {J.}~\bibnamefont {Kelly}}, \bibinfo {author} {\bibfnamefont
  {E.}~\bibnamefont {Lucero}}, \bibinfo {author} {\bibfnamefont
  {A.}~\bibnamefont {Megrant}}, \bibinfo {author} {\bibfnamefont
  {H.}~\bibnamefont {Neven}}, \bibinfo {author} {\bibfnamefont
  {V.}~\bibnamefont {Smelyanskiy}}, \bibinfo {author} {\bibfnamefont
  {C.}~\bibnamefont {Gogolin}}, \bibinfo {author} {\bibfnamefont
  {R.}~\bibnamefont {Babbush}},\ and\ \bibinfo {author} {\bibfnamefont {N.~C.}\
  \bibnamefont {Rubin}},\ }\bibfield  {title} {\enquote {\bibinfo {title}
  {Purification-based quantum error mitigation of pair-correlated electron
  simulations},}\ }\href@noop {} {\bibfield  {journal} {\bibinfo  {journal}
  {arXiv preprint arXiv:2210.10799}\ } (\bibinfo {year} {2022})}\BibitemShut
  {NoStop}%
\bibitem [{\citenamefont {Guo}\ \emph {et~al.}(2022)\citenamefont {Guo},
  \citenamefont {Sun}, \citenamefont {Qian}, \citenamefont {Gong},
  \citenamefont {Zhang}, \citenamefont {Chen}, \citenamefont {Ye},
  \citenamefont {Wu}, \citenamefont {Cao}, \citenamefont {Liu}, \citenamefont
  {Zha}, \citenamefont {Ying}, \citenamefont {Zhu}, \citenamefont {Huang},
  \citenamefont {Zhao}, \citenamefont {Li}, \citenamefont {Wang}, \citenamefont
  {Yu}, \citenamefont {Fan}, \citenamefont {Wu}, \citenamefont {Su},
  \citenamefont {Deng}, \citenamefont {Rong}, \citenamefont {Li}, \citenamefont
  {Zhang}, \citenamefont {Chung}, \citenamefont {Liang}, \citenamefont {Lin},
  \citenamefont {Xu}, \citenamefont {Sun}, \citenamefont {Guo}, \citenamefont
  {Li}, \citenamefont {Huo}, \citenamefont {Peng}, \citenamefont {Lu},
  \citenamefont {Yuan}, \citenamefont {Zhu},\ and\ \citenamefont
  {Pan}}]{guo2022experimental}%
  \BibitemOpen
  \bibfield  {author} {\bibinfo {author} {\bibfnamefont {S.}~\bibnamefont
  {Guo}}, \bibinfo {author} {\bibfnamefont {J.}~\bibnamefont {Sun}}, \bibinfo
  {author} {\bibfnamefont {H.}~\bibnamefont {Qian}}, \bibinfo {author}
  {\bibfnamefont {M.}~\bibnamefont {Gong}}, \bibinfo {author} {\bibfnamefont
  {Y.}~\bibnamefont {Zhang}}, \bibinfo {author} {\bibfnamefont
  {F.}~\bibnamefont {Chen}}, \bibinfo {author} {\bibfnamefont {Y.}~\bibnamefont
  {Ye}}, \bibinfo {author} {\bibfnamefont {Y.}~\bibnamefont {Wu}}, \bibinfo
  {author} {\bibfnamefont {S.}~\bibnamefont {Cao}}, \bibinfo {author}
  {\bibfnamefont {K.}~\bibnamefont {Liu}}, \bibinfo {author} {\bibfnamefont
  {C.}~\bibnamefont {Zha}}, \bibinfo {author} {\bibfnamefont {C.}~\bibnamefont
  {Ying}}, \bibinfo {author} {\bibfnamefont {Q.}~\bibnamefont {Zhu}}, \bibinfo
  {author} {\bibfnamefont {H.-L.}\ \bibnamefont {Huang}}, \bibinfo {author}
  {\bibfnamefont {Y.}~\bibnamefont {Zhao}}, \bibinfo {author} {\bibfnamefont
  {S.}~\bibnamefont {Li}}, \bibinfo {author} {\bibfnamefont {S.}~\bibnamefont
  {Wang}}, \bibinfo {author} {\bibfnamefont {J.}~\bibnamefont {Yu}}, \bibinfo
  {author} {\bibfnamefont {D.}~\bibnamefont {Fan}}, \bibinfo {author}
  {\bibfnamefont {D.}~\bibnamefont {Wu}}, \bibinfo {author} {\bibfnamefont
  {H.}~\bibnamefont {Su}}, \bibinfo {author} {\bibfnamefont {H.}~\bibnamefont
  {Deng}}, \bibinfo {author} {\bibfnamefont {H.}~\bibnamefont {Rong}}, \bibinfo
  {author} {\bibfnamefont {Y.}~\bibnamefont {Li}}, \bibinfo {author}
  {\bibfnamefont {K.}~\bibnamefont {Zhang}}, \bibinfo {author} {\bibfnamefont
  {T.-H.}\ \bibnamefont {Chung}}, \bibinfo {author} {\bibfnamefont
  {F.}~\bibnamefont {Liang}}, \bibinfo {author} {\bibfnamefont
  {J.}~\bibnamefont {Lin}}, \bibinfo {author} {\bibfnamefont {Y.}~\bibnamefont
  {Xu}}, \bibinfo {author} {\bibfnamefont {L.}~\bibnamefont {Sun}}, \bibinfo
  {author} {\bibfnamefont {C.}~\bibnamefont {Guo}}, \bibinfo {author}
  {\bibfnamefont {N.}~\bibnamefont {Li}}, \bibinfo {author} {\bibfnamefont
  {Y.-H.}\ \bibnamefont {Huo}}, \bibinfo {author} {\bibfnamefont {C.-Z.}\
  \bibnamefont {Peng}}, \bibinfo {author} {\bibfnamefont {C.-Y.}\ \bibnamefont
  {Lu}}, \bibinfo {author} {\bibfnamefont {X.}~\bibnamefont {Yuan}}, \bibinfo
  {author} {\bibfnamefont {X.}~\bibnamefont {Zhu}},\ and\ \bibinfo {author}
  {\bibfnamefont {J.-W.}\ \bibnamefont {Pan}},\ }\bibfield  {title} {\enquote
  {\bibinfo {title} {Experimental quantum computational chemistry with
  optimised unitary coupled cluster ansatz},}\ }\href@noop {} {\bibfield
  {journal} {\bibinfo  {journal} {arXiv preprint arXiv:2212.08006}\ } (\bibinfo
  {year} {2022})}\BibitemShut {NoStop}%
\bibitem [{\citenamefont {Barkoutsos}\ \emph {et~al.}(2021)\citenamefont
  {Barkoutsos}, \citenamefont {Gkritsis}, \citenamefont {Ollitrault},
  \citenamefont {Sokolov}, \citenamefont {Woerner},\ and\ \citenamefont
  {Tavernelli}}]{barkoutsos2021quantum}%
  \BibitemOpen
  \bibfield  {author} {\bibinfo {author} {\bibfnamefont {P.~K.}\ \bibnamefont
  {Barkoutsos}}, \bibinfo {author} {\bibfnamefont {F.}~\bibnamefont
  {Gkritsis}}, \bibinfo {author} {\bibfnamefont {P.~J.}\ \bibnamefont
  {Ollitrault}}, \bibinfo {author} {\bibfnamefont {I.~O.}\ \bibnamefont
  {Sokolov}}, \bibinfo {author} {\bibfnamefont {S.}~\bibnamefont {Woerner}},\
  and\ \bibinfo {author} {\bibfnamefont {I.}~\bibnamefont {Tavernelli}},\
  }\bibfield  {title} {\enquote {\bibinfo {title} {Quantum algorithm for
  alchemical optimization in material design},}\ }\href@noop {} {\bibfield
  {journal} {\bibinfo  {journal} {Chem. Sci.}\ }\textbf {\bibinfo {volume}
  {12}},\ \bibinfo {pages} {4345--4352} (\bibinfo {year} {2021})}\BibitemShut
  {NoStop}%
\bibitem [{\citenamefont {Gao}\ \emph {et~al.}(2021{\natexlab{a}})\citenamefont
  {Gao}, \citenamefont {Jones}, \citenamefont {Motta}, \citenamefont
  {Sugawara}, \citenamefont {Watanabe}, \citenamefont {Kobayashi},
  \citenamefont {Watanabe}, \citenamefont {Ohnishi}, \citenamefont {Nakamura},\
  and\ \citenamefont {Yamamoto}}]{gao2021applications}%
  \BibitemOpen
  \bibfield  {author} {\bibinfo {author} {\bibfnamefont {Q.}~\bibnamefont
  {Gao}}, \bibinfo {author} {\bibfnamefont {G.~O.}\ \bibnamefont {Jones}},
  \bibinfo {author} {\bibfnamefont {M.}~\bibnamefont {Motta}}, \bibinfo
  {author} {\bibfnamefont {M.}~\bibnamefont {Sugawara}}, \bibinfo {author}
  {\bibfnamefont {H.~C.}\ \bibnamefont {Watanabe}}, \bibinfo {author}
  {\bibfnamefont {T.}~\bibnamefont {Kobayashi}}, \bibinfo {author}
  {\bibfnamefont {E.}~\bibnamefont {Watanabe}}, \bibinfo {author}
  {\bibfnamefont {Y.-y.}\ \bibnamefont {Ohnishi}}, \bibinfo {author}
  {\bibfnamefont {H.}~\bibnamefont {Nakamura}},\ and\ \bibinfo {author}
  {\bibfnamefont {N.}~\bibnamefont {Yamamoto}},\ }\bibfield  {title} {\enquote
  {\bibinfo {title} {Applications of quantum computing for investigations of
  electronic transitions in phenylsulfonyl-carbazole tadf emitters},}\
  }\href@noop {} {\bibfield  {journal} {\bibinfo  {journal} {npj Comput.
  Mater.}\ }\textbf {\bibinfo {volume} {7}},\ \bibinfo {pages} {70} (\bibinfo
  {year} {2021}{\natexlab{a}})}\BibitemShut {NoStop}%
\bibitem [{\citenamefont {Li}\ \emph {et~al.}(2022)\citenamefont {Li},
  \citenamefont {Huang}, \citenamefont {Cao}, \citenamefont {Huang},
  \citenamefont {Shuai}, \citenamefont {Sun}, \citenamefont {Sun},
  \citenamefont {Yuan},\ and\ \citenamefont {Lv}}]{li2022toward}%
  \BibitemOpen
  \bibfield  {author} {\bibinfo {author} {\bibfnamefont {W.}~\bibnamefont
  {Li}}, \bibinfo {author} {\bibfnamefont {Z.}~\bibnamefont {Huang}}, \bibinfo
  {author} {\bibfnamefont {C.}~\bibnamefont {Cao}}, \bibinfo {author}
  {\bibfnamefont {Y.}~\bibnamefont {Huang}}, \bibinfo {author} {\bibfnamefont
  {Z.}~\bibnamefont {Shuai}}, \bibinfo {author} {\bibfnamefont
  {X.}~\bibnamefont {Sun}}, \bibinfo {author} {\bibfnamefont {J.}~\bibnamefont
  {Sun}}, \bibinfo {author} {\bibfnamefont {X.}~\bibnamefont {Yuan}},\ and\
  \bibinfo {author} {\bibfnamefont {D.}~\bibnamefont {Lv}},\ }\bibfield
  {title} {\enquote {\bibinfo {title} {Toward practical quantum embedding
  simulation of realistic chemical systems on near-term quantum computers},}\
  }\href@noop {} {\bibfield  {journal} {\bibinfo  {journal} {Chem. Sci.}\
  }\textbf {\bibinfo {volume} {13}},\ \bibinfo {pages} {8953--8962} (\bibinfo
  {year} {2022})}\BibitemShut {NoStop}%
\bibitem [{\citenamefont {Malone}\ \emph {et~al.}(2022)\citenamefont {Malone},
  \citenamefont {Parrish}, \citenamefont {Welden}, \citenamefont {Fox},
  \citenamefont {Degroote}, \citenamefont {Kyoseva}, \citenamefont {Moll},
  \citenamefont {Santagati},\ and\ \citenamefont {Streif}}]{malone2022towards}%
  \BibitemOpen
  \bibfield  {author} {\bibinfo {author} {\bibfnamefont {F.~D.}\ \bibnamefont
  {Malone}}, \bibinfo {author} {\bibfnamefont {R.~M.}\ \bibnamefont {Parrish}},
  \bibinfo {author} {\bibfnamefont {A.~R.}\ \bibnamefont {Welden}}, \bibinfo
  {author} {\bibfnamefont {T.}~\bibnamefont {Fox}}, \bibinfo {author}
  {\bibfnamefont {M.}~\bibnamefont {Degroote}}, \bibinfo {author}
  {\bibfnamefont {E.}~\bibnamefont {Kyoseva}}, \bibinfo {author} {\bibfnamefont
  {N.}~\bibnamefont {Moll}}, \bibinfo {author} {\bibfnamefont {R.}~\bibnamefont
  {Santagati}},\ and\ \bibinfo {author} {\bibfnamefont {M.}~\bibnamefont
  {Streif}},\ }\bibfield  {title} {\enquote {\bibinfo {title} {Towards the
  simulation of large scale protein--ligand interactions on nisq-era quantum
  computers},}\ }\href@noop {} {\bibfield  {journal} {\bibinfo  {journal}
  {Chem. Sci.}\ }\textbf {\bibinfo {volume} {13}},\ \bibinfo {pages}
  {3094--3108} (\bibinfo {year} {2022})}\BibitemShut {NoStop}%
\bibitem [{\citenamefont {Motta}\ \emph {et~al.}(2023)\citenamefont {Motta},
  \citenamefont {Jones}, \citenamefont {Rice}, \citenamefont {Gujarati},
  \citenamefont {Sakuma}, \citenamefont {Liepuoniute}, \citenamefont {Garcia},\
  and\ \citenamefont {Ohnishi}}]{motta2023quantum}%
  \BibitemOpen
  \bibfield  {author} {\bibinfo {author} {\bibfnamefont {M.}~\bibnamefont
  {Motta}}, \bibinfo {author} {\bibfnamefont {G.~O.}\ \bibnamefont {Jones}},
  \bibinfo {author} {\bibfnamefont {J.~E.}\ \bibnamefont {Rice}}, \bibinfo
  {author} {\bibfnamefont {T.~P.}\ \bibnamefont {Gujarati}}, \bibinfo {author}
  {\bibfnamefont {R.}~\bibnamefont {Sakuma}}, \bibinfo {author} {\bibfnamefont
  {I.}~\bibnamefont {Liepuoniute}}, \bibinfo {author} {\bibfnamefont {J.~M.}\
  \bibnamefont {Garcia}},\ and\ \bibinfo {author} {\bibfnamefont {Y.-y.}\
  \bibnamefont {Ohnishi}},\ }\bibfield  {title} {\enquote {\bibinfo {title}
  {Quantum chemistry simulation of ground-and excited-state properties of the
  sulfonium cation on a superconducting quantum processor},}\ }\href@noop {}
  {\bibfield  {journal} {\bibinfo  {journal} {Chem. Sci.}\ }\textbf {\bibinfo
  {volume} {14}},\ \bibinfo {pages} {2915--2927} (\bibinfo {year}
  {2023})}\BibitemShut {NoStop}%
\bibitem [{\citenamefont {McClean}\ \emph {et~al.}(2016)\citenamefont
  {McClean}, \citenamefont {Romero}, \citenamefont {Babbush},\ and\
  \citenamefont {Aspuru-Guzik}}]{mcclean2016theory}%
  \BibitemOpen
  \bibfield  {author} {\bibinfo {author} {\bibfnamefont {J.~R.}\ \bibnamefont
  {McClean}}, \bibinfo {author} {\bibfnamefont {J.}~\bibnamefont {Romero}},
  \bibinfo {author} {\bibfnamefont {R.}~\bibnamefont {Babbush}},\ and\ \bibinfo
  {author} {\bibfnamefont {A.}~\bibnamefont {Aspuru-Guzik}},\ }\bibfield
  {title} {\enquote {\bibinfo {title} {The theory of variational hybrid
  quantum-classical algorithms},}\ }\href@noop {} {\bibfield  {journal}
  {\bibinfo  {journal} {New J. Phys.}\ }\textbf {\bibinfo {volume} {18}},\
  \bibinfo {pages} {023023} (\bibinfo {year} {2016})}\BibitemShut {NoStop}%
\bibitem [{\citenamefont {Fedorov}\ \emph {et~al.}(2022)\citenamefont
  {Fedorov}, \citenamefont {Peng}, \citenamefont {Govind},\ and\ \citenamefont
  {Alexeev}}]{fedorov2022vqe}%
  \BibitemOpen
  \bibfield  {author} {\bibinfo {author} {\bibfnamefont {D.~A.}\ \bibnamefont
  {Fedorov}}, \bibinfo {author} {\bibfnamefont {B.}~\bibnamefont {Peng}},
  \bibinfo {author} {\bibfnamefont {N.}~\bibnamefont {Govind}},\ and\ \bibinfo
  {author} {\bibfnamefont {Y.}~\bibnamefont {Alexeev}},\ }\bibfield  {title}
  {\enquote {\bibinfo {title} {Vqe method: a short survey and recent
  developments},}\ }\href@noop {} {\bibfield  {journal} {\bibinfo  {journal}
  {Mater. Theory}\ }\textbf {\bibinfo {volume} {6}},\ \bibinfo {pages} {1--21}
  (\bibinfo {year} {2022})}\BibitemShut {NoStop}%
\bibitem [{\citenamefont {Lee}\ \emph {et~al.}(2018)\citenamefont {Lee},
  \citenamefont {Huggins}, \citenamefont {Head-Gordon},\ and\ \citenamefont
  {Whaley}}]{lee2018generalized}%
  \BibitemOpen
  \bibfield  {author} {\bibinfo {author} {\bibfnamefont {J.}~\bibnamefont
  {Lee}}, \bibinfo {author} {\bibfnamefont {W.~J.}\ \bibnamefont {Huggins}},
  \bibinfo {author} {\bibfnamefont {M.}~\bibnamefont {Head-Gordon}},\ and\
  \bibinfo {author} {\bibfnamefont {K.~B.}\ \bibnamefont {Whaley}},\ }\bibfield
   {title} {\enquote {\bibinfo {title} {Generalized unitary coupled cluster
  wave functions for quantum computation},}\ }\href@noop {} {\bibfield
  {journal} {\bibinfo  {journal} {J. Chem. Theory and Comput.}\ }\textbf
  {\bibinfo {volume} {15}},\ \bibinfo {pages} {311--324} (\bibinfo {year}
  {2018})}\BibitemShut {NoStop}%
\bibitem [{\citenamefont {Gard}\ \emph {et~al.}(2020)\citenamefont {Gard},
  \citenamefont {Zhu}, \citenamefont {Barron}, \citenamefont {Mayhall},
  \citenamefont {Economou},\ and\ \citenamefont {Barnes}}]{gard2020efficient}%
  \BibitemOpen
  \bibfield  {author} {\bibinfo {author} {\bibfnamefont {B.~T.}\ \bibnamefont
  {Gard}}, \bibinfo {author} {\bibfnamefont {L.}~\bibnamefont {Zhu}}, \bibinfo
  {author} {\bibfnamefont {G.~S.}\ \bibnamefont {Barron}}, \bibinfo {author}
  {\bibfnamefont {N.~J.}\ \bibnamefont {Mayhall}}, \bibinfo {author}
  {\bibfnamefont {S.~E.}\ \bibnamefont {Economou}},\ and\ \bibinfo {author}
  {\bibfnamefont {E.}~\bibnamefont {Barnes}},\ }\bibfield  {title} {\enquote
  {\bibinfo {title} {Efficient symmetry-preserving state preparation circuits
  for the variational quantum eigensolver algorithm},}\ }\href@noop {}
  {\bibfield  {journal} {\bibinfo  {journal} {npj Quantum Inf.}\ }\textbf
  {\bibinfo {volume} {6}},\ \bibinfo {pages} {10} (\bibinfo {year}
  {2020})}\BibitemShut {NoStop}%
\bibitem [{\citenamefont {Anselmetti}\ \emph {et~al.}(2021)\citenamefont
  {Anselmetti}, \citenamefont {Wierichs}, \citenamefont {Gogolin},\ and\
  \citenamefont {Parrish}}]{anselmetti2021local}%
  \BibitemOpen
  \bibfield  {author} {\bibinfo {author} {\bibfnamefont {G.-L.~R.}\
  \bibnamefont {Anselmetti}}, \bibinfo {author} {\bibfnamefont
  {D.}~\bibnamefont {Wierichs}}, \bibinfo {author} {\bibfnamefont
  {C.}~\bibnamefont {Gogolin}},\ and\ \bibinfo {author} {\bibfnamefont {R.~M.}\
  \bibnamefont {Parrish}},\ }\bibfield  {title} {\enquote {\bibinfo {title}
  {Local, expressive, quantum-number-preserving vqe ans{\"a}tze for fermionic
  systems},}\ }\href@noop {} {\bibfield  {journal} {\bibinfo  {journal} {New J.
  Phys.}\ }\textbf {\bibinfo {volume} {23}},\ \bibinfo {pages} {113010}
  (\bibinfo {year} {2021})}\BibitemShut {NoStop}%
\bibitem [{\citenamefont {Xiao}\ \emph {et~al.}(2023)\citenamefont {Xiao},
  \citenamefont {Zhao}, \citenamefont {Ren}, \citenamefont {Fang},\ and\
  \citenamefont {Li}}]{xiao2023physics}%
  \BibitemOpen
  \bibfield  {author} {\bibinfo {author} {\bibfnamefont {X.}~\bibnamefont
  {Xiao}}, \bibinfo {author} {\bibfnamefont {H.}~\bibnamefont {Zhao}}, \bibinfo
  {author} {\bibfnamefont {J.}~\bibnamefont {Ren}}, \bibinfo {author}
  {\bibfnamefont {W.-h.}\ \bibnamefont {Fang}},\ and\ \bibinfo {author}
  {\bibfnamefont {Z.}~\bibnamefont {Li}},\ }\bibfield  {title} {\enquote
  {\bibinfo {title} {Physics-constrained hardware-efficient ansatz on quantum
  computers that is universal, systematically improvable, and
  size-consistent},}\ }\href@noop {} {\bibfield  {journal} {\bibinfo  {journal}
  {arXiv preprint arXiv:2307.03563}\ } (\bibinfo {year} {2023})}\BibitemShut
  {NoStop}%
\bibitem [{\citenamefont {Anand}\ \emph {et~al.}(2022)\citenamefont {Anand},
  \citenamefont {Schleich}, \citenamefont {Alperin-Lea}, \citenamefont
  {Jensen}, \citenamefont {Sim}, \citenamefont {D{\'\i}az-Tinoco},
  \citenamefont {Kottmann}, \citenamefont {Degroote}, \citenamefont
  {Izmaylov},\ and\ \citenamefont {Aspuru-Guzik}}]{anand2022quantum}%
  \BibitemOpen
  \bibfield  {author} {\bibinfo {author} {\bibfnamefont {A.}~\bibnamefont
  {Anand}}, \bibinfo {author} {\bibfnamefont {P.}~\bibnamefont {Schleich}},
  \bibinfo {author} {\bibfnamefont {S.}~\bibnamefont {Alperin-Lea}}, \bibinfo
  {author} {\bibfnamefont {P.~W.}\ \bibnamefont {Jensen}}, \bibinfo {author}
  {\bibfnamefont {S.}~\bibnamefont {Sim}}, \bibinfo {author} {\bibfnamefont
  {M.}~\bibnamefont {D{\'\i}az-Tinoco}}, \bibinfo {author} {\bibfnamefont
  {J.~S.}\ \bibnamefont {Kottmann}}, \bibinfo {author} {\bibfnamefont
  {M.}~\bibnamefont {Degroote}}, \bibinfo {author} {\bibfnamefont {A.~F.}\
  \bibnamefont {Izmaylov}},\ and\ \bibinfo {author} {\bibfnamefont
  {A.}~\bibnamefont {Aspuru-Guzik}},\ }\bibfield  {title} {\enquote {\bibinfo
  {title} {A quantum computing view on unitary coupled cluster theory},}\
  }\href@noop {} {\bibfield  {journal} {\bibinfo  {journal} {Chem. Soc. Rev.}\
  ,\ \bibinfo {pages} {1659--1684}} (\bibinfo {year} {2022})}\BibitemShut
  {NoStop}%
\bibitem [{\citenamefont {Kandala}\ \emph {et~al.}(2019)\citenamefont
  {Kandala}, \citenamefont {Temme}, \citenamefont {C{\'o}rcoles}, \citenamefont
  {Mezzacapo}, \citenamefont {Chow},\ and\ \citenamefont
  {Gambetta}}]{kandala2019error}%
  \BibitemOpen
  \bibfield  {author} {\bibinfo {author} {\bibfnamefont {A.}~\bibnamefont
  {Kandala}}, \bibinfo {author} {\bibfnamefont {K.}~\bibnamefont {Temme}},
  \bibinfo {author} {\bibfnamefont {A.~D.}\ \bibnamefont {C{\'o}rcoles}},
  \bibinfo {author} {\bibfnamefont {A.}~\bibnamefont {Mezzacapo}}, \bibinfo
  {author} {\bibfnamefont {J.~M.}\ \bibnamefont {Chow}},\ and\ \bibinfo
  {author} {\bibfnamefont {J.~M.}\ \bibnamefont {Gambetta}},\ }\bibfield
  {title} {\enquote {\bibinfo {title} {Error mitigation extends the
  computational reach of a noisy quantum processor},}\ }\href@noop {}
  {\bibfield  {journal} {\bibinfo  {journal} {Nature}\ }\textbf {\bibinfo
  {volume} {567}},\ \bibinfo {pages} {491--495} (\bibinfo {year}
  {2019})}\BibitemShut {NoStop}%
\bibitem [{\citenamefont {Gao}\ \emph {et~al.}(2021{\natexlab{b}})\citenamefont
  {Gao}, \citenamefont {Nakamura}, \citenamefont {Gujarati}, \citenamefont
  {Jones}, \citenamefont {Rice}, \citenamefont {Wood}, \citenamefont {Pistoia},
  \citenamefont {Garcia},\ and\ \citenamefont
  {Yamamoto}}]{gao2021computational}%
  \BibitemOpen
  \bibfield  {author} {\bibinfo {author} {\bibfnamefont {Q.}~\bibnamefont
  {Gao}}, \bibinfo {author} {\bibfnamefont {H.}~\bibnamefont {Nakamura}},
  \bibinfo {author} {\bibfnamefont {T.~P.}\ \bibnamefont {Gujarati}}, \bibinfo
  {author} {\bibfnamefont {G.~O.}\ \bibnamefont {Jones}}, \bibinfo {author}
  {\bibfnamefont {J.~E.}\ \bibnamefont {Rice}}, \bibinfo {author}
  {\bibfnamefont {S.~P.}\ \bibnamefont {Wood}}, \bibinfo {author}
  {\bibfnamefont {M.}~\bibnamefont {Pistoia}}, \bibinfo {author} {\bibfnamefont
  {J.~M.}\ \bibnamefont {Garcia}},\ and\ \bibinfo {author} {\bibfnamefont
  {N.}~\bibnamefont {Yamamoto}},\ }\bibfield  {title} {\enquote {\bibinfo
  {title} {Computational investigations of the lithium superoxide dimer
  rearrangement on noisy quantum devices},}\ }\href@noop {} {\bibfield
  {journal} {\bibinfo  {journal} {J. Phys. Chem. A}\ }\textbf {\bibinfo
  {volume} {125}},\ \bibinfo {pages} {1827--1836} (\bibinfo {year}
  {2021}{\natexlab{b}})}\BibitemShut {NoStop}%
\bibitem [{\citenamefont {Kirsopp}\ \emph {et~al.}(2022)\citenamefont
  {Kirsopp}, \citenamefont {Di~Paola}, \citenamefont {Manrique}, \citenamefont
  {Krompiec}, \citenamefont {Greene-Diniz}, \citenamefont {Guba}, \citenamefont
  {Meyder}, \citenamefont {Wolf}, \citenamefont {Strahm},\ and\ \citenamefont
  {Mu{\~n}oz~Ramo}}]{kirsopp2022quantum}%
  \BibitemOpen
  \bibfield  {author} {\bibinfo {author} {\bibfnamefont {J.~J.}\ \bibnamefont
  {Kirsopp}}, \bibinfo {author} {\bibfnamefont {C.}~\bibnamefont {Di~Paola}},
  \bibinfo {author} {\bibfnamefont {D.~Z.}\ \bibnamefont {Manrique}}, \bibinfo
  {author} {\bibfnamefont {M.}~\bibnamefont {Krompiec}}, \bibinfo {author}
  {\bibfnamefont {G.}~\bibnamefont {Greene-Diniz}}, \bibinfo {author}
  {\bibfnamefont {W.}~\bibnamefont {Guba}}, \bibinfo {author} {\bibfnamefont
  {A.}~\bibnamefont {Meyder}}, \bibinfo {author} {\bibfnamefont
  {D.}~\bibnamefont {Wolf}}, \bibinfo {author} {\bibfnamefont {M.}~\bibnamefont
  {Strahm}},\ and\ \bibinfo {author} {\bibfnamefont {D.}~\bibnamefont
  {Mu{\~n}oz~Ramo}},\ }\bibfield  {title} {\enquote {\bibinfo {title} {Quantum
  computational quantification of protein-ligand interactions},}\ }\href@noop
  {} {\bibfield  {journal} {\bibinfo  {journal} {Int. J. Quantum Chem.}\
  }\textbf {\bibinfo {volume} {122}},\ \bibinfo {pages} {e26975} (\bibinfo
  {year} {2022})}\BibitemShut {NoStop}%
\bibitem [{\citenamefont {O'Malley}\ \emph {et~al.}(2016)\citenamefont
  {O'Malley}, \citenamefont {Babbush}, \citenamefont {Kivlichan}, \citenamefont
  {Romero}, \citenamefont {McClean}, \citenamefont {Barends}, \citenamefont
  {Kelly}, \citenamefont {Roushan}, \citenamefont {Tranter}, \citenamefont
  {Ding}, \citenamefont {Campbell}, \citenamefont {Chen}, \citenamefont {Chen},
  \citenamefont {Chiaro}, \citenamefont {Dunsworth}, \citenamefont {Fowler},
  \citenamefont {Jeffrey}, \citenamefont {Lucero}, \citenamefont {Megrant},
  \citenamefont {Mutus}, \citenamefont {Neeley}, \citenamefont {Neill},
  \citenamefont {Quintana}, \citenamefont {Sank}, \citenamefont {Vainsencher},
  \citenamefont {Wenner}, \citenamefont {White}, \citenamefont {Coveney},
  \citenamefont {Love}, \citenamefont {Neven}, \citenamefont {Aspuru-Guzik},\
  and\ \citenamefont {Martinis}}]{o2016scalable}%
  \BibitemOpen
  \bibfield  {author} {\bibinfo {author} {\bibfnamefont {P.~J.~J.}\
  \bibnamefont {O'Malley}}, \bibinfo {author} {\bibfnamefont {R.}~\bibnamefont
  {Babbush}}, \bibinfo {author} {\bibfnamefont {I.~D.}\ \bibnamefont
  {Kivlichan}}, \bibinfo {author} {\bibfnamefont {J.}~\bibnamefont {Romero}},
  \bibinfo {author} {\bibfnamefont {J.~R.}\ \bibnamefont {McClean}}, \bibinfo
  {author} {\bibfnamefont {R.}~\bibnamefont {Barends}}, \bibinfo {author}
  {\bibfnamefont {J.}~\bibnamefont {Kelly}}, \bibinfo {author} {\bibfnamefont
  {P.}~\bibnamefont {Roushan}}, \bibinfo {author} {\bibfnamefont
  {A.}~\bibnamefont {Tranter}}, \bibinfo {author} {\bibfnamefont
  {N.}~\bibnamefont {Ding}}, \bibinfo {author} {\bibfnamefont {B.}~\bibnamefont
  {Campbell}}, \bibinfo {author} {\bibfnamefont {Y.}~\bibnamefont {Chen}},
  \bibinfo {author} {\bibfnamefont {Z.}~\bibnamefont {Chen}}, \bibinfo {author}
  {\bibfnamefont {B.}~\bibnamefont {Chiaro}}, \bibinfo {author} {\bibfnamefont
  {A.}~\bibnamefont {Dunsworth}}, \bibinfo {author} {\bibfnamefont {A.~G.}\
  \bibnamefont {Fowler}}, \bibinfo {author} {\bibfnamefont {E.}~\bibnamefont
  {Jeffrey}}, \bibinfo {author} {\bibfnamefont {E.}~\bibnamefont {Lucero}},
  \bibinfo {author} {\bibfnamefont {A.}~\bibnamefont {Megrant}}, \bibinfo
  {author} {\bibfnamefont {J.~Y.}\ \bibnamefont {Mutus}}, \bibinfo {author}
  {\bibfnamefont {M.}~\bibnamefont {Neeley}}, \bibinfo {author} {\bibfnamefont
  {C.}~\bibnamefont {Neill}}, \bibinfo {author} {\bibfnamefont
  {C.}~\bibnamefont {Quintana}}, \bibinfo {author} {\bibfnamefont
  {D.}~\bibnamefont {Sank}}, \bibinfo {author} {\bibfnamefont {A.}~\bibnamefont
  {Vainsencher}}, \bibinfo {author} {\bibfnamefont {J.}~\bibnamefont {Wenner}},
  \bibinfo {author} {\bibfnamefont {T.~C.}\ \bibnamefont {White}}, \bibinfo
  {author} {\bibfnamefont {P.~V.}\ \bibnamefont {Coveney}}, \bibinfo {author}
  {\bibfnamefont {P.~J.}\ \bibnamefont {Love}}, \bibinfo {author}
  {\bibfnamefont {H.}~\bibnamefont {Neven}}, \bibinfo {author} {\bibfnamefont
  {A.}~\bibnamefont {Aspuru-Guzik}},\ and\ \bibinfo {author} {\bibfnamefont
  {J.~M.}\ \bibnamefont {Martinis}},\ }\bibfield  {title} {\enquote {\bibinfo
  {title} {Scalable quantum simulation of molecular energies},}\ }\href@noop {}
  {\bibfield  {journal} {\bibinfo  {journal} {Phys. Rev. X.}\ }\textbf
  {\bibinfo {volume} {6}},\ \bibinfo {pages} {031007} (\bibinfo {year}
  {2016})}\BibitemShut {NoStop}%
\bibitem [{\citenamefont {Nam}\ \emph {et~al.}(2020)\citenamefont {Nam},
  \citenamefont {Chen}, \citenamefont {Pisenti}, \citenamefont {Wright},
  \citenamefont {Delaney}, \citenamefont {Maslov}, \citenamefont {Brown},
  \citenamefont {Allen}, \citenamefont {Amini}, \citenamefont {Apisdorf},
  \citenamefont {Beck}, \citenamefont {Blinov}, \citenamefont {Chaplin},
  \citenamefont {Chmielewski}, \citenamefont {Collins}, \citenamefont
  {Debnath}, \citenamefont {Hudek}, \citenamefont {Ducore}, \citenamefont
  {Keesan}, \citenamefont {Kreikemeier}, \citenamefont {Mizrahi}, \citenamefont
  {Solomon}, \citenamefont {Williams}, \citenamefont {Wong-Campos},
  \citenamefont {Moehring}, \citenamefont {Monroe},\ and\ \citenamefont
  {Kim}}]{nam2020ground}%
  \BibitemOpen
  \bibfield  {author} {\bibinfo {author} {\bibfnamefont {Y.}~\bibnamefont
  {Nam}}, \bibinfo {author} {\bibfnamefont {J.-S.}\ \bibnamefont {Chen}},
  \bibinfo {author} {\bibfnamefont {N.~C.}\ \bibnamefont {Pisenti}}, \bibinfo
  {author} {\bibfnamefont {K.}~\bibnamefont {Wright}}, \bibinfo {author}
  {\bibfnamefont {C.}~\bibnamefont {Delaney}}, \bibinfo {author} {\bibfnamefont
  {D.}~\bibnamefont {Maslov}}, \bibinfo {author} {\bibfnamefont {K.~R.}\
  \bibnamefont {Brown}}, \bibinfo {author} {\bibfnamefont {S.}~\bibnamefont
  {Allen}}, \bibinfo {author} {\bibfnamefont {J.~M.}\ \bibnamefont {Amini}},
  \bibinfo {author} {\bibfnamefont {J.}~\bibnamefont {Apisdorf}}, \bibinfo
  {author} {\bibfnamefont {K.~M.}\ \bibnamefont {Beck}}, \bibinfo {author}
  {\bibfnamefont {A.}~\bibnamefont {Blinov}}, \bibinfo {author} {\bibfnamefont
  {V.}~\bibnamefont {Chaplin}}, \bibinfo {author} {\bibfnamefont
  {M.}~\bibnamefont {Chmielewski}}, \bibinfo {author} {\bibfnamefont
  {C.}~\bibnamefont {Collins}}, \bibinfo {author} {\bibfnamefont
  {S.}~\bibnamefont {Debnath}}, \bibinfo {author} {\bibfnamefont {K.~M.}\
  \bibnamefont {Hudek}}, \bibinfo {author} {\bibfnamefont {A.~M.}\ \bibnamefont
  {Ducore}}, \bibinfo {author} {\bibfnamefont {M.}~\bibnamefont {Keesan}},
  \bibinfo {author} {\bibfnamefont {S.~M.}\ \bibnamefont {Kreikemeier}},
  \bibinfo {author} {\bibfnamefont {J.}~\bibnamefont {Mizrahi}}, \bibinfo
  {author} {\bibfnamefont {P.}~\bibnamefont {Solomon}}, \bibinfo {author}
  {\bibfnamefont {M.}~\bibnamefont {Williams}}, \bibinfo {author}
  {\bibfnamefont {J.~D.}\ \bibnamefont {Wong-Campos}}, \bibinfo {author}
  {\bibfnamefont {D.}~\bibnamefont {Moehring}}, \bibinfo {author}
  {\bibfnamefont {C.}~\bibnamefont {Monroe}},\ and\ \bibinfo {author}
  {\bibfnamefont {J.}~\bibnamefont {Kim}},\ }\bibfield  {title} {\enquote
  {\bibinfo {title} {Ground-state energy estimation of the water molecule on a
  trapped-ion quantum computer},}\ }\href@noop {} {\bibfield  {journal}
  {\bibinfo  {journal} {npj Quantum Inf.}\ }\textbf {\bibinfo {volume} {6}},\
  \bibinfo {pages} {33} (\bibinfo {year} {2020})}\BibitemShut {NoStop}%
\bibitem [{\citenamefont {McClean}\ \emph {et~al.}(2018)\citenamefont
  {McClean}, \citenamefont {Boixo}, \citenamefont {Smelyanskiy}, \citenamefont
  {Babbush},\ and\ \citenamefont {Neven}}]{mcclean2018barren}%
  \BibitemOpen
  \bibfield  {author} {\bibinfo {author} {\bibfnamefont {J.~R.}\ \bibnamefont
  {McClean}}, \bibinfo {author} {\bibfnamefont {S.}~\bibnamefont {Boixo}},
  \bibinfo {author} {\bibfnamefont {V.~N.}\ \bibnamefont {Smelyanskiy}},
  \bibinfo {author} {\bibfnamefont {R.}~\bibnamefont {Babbush}},\ and\ \bibinfo
  {author} {\bibfnamefont {H.}~\bibnamefont {Neven}},\ }\bibfield  {title}
  {\enquote {\bibinfo {title} {Barren plateaus in quantum neural network
  training landscapes},}\ }\href@noop {} {\bibfield  {journal} {\bibinfo
  {journal} {Nat. Commun.}\ }\textbf {\bibinfo {volume} {9}},\ \bibinfo {pages}
  {4812} (\bibinfo {year} {2018})}\BibitemShut {NoStop}%
\bibitem [{\citenamefont {Choy}\ and\ \citenamefont
  {Wales}(2023)}]{choy2023molecular}%
  \BibitemOpen
  \bibfield  {author} {\bibinfo {author} {\bibfnamefont {B.}~\bibnamefont
  {Choy}}\ and\ \bibinfo {author} {\bibfnamefont {D.~J.}\ \bibnamefont
  {Wales}},\ }\bibfield  {title} {\enquote {\bibinfo {title} {Molecular energy
  landscapes of hardware-efficient ansatz in quantum computing},}\ }\href@noop
  {} {\bibfield  {journal} {\bibinfo  {journal} {J. Chem. Theory and Comput.}\
  }\textbf {\bibinfo {volume} {19}},\ \bibinfo {pages} {1197--1206} (\bibinfo
  {year} {2023})}\BibitemShut {NoStop}%
\bibitem [{\citenamefont {Grimsley}\ \emph {et~al.}(2019)\citenamefont
  {Grimsley}, \citenamefont {Economou}, \citenamefont {Barnes},\ and\
  \citenamefont {Mayhall}}]{grimsley2019adaptive}%
  \BibitemOpen
  \bibfield  {author} {\bibinfo {author} {\bibfnamefont {H.~R.}\ \bibnamefont
  {Grimsley}}, \bibinfo {author} {\bibfnamefont {S.~E.}\ \bibnamefont
  {Economou}}, \bibinfo {author} {\bibfnamefont {E.}~\bibnamefont {Barnes}},\
  and\ \bibinfo {author} {\bibfnamefont {N.~J.}\ \bibnamefont {Mayhall}},\
  }\bibfield  {title} {\enquote {\bibinfo {title} {An adaptive variational
  algorithm for exact molecular simulations on a quantum computer},}\
  }\href@noop {} {\bibfield  {journal} {\bibinfo  {journal} {Nat. Commun.}\
  }\textbf {\bibinfo {volume} {10}},\ \bibinfo {pages} {1--9} (\bibinfo {year}
  {2019})}\BibitemShut {NoStop}%
\bibitem [{\citenamefont {Yordanov}\ \emph {et~al.}(2021)\citenamefont
  {Yordanov}, \citenamefont {Armaos}, \citenamefont {Barnes},\ and\
  \citenamefont {Arvidsson-Shukur}}]{yordanov2021qubit}%
  \BibitemOpen
  \bibfield  {author} {\bibinfo {author} {\bibfnamefont {Y.~S.}\ \bibnamefont
  {Yordanov}}, \bibinfo {author} {\bibfnamefont {V.}~\bibnamefont {Armaos}},
  \bibinfo {author} {\bibfnamefont {C.~H.}\ \bibnamefont {Barnes}},\ and\
  \bibinfo {author} {\bibfnamefont {D.~R.}\ \bibnamefont {Arvidsson-Shukur}},\
  }\bibfield  {title} {\enquote {\bibinfo {title} {Qubit-excitation-based
  adaptive variational quantum eigensolver},}\ }\href@noop {} {\bibfield
  {journal} {\bibinfo  {journal} {Commun. Phys.}\ }\textbf {\bibinfo {volume}
  {4}},\ \bibinfo {pages} {228} (\bibinfo {year} {2021})}\BibitemShut {NoStop}%
\bibitem [{\citenamefont {Tang}\ \emph {et~al.}(2021)\citenamefont {Tang},
  \citenamefont {Shkolnikov}, \citenamefont {Barron}, \citenamefont {Grimsley},
  \citenamefont {Mayhall}, \citenamefont {Barnes},\ and\ \citenamefont
  {Economou}}]{tang2021qubit}%
  \BibitemOpen
  \bibfield  {author} {\bibinfo {author} {\bibfnamefont {H.~L.}\ \bibnamefont
  {Tang}}, \bibinfo {author} {\bibfnamefont {V.}~\bibnamefont {Shkolnikov}},
  \bibinfo {author} {\bibfnamefont {G.~S.}\ \bibnamefont {Barron}}, \bibinfo
  {author} {\bibfnamefont {H.~R.}\ \bibnamefont {Grimsley}}, \bibinfo {author}
  {\bibfnamefont {N.~J.}\ \bibnamefont {Mayhall}}, \bibinfo {author}
  {\bibfnamefont {E.}~\bibnamefont {Barnes}},\ and\ \bibinfo {author}
  {\bibfnamefont {S.~E.}\ \bibnamefont {Economou}},\ }\bibfield  {title}
  {\enquote {\bibinfo {title} {qubit-{ADAPT-VQE}: An adaptive algorithm for
  constructing hardware-efficient ans{\"a}tze on a quantum processor},}\
  }\href@noop {} {\bibfield  {journal} {\bibinfo  {journal} {PRX Quantum}\
  }\textbf {\bibinfo {volume} {2}},\ \bibinfo {pages} {020310} (\bibinfo {year}
  {2021})}\BibitemShut {NoStop}%
\bibitem [{\citenamefont {Mizukami}\ \emph {et~al.}(2020)\citenamefont
  {Mizukami}, \citenamefont {Mitarai}, \citenamefont {Nakagawa}, \citenamefont
  {Yamamoto}, \citenamefont {Yan},\ and\ \citenamefont
  {Ohnishi}}]{mizukami2020orbital}%
  \BibitemOpen
  \bibfield  {author} {\bibinfo {author} {\bibfnamefont {W.}~\bibnamefont
  {Mizukami}}, \bibinfo {author} {\bibfnamefont {K.}~\bibnamefont {Mitarai}},
  \bibinfo {author} {\bibfnamefont {Y.~O.}\ \bibnamefont {Nakagawa}}, \bibinfo
  {author} {\bibfnamefont {T.}~\bibnamefont {Yamamoto}}, \bibinfo {author}
  {\bibfnamefont {T.}~\bibnamefont {Yan}},\ and\ \bibinfo {author}
  {\bibfnamefont {Y.-y.}\ \bibnamefont {Ohnishi}},\ }\bibfield  {title}
  {\enquote {\bibinfo {title} {Orbital optimized unitary coupled cluster theory
  for quantum computer},}\ }\href@noop {} {\bibfield  {journal} {\bibinfo
  {journal} {Phys. Rev. Res.}\ }\textbf {\bibinfo {volume} {2}},\ \bibinfo
  {pages} {033421} (\bibinfo {year} {2020})}\BibitemShut {NoStop}%
\bibitem [{\citenamefont {Sokolov}\ \emph {et~al.}(2020)\citenamefont
  {Sokolov}, \citenamefont {Barkoutsos}, \citenamefont {Ollitrault},
  \citenamefont {Greenberg}, \citenamefont {Rice}, \citenamefont {Pistoia},\
  and\ \citenamefont {Tavernelli}}]{sokolov2020quantum}%
  \BibitemOpen
  \bibfield  {author} {\bibinfo {author} {\bibfnamefont {I.~O.}\ \bibnamefont
  {Sokolov}}, \bibinfo {author} {\bibfnamefont {P.~K.}\ \bibnamefont
  {Barkoutsos}}, \bibinfo {author} {\bibfnamefont {P.~J.}\ \bibnamefont
  {Ollitrault}}, \bibinfo {author} {\bibfnamefont {D.}~\bibnamefont
  {Greenberg}}, \bibinfo {author} {\bibfnamefont {J.}~\bibnamefont {Rice}},
  \bibinfo {author} {\bibfnamefont {M.}~\bibnamefont {Pistoia}},\ and\ \bibinfo
  {author} {\bibfnamefont {I.}~\bibnamefont {Tavernelli}},\ }\bibfield  {title}
  {\enquote {\bibinfo {title} {Quantum orbital-optimized unitary coupled
  cluster methods in the strongly correlated regime: Can quantum algorithms
  outperform their classical equivalents?}}\ }\href@noop {} {\bibfield
  {journal} {\bibinfo  {journal} {J. Chem. Phys.}\ }\textbf {\bibinfo {volume}
  {152}},\ \bibinfo {pages} {124107} (\bibinfo {year} {2020})}\BibitemShut
  {NoStop}%
\bibitem [{\citenamefont {Bierman}, \citenamefont {Li},\ and\ \citenamefont
  {Lu}(2023)}]{bierman2023improving}%
  \BibitemOpen
  \bibfield  {author} {\bibinfo {author} {\bibfnamefont {J.}~\bibnamefont
  {Bierman}}, \bibinfo {author} {\bibfnamefont {Y.}~\bibnamefont {Li}},\ and\
  \bibinfo {author} {\bibfnamefont {J.}~\bibnamefont {Lu}},\ }\bibfield
  {title} {\enquote {\bibinfo {title} {Improving the accuracy of variational
  quantum eigensolvers with fewer qubits using orbital optimization},}\
  }\href@noop {} {\bibfield  {journal} {\bibinfo  {journal} {J. Chem. Theory
  Comput.}\ }\textbf {\bibinfo {volume} {19}},\ \bibinfo {pages} {790--798}
  (\bibinfo {year} {2023})}\BibitemShut {NoStop}%
\bibitem [{\citenamefont {Gottesman}(1998)}]{gottesman1998heisenberg}%
  \BibitemOpen
  \bibfield  {author} {\bibinfo {author} {\bibfnamefont {D.}~\bibnamefont
  {Gottesman}},\ }\bibfield  {title} {\enquote {\bibinfo {title} {The
  heisenberg representation of quantum computers},}\ }\href@noop {} {\bibfield
  {journal} {\bibinfo  {journal} {arXiv preprint quant-ph/9807006}\ } (\bibinfo
  {year} {1998})}\BibitemShut {NoStop}%
\bibitem [{\citenamefont {Aaronson}\ and\ \citenamefont
  {Gottesman}(2004)}]{aaronson2004improved}%
  \BibitemOpen
  \bibfield  {author} {\bibinfo {author} {\bibfnamefont {S.}~\bibnamefont
  {Aaronson}}\ and\ \bibinfo {author} {\bibfnamefont {D.}~\bibnamefont
  {Gottesman}},\ }\bibfield  {title} {\enquote {\bibinfo {title} {Improved
  simulation of stabilizer circuits},}\ }\href@noop {} {\bibfield  {journal}
  {\bibinfo  {journal} {Phys. Rev. A}\ }\textbf {\bibinfo {volume} {70}},\
  \bibinfo {pages} {052328} (\bibinfo {year} {2004})}\BibitemShut {NoStop}%
\bibitem [{\citenamefont {Leone}\ \emph {et~al.}(2022)\citenamefont {Leone},
  \citenamefont {Oliviero}, \citenamefont {Cincio},\ and\ \citenamefont
  {Cerezo}}]{leone2022practical}%
  \BibitemOpen
  \bibfield  {author} {\bibinfo {author} {\bibfnamefont {L.}~\bibnamefont
  {Leone}}, \bibinfo {author} {\bibfnamefont {S.~F.}\ \bibnamefont {Oliviero}},
  \bibinfo {author} {\bibfnamefont {L.}~\bibnamefont {Cincio}},\ and\ \bibinfo
  {author} {\bibfnamefont {M.}~\bibnamefont {Cerezo}},\ }\bibfield  {title}
  {\enquote {\bibinfo {title} {On the practical usefulness of the hardware
  efficient ansatz},}\ }\href@noop {} {\bibfield  {journal} {\bibinfo
  {journal} {arXiv preprint arXiv:2211.01477}\ } (\bibinfo {year}
  {2022})}\BibitemShut {NoStop}%
\bibitem [{\citenamefont {Ryabinkin}\ \emph {et~al.}(2018)\citenamefont
  {Ryabinkin}, \citenamefont {Yen}, \citenamefont {Genin},\ and\ \citenamefont
  {Izmaylov}}]{ryabinkin2018qubit}%
  \BibitemOpen
  \bibfield  {author} {\bibinfo {author} {\bibfnamefont {I.~G.}\ \bibnamefont
  {Ryabinkin}}, \bibinfo {author} {\bibfnamefont {T.-C.}\ \bibnamefont {Yen}},
  \bibinfo {author} {\bibfnamefont {S.~N.}\ \bibnamefont {Genin}},\ and\
  \bibinfo {author} {\bibfnamefont {A.~F.}\ \bibnamefont {Izmaylov}},\
  }\bibfield  {title} {\enquote {\bibinfo {title} {Qubit coupled cluster
  method: a systematic approach to quantum chemistry on a quantum computer},}\
  }\href@noop {} {\bibfield  {journal} {\bibinfo  {journal} {J. Chem. Theory
  and Comput.}\ }\textbf {\bibinfo {volume} {14}},\ \bibinfo {pages}
  {6317--6326} (\bibinfo {year} {2018})}\BibitemShut {NoStop}%
\bibitem [{\citenamefont {Ryabinkin}\ \emph {et~al.}(2020)\citenamefont
  {Ryabinkin}, \citenamefont {Lang}, \citenamefont {Genin},\ and\ \citenamefont
  {Izmaylov}}]{ryabinkin2020iterative}%
  \BibitemOpen
  \bibfield  {author} {\bibinfo {author} {\bibfnamefont {I.~G.}\ \bibnamefont
  {Ryabinkin}}, \bibinfo {author} {\bibfnamefont {R.~A.}\ \bibnamefont {Lang}},
  \bibinfo {author} {\bibfnamefont {S.~N.}\ \bibnamefont {Genin}},\ and\
  \bibinfo {author} {\bibfnamefont {A.~F.}\ \bibnamefont {Izmaylov}},\
  }\bibfield  {title} {\enquote {\bibinfo {title} {Iterative qubit coupled
  cluster approach with efficient screening of generators},}\ }\href@noop {}
  {\bibfield  {journal} {\bibinfo  {journal} {J. Chem. Theory Comput.}\
  }\textbf {\bibinfo {volume} {16}},\ \bibinfo {pages} {1055--1063} (\bibinfo
  {year} {2020})}\BibitemShut {NoStop}%
\bibitem [{\citenamefont {Ryabinkin}, \citenamefont {Izmaylov},\ and\
  \citenamefont {Genin}(2021)}]{ryabinkin2021posteriori}%
  \BibitemOpen
  \bibfield  {author} {\bibinfo {author} {\bibfnamefont {I.~G.}\ \bibnamefont
  {Ryabinkin}}, \bibinfo {author} {\bibfnamefont {A.~F.}\ \bibnamefont
  {Izmaylov}},\ and\ \bibinfo {author} {\bibfnamefont {S.~N.}\ \bibnamefont
  {Genin}},\ }\bibfield  {title} {\enquote {\bibinfo {title} {A posteriori
  corrections to the iterative qubit coupled cluster method to minimize the use
  of quantum resources in large-scale calculations},}\ }\href@noop {}
  {\bibfield  {journal} {\bibinfo  {journal} {Quantum Sci. Technol.}\ }\textbf
  {\bibinfo {volume} {6}},\ \bibinfo {pages} {024012} (\bibinfo {year}
  {2021})}\BibitemShut {NoStop}%
\bibitem [{\citenamefont {Lang}, \citenamefont {Ryabinkin},\ and\ \citenamefont
  {Izmaylov}(2020)}]{lang2020unitary}%
  \BibitemOpen
  \bibfield  {author} {\bibinfo {author} {\bibfnamefont {R.~A.}\ \bibnamefont
  {Lang}}, \bibinfo {author} {\bibfnamefont {I.~G.}\ \bibnamefont
  {Ryabinkin}},\ and\ \bibinfo {author} {\bibfnamefont {A.~F.}\ \bibnamefont
  {Izmaylov}},\ }\bibfield  {title} {\enquote {\bibinfo {title} {Unitary
  transformation of the electronic {Hamiltonian} with an exact quadratic
  truncation of the {Baker-Campbell-Hausdorff} expansion},}\ }\href@noop {}
  {\bibfield  {journal} {\bibinfo  {journal} {J. Chem. Theory and Comput.}\
  }\textbf {\bibinfo {volume} {17}},\ \bibinfo {pages} {66--78} (\bibinfo
  {year} {2020})}\BibitemShut {NoStop}%
\bibitem [{\citenamefont {Genin}\ \emph {et~al.}(2022)\citenamefont {Genin},
  \citenamefont {Ryabinkin}, \citenamefont {Paisley}, \citenamefont {Whelan},
  \citenamefont {Helander},\ and\ \citenamefont
  {Hudson}}]{genin2022estimating}%
  \BibitemOpen
  \bibfield  {author} {\bibinfo {author} {\bibfnamefont {S.~N.}\ \bibnamefont
  {Genin}}, \bibinfo {author} {\bibfnamefont {I.~G.}\ \bibnamefont
  {Ryabinkin}}, \bibinfo {author} {\bibfnamefont {N.~R.}\ \bibnamefont
  {Paisley}}, \bibinfo {author} {\bibfnamefont {S.~O.}\ \bibnamefont {Whelan}},
  \bibinfo {author} {\bibfnamefont {M.~G.}\ \bibnamefont {Helander}},\ and\
  \bibinfo {author} {\bibfnamefont {Z.~M.}\ \bibnamefont {Hudson}},\ }\bibfield
   {title} {\enquote {\bibinfo {title} {Estimating phosphorescent emission
  energies in ir$^\textrm{III}$ complexes using large-scale quantum computing
  simulations},}\ }\href@noop {} {\bibfield  {journal} {\bibinfo  {journal}
  {Angew. Chem. Int. Ed.}\ }\textbf {\bibinfo {volume} {61}},\ \bibinfo {pages}
  {e202116175} (\bibinfo {year} {2022})}\BibitemShut {NoStop}%
\bibitem [{\citenamefont {Ryabinkin}, \citenamefont {Jena},\ and\ \citenamefont
  {Genin}(2023)}]{ryabinkin2023efficient}%
  \BibitemOpen
  \bibfield  {author} {\bibinfo {author} {\bibfnamefont {I.~G.}\ \bibnamefont
  {Ryabinkin}}, \bibinfo {author} {\bibfnamefont {A.~J.}\ \bibnamefont
  {Jena}},\ and\ \bibinfo {author} {\bibfnamefont {S.~N.}\ \bibnamefont
  {Genin}},\ }\bibfield  {title} {\enquote {\bibinfo {title} {Efficient
  construction of involutory linear combinations of anticommuting pauli
  generators for large-scale iterative qubit coupled cluster calculations},}\
  }\href@noop {} {\bibfield  {journal} {\bibinfo  {journal} {J. Chem. Theory
  and Comput.}\ }\textbf {\bibinfo {volume} {19}},\ \bibinfo {pages}
  {1722--1733} (\bibinfo {year} {2023})}\BibitemShut {NoStop}%
\bibitem [{\citenamefont {Mishmash}\ \emph {et~al.}(2023)\citenamefont
  {Mishmash}, \citenamefont {Gujarati}, \citenamefont {Motta}, \citenamefont
  {Zhai}, \citenamefont {Chan},\ and\ \citenamefont
  {Mezzacapo}}]{mishmash2023hierarchical}%
  \BibitemOpen
  \bibfield  {author} {\bibinfo {author} {\bibfnamefont {R.~V.}\ \bibnamefont
  {Mishmash}}, \bibinfo {author} {\bibfnamefont {T.~P.}\ \bibnamefont
  {Gujarati}}, \bibinfo {author} {\bibfnamefont {M.}~\bibnamefont {Motta}},
  \bibinfo {author} {\bibfnamefont {H.}~\bibnamefont {Zhai}}, \bibinfo {author}
  {\bibfnamefont {G.~K.}\ \bibnamefont {Chan}},\ and\ \bibinfo {author}
  {\bibfnamefont {A.}~\bibnamefont {Mezzacapo}},\ }\bibfield  {title} {\enquote
  {\bibinfo {title} {Hierarchical clifford transformations to reduce
  entanglement in quantum chemistry wavefunctions},}\ }\href@noop {} {\bibfield
   {journal} {\bibinfo  {journal} {arXiv preprint arXiv:2301.07726}\ }
  (\bibinfo {year} {2023})}\BibitemShut {NoStop}%
\bibitem [{\citenamefont {Schleich}\ \emph {et~al.}(2023)\citenamefont
  {Schleich}, \citenamefont {Boen}, \citenamefont {Cincio}, \citenamefont
  {Anand}, \citenamefont {Kottmann}, \citenamefont {Tretiak}, \citenamefont
  {Dub},\ and\ \citenamefont {Aspuru-Guzik}}]{schleich2023partitioning}%
  \BibitemOpen
  \bibfield  {author} {\bibinfo {author} {\bibfnamefont {P.}~\bibnamefont
  {Schleich}}, \bibinfo {author} {\bibfnamefont {J.}~\bibnamefont {Boen}},
  \bibinfo {author} {\bibfnamefont {L.}~\bibnamefont {Cincio}}, \bibinfo
  {author} {\bibfnamefont {A.}~\bibnamefont {Anand}}, \bibinfo {author}
  {\bibfnamefont {J.~S.}\ \bibnamefont {Kottmann}}, \bibinfo {author}
  {\bibfnamefont {S.}~\bibnamefont {Tretiak}}, \bibinfo {author} {\bibfnamefont
  {P.~A.}\ \bibnamefont {Dub}},\ and\ \bibinfo {author} {\bibfnamefont
  {A.}~\bibnamefont {Aspuru-Guzik}},\ }\bibfield  {title} {\enquote {\bibinfo
  {title} {Partitioning quantum chemistry simulations with clifford
  circuits},}\ }\href@noop {} {\bibfield  {journal} {\bibinfo  {journal} {arXiv
  preprint arXiv:2303.01221}\ } (\bibinfo {year} {2023})}\BibitemShut {NoStop}%
\bibitem [{\citenamefont {Cheng}\ \emph {et~al.}(2022)\citenamefont {Cheng},
  \citenamefont {Khosla}, \citenamefont {Self}, \citenamefont {Lin},
  \citenamefont {Li}, \citenamefont {Medina},\ and\ \citenamefont
  {Kim}}]{cheng2022clifford}%
  \BibitemOpen
  \bibfield  {author} {\bibinfo {author} {\bibfnamefont {M.}~\bibnamefont
  {Cheng}}, \bibinfo {author} {\bibfnamefont {K.}~\bibnamefont {Khosla}},
  \bibinfo {author} {\bibfnamefont {C.}~\bibnamefont {Self}}, \bibinfo {author}
  {\bibfnamefont {M.}~\bibnamefont {Lin}}, \bibinfo {author} {\bibfnamefont
  {B.}~\bibnamefont {Li}}, \bibinfo {author} {\bibfnamefont {A.}~\bibnamefont
  {Medina}},\ and\ \bibinfo {author} {\bibfnamefont {M.}~\bibnamefont {Kim}},\
  }\bibfield  {title} {\enquote {\bibinfo {title} {Clifford circuit
  initialisation for variational quantum algorithms},}\ }\href@noop {}
  {\bibfield  {journal} {\bibinfo  {journal} {arXiv preprint arXiv:2207.01539}\
  } (\bibinfo {year} {2022})}\BibitemShut {NoStop}%
\bibitem [{\citenamefont {Mitarai}\ \emph {et~al.}(2022)\citenamefont
  {Mitarai}, \citenamefont {Suzuki}, \citenamefont {Mizukami}, \citenamefont
  {Nakagawa},\ and\ \citenamefont {Fujii}}]{mitarai2022quadratic}%
  \BibitemOpen
  \bibfield  {author} {\bibinfo {author} {\bibfnamefont {K.}~\bibnamefont
  {Mitarai}}, \bibinfo {author} {\bibfnamefont {Y.}~\bibnamefont {Suzuki}},
  \bibinfo {author} {\bibfnamefont {W.}~\bibnamefont {Mizukami}}, \bibinfo
  {author} {\bibfnamefont {Y.~O.}\ \bibnamefont {Nakagawa}},\ and\ \bibinfo
  {author} {\bibfnamefont {K.}~\bibnamefont {Fujii}},\ }\bibfield  {title}
  {\enquote {\bibinfo {title} {Quadratic clifford expansion for efficient
  benchmarking and initialization of variational quantum algorithms},}\
  }\href@noop {} {\bibfield  {journal} {\bibinfo  {journal} {Phys. Rev. Res.}\
  }\textbf {\bibinfo {volume} {4}},\ \bibinfo {pages} {033012} (\bibinfo {year}
  {2022})}\BibitemShut {NoStop}%
\bibitem [{\citenamefont {Li}\ \emph {et~al.}(2023)\citenamefont {Li},
  \citenamefont {Allcock}, \citenamefont {Cheng}, \citenamefont {Zhang},
  \citenamefont {Chen}, \citenamefont {Mailoa}, \citenamefont {Shuai},\ and\
  \citenamefont {Zhang}}]{li2023tencirchem}%
  \BibitemOpen
  \bibfield  {author} {\bibinfo {author} {\bibfnamefont {W.}~\bibnamefont
  {Li}}, \bibinfo {author} {\bibfnamefont {J.}~\bibnamefont {Allcock}},
  \bibinfo {author} {\bibfnamefont {L.}~\bibnamefont {Cheng}}, \bibinfo
  {author} {\bibfnamefont {S.-X.}\ \bibnamefont {Zhang}}, \bibinfo {author}
  {\bibfnamefont {Y.-Q.}\ \bibnamefont {Chen}}, \bibinfo {author}
  {\bibfnamefont {J.~P.}\ \bibnamefont {Mailoa}}, \bibinfo {author}
  {\bibfnamefont {Z.}~\bibnamefont {Shuai}},\ and\ \bibinfo {author}
  {\bibfnamefont {S.}~\bibnamefont {Zhang}},\ }\bibfield  {title} {\enquote
  {\bibinfo {title} {Tencirchem: An efficient quantum computational chemistry
  package for the nisq era},}\ }\href@noop {} {\bibfield  {journal} {\bibinfo
  {journal} {J. Chem. Theory Comput.}\ } (\bibinfo {year} {2023})}\BibitemShut
  {NoStop}%
\bibitem [{\citenamefont {Sun}\ \emph {et~al.}(2018)\citenamefont {Sun},
  \citenamefont {Berkelbach}, \citenamefont {Blunt}, \citenamefont {Booth},
  \citenamefont {Guo}, \citenamefont {Li}, \citenamefont {Liu}, \citenamefont
  {McClain}, \citenamefont {Sayfutyarova}, \citenamefont {Sharma},
  \citenamefont {Wouters},\ and\ \citenamefont {Chan}}]{sun2018pyscf}%
  \BibitemOpen
  \bibfield  {author} {\bibinfo {author} {\bibfnamefont {Q.}~\bibnamefont
  {Sun}}, \bibinfo {author} {\bibfnamefont {T.~C.}\ \bibnamefont {Berkelbach}},
  \bibinfo {author} {\bibfnamefont {N.~S.}\ \bibnamefont {Blunt}}, \bibinfo
  {author} {\bibfnamefont {G.~H.}\ \bibnamefont {Booth}}, \bibinfo {author}
  {\bibfnamefont {S.}~\bibnamefont {Guo}}, \bibinfo {author} {\bibfnamefont
  {Z.}~\bibnamefont {Li}}, \bibinfo {author} {\bibfnamefont {J.}~\bibnamefont
  {Liu}}, \bibinfo {author} {\bibfnamefont {J.}~\bibnamefont {McClain}},
  \bibinfo {author} {\bibfnamefont {E.~R.}\ \bibnamefont {Sayfutyarova}},
  \bibinfo {author} {\bibfnamefont {S.}~\bibnamefont {Sharma}}, \bibinfo
  {author} {\bibfnamefont {S.}~\bibnamefont {Wouters}},\ and\ \bibinfo {author}
  {\bibfnamefont {G.~K.-L.}\ \bibnamefont {Chan}},\ }\bibfield  {title}
  {\enquote {\bibinfo {title} {{PySCF}: the {Python}-based simulations of
  chemistry framework},}\ }\href@noop {} {\bibfield  {journal} {\bibinfo
  {journal} {Wiley Interdiscip. Rev. Comput. Mol. Sci.}\ }\textbf {\bibinfo
  {volume} {8}},\ \bibinfo {pages} {e1340} (\bibinfo {year}
  {2018})}\BibitemShut {NoStop}%
\bibitem [{\citenamefont {Bravyi}\ and\ \citenamefont
  {Kitaev}(2002)}]{bravyi2002fermionic}%
  \BibitemOpen
  \bibfield  {author} {\bibinfo {author} {\bibfnamefont {S.~B.}\ \bibnamefont
  {Bravyi}}\ and\ \bibinfo {author} {\bibfnamefont {A.~Y.}\ \bibnamefont
  {Kitaev}},\ }\bibfield  {title} {\enquote {\bibinfo {title} {Fermionic
  quantum computation},}\ }\href@noop {} {\bibfield  {journal} {\bibinfo
  {journal} {Ann. Phys.}\ }\textbf {\bibinfo {volume} {298}},\ \bibinfo {pages}
  {210--226} (\bibinfo {year} {2002})}\BibitemShut {NoStop}%
\bibitem [{\citenamefont {Seeley}, \citenamefont {Richard},\ and\ \citenamefont
  {Love}(2012)}]{seeley2012bravyi}%
  \BibitemOpen
  \bibfield  {author} {\bibinfo {author} {\bibfnamefont {J.~T.}\ \bibnamefont
  {Seeley}}, \bibinfo {author} {\bibfnamefont {M.~J.}\ \bibnamefont
  {Richard}},\ and\ \bibinfo {author} {\bibfnamefont {P.~J.}\ \bibnamefont
  {Love}},\ }\bibfield  {title} {\enquote {\bibinfo {title} {The
  {Bravyi-Kitaev} transformation for quantum computation of electronic
  structure},}\ }\href@noop {} {\bibfield  {journal} {\bibinfo  {journal} {J.
  Chem. Phys.}\ }\textbf {\bibinfo {volume} {137}},\ \bibinfo {pages} {224109}
  (\bibinfo {year} {2012})}\BibitemShut {NoStop}%
\bibitem [{\citenamefont {Bradbury}\ \emph {et~al.}(2018)\citenamefont
  {Bradbury}, \citenamefont {Frostig}, \citenamefont {Hawkins}, \citenamefont
  {Johnson}, \citenamefont {Leary}, \citenamefont {Maclaurin}, \citenamefont
  {Necula}, \citenamefont {Paszke}, \citenamefont {Vander{P}las}, \citenamefont
  {Wanderman-{M}ilne},\ and\ \citenamefont {Zhang}}]{jax2018github}%
  \BibitemOpen
  \bibfield  {author} {\bibinfo {author} {\bibfnamefont {J.}~\bibnamefont
  {Bradbury}}, \bibinfo {author} {\bibfnamefont {R.}~\bibnamefont {Frostig}},
  \bibinfo {author} {\bibfnamefont {P.}~\bibnamefont {Hawkins}}, \bibinfo
  {author} {\bibfnamefont {M.~J.}\ \bibnamefont {Johnson}}, \bibinfo {author}
  {\bibfnamefont {C.}~\bibnamefont {Leary}}, \bibinfo {author} {\bibfnamefont
  {D.}~\bibnamefont {Maclaurin}}, \bibinfo {author} {\bibfnamefont
  {G.}~\bibnamefont {Necula}}, \bibinfo {author} {\bibfnamefont
  {A.}~\bibnamefont {Paszke}}, \bibinfo {author} {\bibfnamefont
  {J.}~\bibnamefont {Vander{P}las}}, \bibinfo {author} {\bibfnamefont
  {S.}~\bibnamefont {Wanderman-{M}ilne}},\ and\ \bibinfo {author}
  {\bibfnamefont {Q.}~\bibnamefont {Zhang}},\ }\href
  {http://github.com/google/jax} {\enquote {\bibinfo {title} {{JAX}: composable
  transformations of {P}ython+{N}um{P}y programs},}\ } (\bibinfo {year}
  {2018}),\ \bibinfo {note} {accessed in 2023.12.02}\BibitemShut {NoStop}%
\bibitem [{\citenamefont {Virtanen}\ \emph {et~al.}(2020)\citenamefont
  {Virtanen}, \citenamefont {Gommers}, \citenamefont {Oliphant}, \citenamefont
  {Haberland}, \citenamefont {Reddy}, \citenamefont {Cournapeau}, \citenamefont
  {Burovski}, \citenamefont {Peterson}, \citenamefont {Weckesser},
  \citenamefont {Bright}, \citenamefont {{van der Walt}}, \citenamefont
  {Brett}, \citenamefont {Wilson}, \citenamefont {Millman}, \citenamefont
  {Mayorov}, \citenamefont {Nelson}, \citenamefont {Jones}, \citenamefont
  {Kern}, \citenamefont {Larson}, \citenamefont {Carey}, \citenamefont {Polat},
  \citenamefont {Feng}, \citenamefont {Moore}, \citenamefont {{VanderPlas}},
  \citenamefont {Laxalde}, \citenamefont {Perktold}, \citenamefont {Cimrman},
  \citenamefont {Henriksen}, \citenamefont {Quintero}, \citenamefont {Harris},
  \citenamefont {Archibald}, \citenamefont {Ribeiro}, \citenamefont
  {Pedregosa}, \citenamefont {{van Mulbregt}},\ and\ \citenamefont {{SciPy 1.0
  Contributors}}}]{scipy}%
  \BibitemOpen
  \bibfield  {author} {\bibinfo {author} {\bibfnamefont {P.}~\bibnamefont
  {Virtanen}}, \bibinfo {author} {\bibfnamefont {R.}~\bibnamefont {Gommers}},
  \bibinfo {author} {\bibfnamefont {T.~E.}\ \bibnamefont {Oliphant}}, \bibinfo
  {author} {\bibfnamefont {M.}~\bibnamefont {Haberland}}, \bibinfo {author}
  {\bibfnamefont {T.}~\bibnamefont {Reddy}}, \bibinfo {author} {\bibfnamefont
  {D.}~\bibnamefont {Cournapeau}}, \bibinfo {author} {\bibfnamefont
  {E.}~\bibnamefont {Burovski}}, \bibinfo {author} {\bibfnamefont
  {P.}~\bibnamefont {Peterson}}, \bibinfo {author} {\bibfnamefont
  {W.}~\bibnamefont {Weckesser}}, \bibinfo {author} {\bibfnamefont
  {J.}~\bibnamefont {Bright}}, \bibinfo {author} {\bibfnamefont {S.~J.}\
  \bibnamefont {{van der Walt}}}, \bibinfo {author} {\bibfnamefont
  {M.}~\bibnamefont {Brett}}, \bibinfo {author} {\bibfnamefont
  {J.}~\bibnamefont {Wilson}}, \bibinfo {author} {\bibfnamefont {K.~J.}\
  \bibnamefont {Millman}}, \bibinfo {author} {\bibfnamefont {N.}~\bibnamefont
  {Mayorov}}, \bibinfo {author} {\bibfnamefont {A.~R.~J.}\ \bibnamefont
  {Nelson}}, \bibinfo {author} {\bibfnamefont {E.}~\bibnamefont {Jones}},
  \bibinfo {author} {\bibfnamefont {R.}~\bibnamefont {Kern}}, \bibinfo {author}
  {\bibfnamefont {E.}~\bibnamefont {Larson}}, \bibinfo {author} {\bibfnamefont
  {C.~J.}\ \bibnamefont {Carey}}, \bibinfo {author} {\bibfnamefont
  {{\.I}.}~\bibnamefont {Polat}}, \bibinfo {author} {\bibfnamefont
  {Y.}~\bibnamefont {Feng}}, \bibinfo {author} {\bibfnamefont {E.~W.}\
  \bibnamefont {Moore}}, \bibinfo {author} {\bibfnamefont {J.}~\bibnamefont
  {{VanderPlas}}}, \bibinfo {author} {\bibfnamefont {D.}~\bibnamefont
  {Laxalde}}, \bibinfo {author} {\bibfnamefont {J.}~\bibnamefont {Perktold}},
  \bibinfo {author} {\bibfnamefont {R.}~\bibnamefont {Cimrman}}, \bibinfo
  {author} {\bibfnamefont {I.}~\bibnamefont {Henriksen}}, \bibinfo {author}
  {\bibfnamefont {E.~A.}\ \bibnamefont {Quintero}}, \bibinfo {author}
  {\bibfnamefont {C.~R.}\ \bibnamefont {Harris}}, \bibinfo {author}
  {\bibfnamefont {A.~M.}\ \bibnamefont {Archibald}}, \bibinfo {author}
  {\bibfnamefont {A.~H.}\ \bibnamefont {Ribeiro}}, \bibinfo {author}
  {\bibfnamefont {F.}~\bibnamefont {Pedregosa}}, \bibinfo {author}
  {\bibfnamefont {P.}~\bibnamefont {{van Mulbregt}}},\ and\ \bibinfo {author}
  {\bibnamefont {{SciPy 1.0 Contributors}}},\ }\bibfield  {title} {\enquote
  {\bibinfo {title} {{{SciPy} 1.0: Fundamental algorithms for scientific
  computing in python}},}\ }\href {https://doi.org/10.1038/s41592-019-0686-2}
  {\bibfield  {journal} {\bibinfo  {journal} {Nat. Methods}\ }\textbf {\bibinfo
  {volume} {17}},\ \bibinfo {pages} {261--272} (\bibinfo {year}
  {2020})}\BibitemShut {NoStop}%
\bibitem [{\citenamefont {Cheng}\ \emph {et~al.}(2023)\citenamefont {Cheng},
  \citenamefont {Chen}, \citenamefont {Zhang},\ and\ \citenamefont
  {Zhang}}]{cheng2023error}%
  \BibitemOpen
  \bibfield  {author} {\bibinfo {author} {\bibfnamefont {L.}~\bibnamefont
  {Cheng}}, \bibinfo {author} {\bibfnamefont {Y.-Q.}\ \bibnamefont {Chen}},
  \bibinfo {author} {\bibfnamefont {S.-X.}\ \bibnamefont {Zhang}},\ and\
  \bibinfo {author} {\bibfnamefont {S.}~\bibnamefont {Zhang}},\ }\bibfield
  {title} {\enquote {\bibinfo {title} {Error-mitigated quantum approximate
  optimization via learning-based adaptive optimization},}\ }\href@noop {}
  {\bibfield  {journal} {\bibinfo  {journal} {arXiv preprint arXiv:2303.14877}\
  } (\bibinfo {year} {2023})}\BibitemShut {NoStop}%
\bibitem [{\citenamefont {Mih{\'a}likov{\'a}}\ \emph
  {et~al.}(2022)\citenamefont {Mih{\'a}likov{\'a}}, \citenamefont {Pivoluska},
  \citenamefont {Plesch}, \citenamefont {Fri{\'a}k}, \citenamefont {Nagaj},\
  and\ \citenamefont {{\v{S}}ob}}]{mihalikova2022cost}%
  \BibitemOpen
  \bibfield  {author} {\bibinfo {author} {\bibfnamefont {I.}~\bibnamefont
  {Mih{\'a}likov{\'a}}}, \bibinfo {author} {\bibfnamefont {M.}~\bibnamefont
  {Pivoluska}}, \bibinfo {author} {\bibfnamefont {M.}~\bibnamefont {Plesch}},
  \bibinfo {author} {\bibfnamefont {M.}~\bibnamefont {Fri{\'a}k}}, \bibinfo
  {author} {\bibfnamefont {D.}~\bibnamefont {Nagaj}},\ and\ \bibinfo {author}
  {\bibfnamefont {M.}~\bibnamefont {{\v{S}}ob}},\ }\bibfield  {title} {\enquote
  {\bibinfo {title} {The cost of improving the precision of the variational
  quantum eigensolver for quantum chemistry},}\ }\href@noop {} {\bibfield
  {journal} {\bibinfo  {journal} {Nanomater.}\ }\textbf {\bibinfo {volume}
  {12}},\ \bibinfo {pages} {243} (\bibinfo {year} {2022})}\BibitemShut
  {NoStop}%
\end{thebibliography}%

\clearpage

\end{document}